\documentclass[10pt]{article}

\usepackage[table]{xcolor} 
\usepackage[T1]{fontenc}

\usepackage{authblk}
\usepackage{orcidlink}

\usepackage{algorithm,algorithmic}
\usepackage[ruled,vlined,algo2e,linesnumbered]{algorithm2e} 
\usepackage{amsfonts}
\usepackage{amsmath}
\usepackage{amssymb}
\usepackage{amsthm}
\usepackage{fontawesome5}
\usepackage{arcs}
\usepackage[small]{caption}
\usepackage{cases}
\usepackage{cite}
\usepackage{color}
\usepackage{enumitem}
\usepackage{float}
\usepackage{fullpage}
\usepackage{graphicx}
\usepackage{hyperref}
\usepackage{lscape}
\usepackage{marvosym}
\usepackage{microtype}
\usepackage{multirow}
\usepackage{pdflscape}
\usepackage{pgfplots} 
\usepackage{pifont}
\usepackage{rotating}
\usepackage{subcaption}
\usepackage{tcolorbox}
\usepackage{tikz}
\usepackage{url}

\pgfplotsset{compat=1.15} 
\usetikzlibrary{shapes}

\newcommand{\etal}{{et~al.}}

\definecolor{ffffff}{RGB}{ 255, 255, 255 }
\definecolor{264653}{RGB}{ 38, 70, 83 }
\definecolor{1789a6}{RGB}{ 23, 137, 166 }
\definecolor{cccccc}{RGB}{ 204, 204, 204 }
\definecolor{e76f51}{RGB}{ 231, 111, 81 }
\definecolor{33c1b1}{RGB}{ 51, 193, 177 }

\definecolor{FEFEF6}{RGB}{ 254, 254, 246 }
\definecolor{bbbbbb}{RGB}{ 187, 187, 187 }
\definecolor{000000}{RGB}{ 0, 0, 0 }
\definecolor{ff0000}{RGB}{ 255, 0, 0 }
\definecolor{0000ff}{RGB}{ 0, 0, 255 }
\definecolor{dddddd}{RGB}{ 221, 221, 221 }
\definecolor{f7b267}{RGB}{ 247, 178, 103 }

\SetKw{Continue}{continue}
\SetKwRepeat{Do}{do}{while}

\makeatletter
\DeclareRobustCommand{\iscircle}{\mathord{\mathpalette\is@circle\relax}}
\newcommand\is@circle[2]{%
	\begingroup
	\sbox\z@{\raisebox{\depth}{$\m@th#1\bigcirc$}}%
	\sbox\tw@{$#1\square$}%
	\resizebox{!}{\ht\tw@}{\usebox{\z@}}%
	\endgroup
}
\makeatother

\graphicspath{{./resources/graphics/}{./spanner_tables2/}{./small_exp_plots/}{./sfcomp/}}

\begin{document}

\title{\text{Bounded-degree plane geometric spanners in practice}\thanks{Research supported by the University of North Florida Academic Technology Grant and NSF Award CCF-1947887. Visualizations of some of the implemented algorithms were demonstrated at the 37th International Symposium on Computational Geometry (SoCG 2021) in Media Exposition.}}

\author{Frederick Anderson\,\orcidlink{0000-0002-0674-7223}}
\author{Anirban Ghosh\,\orcidlink{0000-0003-0130-5968}}
\author{Matthew Graham\,\orcidlink{0000-0002-7742-0834}}
\author{\hspace{40pt}\\\vspace*{-12pt}Lucas Mougeot\,\orcidlink{0000-0002-4598-3704}}
\author{David Wisnosky\,\orcidlink{0000-0001-5463-1949}}
\affil{School of Computing\\University of North Florida\\Jacksonville, FL, USA\\
	{\small \textcolor{black}{\faIcon{envelope}} \texttt{\{n01451351,~anirban.ghosh,~n00612546,~n01398041,~n01153911\}@unf.edu}}
}

\date{}

\providecommand{\keywords}[1]{\textbf{{Keywords ---}} #1}

\maketitle

\vspace{-40pt}
\begin{abstract}
	The construction of bounded-degree plane geometric spanners has been a focus of interest  since 2002 when Bose, Gudmundsson, and Smid proposed the first algorithm to construct such spanners. 
	To date, eleven algorithms have been designed with various trade-offs in degree and stretch factor. We have implemented these sophisticated algorithms in \textsf{C}\texttt{++} using the  \textsf{CGAL} library and experimented with them using large synthetic and real-world pointsets. Our  experiments  have revealed their practical behavior and real-world efficacy. 
	We share the implementations via \textsf{GitHub}\footnote{\url{https://github.com/ghoshanirban/BoundedDegreePlaneSpannersCppCode}} for broader uses and future research.

	We present a simple practical algorithm, named \textsc{AppxStretchFactor}, that can estimate  stretch factors (obtains  lower bounds on the exact stretch factors) of geometric spanners -- a challenging problem for which no practical algorithm is known yet. In our experiments with bounded-degree plane geometric spanners, we find that \textsc{AppxStretchFactor} estimates stretch factors almost precisely. Further, it gives linear runtime performance in practice for the pointset distributions considered in this work, making it much faster than the naive Dijkstra-based algorithm for calculating stretch factors. 
	
\end{abstract}

\section{Introduction}
\label{chap:introduction}

\begin{figure}
	\begin{center}
\includegraphics[scale=0.5]{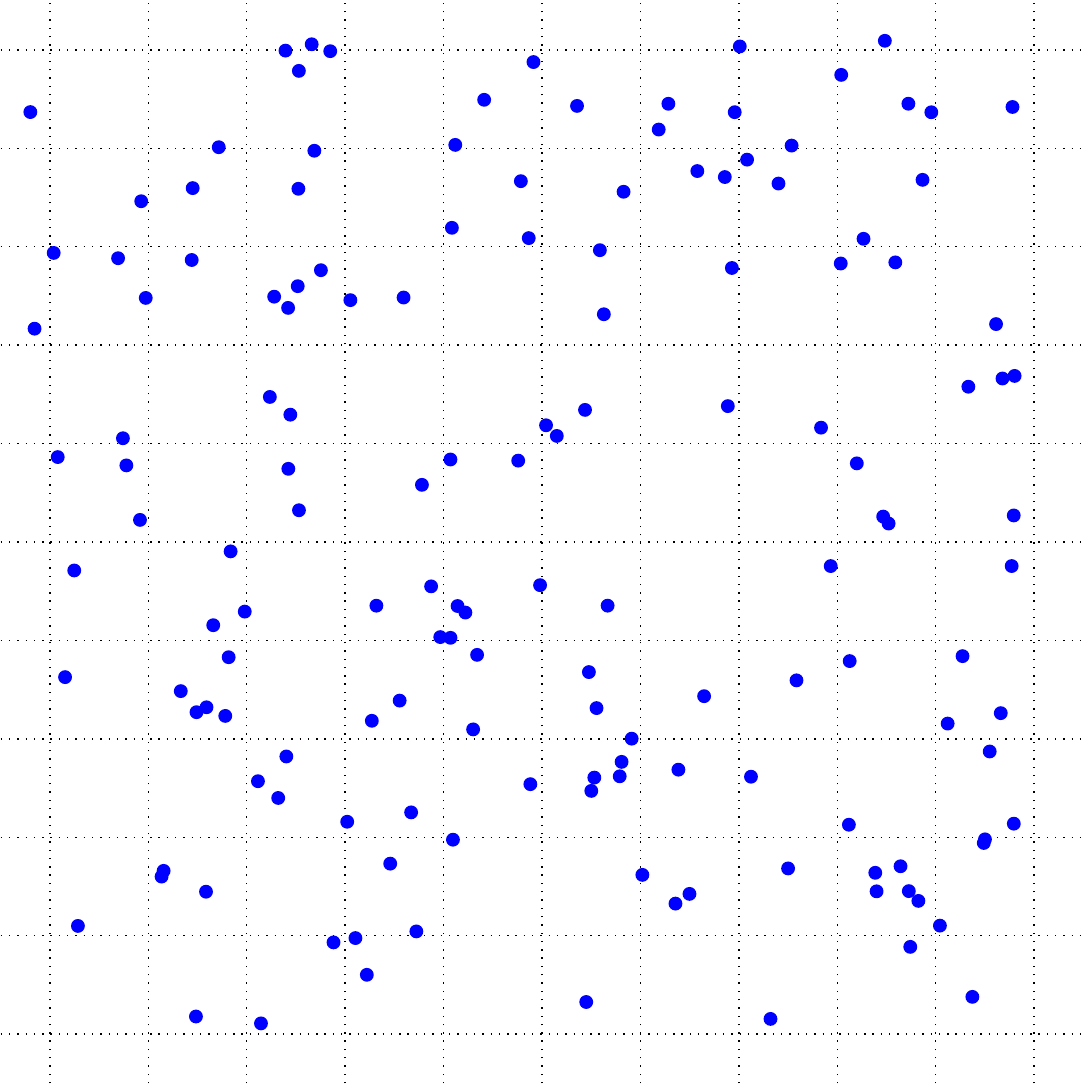} \hspace{20pt}
\includegraphics[scale=0.5]{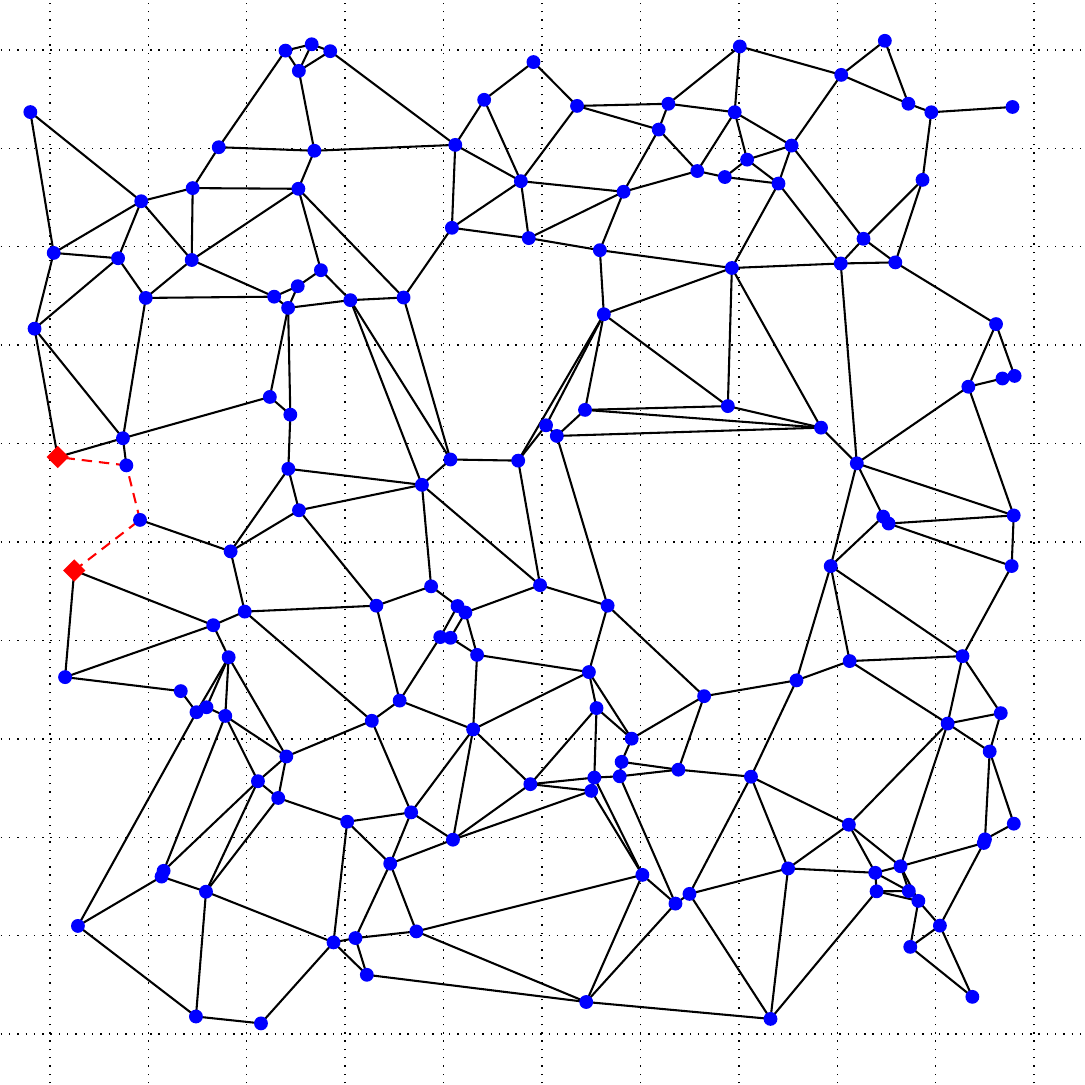} 
	\end{center}
\caption{Left: A set $P$ of $150$ points, generated randomly within a square. Right: A plane degree-$6$  spanner on $P$ with stretch factor $\approx 1.82$. The pair of points for which the spanner achieves a stretch factor of $\approx 1.82$ is shown in red along with the shortest path between them.}
\label{fig: example}
\end{figure}

Let $G$ be the complete Euclidean graph on a given set $P$ of $n$ points embedded in the Euclidean plane. A \emph{geometric $t$-spanner} on $P$ is a geometric graph $G':=(P,E)$, a subgraph of $G$ such that for every pair of points $u,v\in P$, the distance between them in $G'$ (the Euclidean length of a shortest path between $u,v$ in $G'$) is at most $t$ times their Euclidean distance $|uv|$, for some $t\geq 1$. It is easy to check that $G$ itself is a $1$-spanner with $\Theta(n^2)$ edges. The quantity $t$ is referred to as the \emph{stretch factor} of $G'$. If there is no need to specify $t$, we simply use the term \emph{geometric spanner} and assume that there exists some $t$ for $G'$. We say that $G'$ is \emph{plane} if it is crossing-free.  $G'$ is \emph{degree-$k$} or is said to have \emph{bounded-degree} if its degree is at most $k$. In this work, we experiment with bounded-degree plane geometric spanners. See Figure~\ref{fig: example} for an example of such a  spanner.

Bounded-degree plane geometric spanners have been an area of interest in computational geometry for a long time. Non-crossing edges make them suitable for wireless network applications where edge crossings create communication interference. The absence of crossing edges also makes them useful for the design of road networks to eliminate high-budget flyovers. Such spanners have $O(n)$ edges (at most $3n-6$ edges); as a result, they are less expensive to store and navigate. Further, shortest-path algorithms run quicker on them since they are sparse. Bounded-degree helps in reducing on-site equipment costs. 

Bose, Gudmundsson, and Smid~\cite{bose2005constructing} were the first to show that there always exists a plane geometric $\sigma(\pi+1)$-spanner of degree at most $27$ on any pointset, where $\sigma$ denotes an upper bound for the stretch factor of $L_2$-Delaunay triangulations\footnote{A triangulation $T$ for a pointset $P$ is referred to as a $L_2$-Delaunay triangulation  if no point in $P$ lies inside the circumcircle of any triangle in $T$.} (the current best known value is $\sigma=1.998$ due to Xia~\cite{xia2013stretch}).
This result was subsequently improved
in a long series of papers~\cite{li2004efficient, bose2009delaunay, kanj2010spanners, bonichon2010plane,  bose2012bounded, kanj2012improved, bose2018improved} in terms of degree and/or stretch factor. Bonichon et al.~\cite{bonichon2015there} reduced the degree to $4$ with $t \approx 156.8$. Soon after this, Kanj et al. improved this stretch factor upper bound to $20$  in~\cite{kanj2017degree}. A summary of these results is presented in Table~\ref{tab:summary}. 
This family of spanner construction algorithms has turned out to be a fascinating application of the Delaunay triangulation.  Note that all these algorithms produce bounded-degree plane subgraphs of the complete Euclidean graph on $P$ with constant stretch factors.

\begin{table}[htb]

	\centering
	\footnotesize
	\begin{tabular}{|c|c|c|}
		\hline
		\textsc{Reference}                                                  & \textsc{Degree}                                                                       & \textsc{Stretch factor} \\
		\hline
		\hline
		{ Bose, Gudmundsson, and Smid}~\cite{bose2005constructing}          &
		$27$                                                                & $\sigma(\pi+1)\approx 8.3$                                                                                      \\
		\hline
		{ Li and Wang}~\cite{li2004efficient}                               &
		$23$                                                                & $\sigma(1 + \frac{\pi}{\sqrt{2}})\approx 6.4$                                                                   \\
		\hline
		{ Bose, Smid, and Xu}~\cite{bose2009delaunay}                       &
		$17$                                                                & $\sigma(2+2\sqrt{3}+\frac{3\pi}{2} + 2\pi \sin \frac{\pi}{12}) \approx 23.6$                                    \\
		\hline
		{ Kanj,  Perkovi{\'c}, and Xia}~\cite{kanj2010spanners}             &
		$14$                                                                & $\sigma (1 + \frac{2\pi}{14 \cos (\pi/14)})\approx 2.9$                                                         \\
		\hline
		{ Kanj and Xia}~\cite{kanj2012improved}                             &
		$11$                                                                & $\sigma (\frac{2\sin (2\pi/5)\cos (\pi/5)}{2\sin (2\pi/5)\cos (\pi/5)-1})\approx 5.7$                           \\
		\hline
		{Bose,  Hill, and Smid}~\cite{bose2018improved}                     &
		$8$                                                                 & $\sigma \left( 1 + \frac{2\pi}{6 \cos (\pi/6) } \right) \approx 4.4$                                            \\
		\hline
		{Bose,  Carmi, and Chaitman-Yerushalmi}~\cite{bose2012bounded}                               &
		$7$                                                                 & $\sigma (1 + \sqrt{2})^2 \approx 11.6$                                                                          \\
		\hline
		{Bose,  Carmi, and Chaitman-Yerushalmi}~\cite{bose2012bounded}                               &
		$6$                                                                 & $\sigma\left(\frac{1}{1-\tan (\pi/7)(1 + 1/\cos (\pi/14))} \right)\approx 81.7$                                 \\
		\hline
		{Bonichon, Gavoille, Hanusse, and Perkovi{\'c}}~\cite{bonichon2010plane}                         &
		$6$                                                                 & $6$                                                                                                             \\
		\hline
		{ Bonichon,  Kanj,  Perkovi{\'c}, and Xia}~\cite{bonichon2015there} &
		$4$                                                                 & $\sqrt{4+2\sqrt{2}}(19+29\sqrt{2})\approx 156.8$                                                                \\
		\hline
		{ Kanj,  Perkovi{\'c}, and T{\"u}rkoǧlu}~\cite{kanj2017degree}      &
		$4$                                                                 & $20$                                                                                                            \\
		\hline
		
	\end{tabular}

\caption{A summary of results on constructions of bounded-degree plane geometric spanners, sorted by the degree they guarantee.  The best known upper bound of $\sigma=1.998$ for the stretch factor of the $L_2$-Delaunay triangulation~\cite{xia2013stretch} is used in this table for expressing the stretch factors. 
}
	\label{tab:summary}
\end{table}

The intriguing question that remains to be answered is whether the degree can be reduced to $3$ while keeping $t$ bounded; refer to~\cite[Problem 14]{bose2013plane} and~\cite[Chapter 32]{toth2017handbook}.
Interestingly, if one does not insist on constructing a \textit{plane} spanner, Das and Heffernan~\cite{das1996constructing} showed that degree 3 is always achievable. Narasimhan and Smid ~\cite[Section 20.1]{narasimhan2007geometric} show that no degree-2 plane spanner of the infinite integer lattice can have a constant stretch factor.  Thus, a minimum degree of 3 is necessary to achieve a constant stretch factor. If $P$ is convex, then it is always possible to construct a degree-3 plane geometric spanners, see~\cite{biniaz2017towards,kanj2017degree,bakhshesh2020degree}.
From the other direction, lower bounds on the stretch factors of plane spanners
for finite pointsets have been investigated in~\cite{dumitrescu2016lower, dumitrescu2016lattice, klein2015most, mulzer2004minimum}. In-browser visualizations of some of the algorithms (those based on the $L_2$-Delaunay triangulation) have been recently presented in~\cite{anderson2021interactive}. In related works, the construction of plane hop spanners (where the number of hops in shortest paths is of interest) for unit disk graphs has been considered in~\cite{catusse2010planar, biniaz2020plane, dumitrescu2020sparse}.

The most notable experimental work for geometric spanners is done by Farshi and Gudmundsson~\cite{farshi2009experimental}. The authors engineered and experimented with some of the well-known geometric spanners construction algorithms published before 2009. However, the authors did not use the algorithms considered in this work in their experiments. Planarity and bounded-degree are important concerns in geometric network design. Hence, we found it motivating to implement the eleven algorithms (refer to Table~\ref{tab:summary}) meant to construct bounded-degree plane geometric spanners. Asymptotic runtimes and various theoretical bounds do not always do justice in explaining the real-world performance of algorithms, especially in computational geometry, because of heavy floating-point operations needed for various geometric calculations. Experiments reveal their real-world performance. A unique aspect of the family of bounded-degree plane spanner construction algorithms is that users cannot specify an arbitrary value of $t$ and/or degree for spanner construction. It is a deviation from many standard spanner algorithms; see~\cite{narasimhan2007geometric,bose2012bounded} for a review of such algorithms. This makes experiments with them even more interesting.

\textbf{Our contributions.} (i) We experimentally compare the aforementioned eleven bounded-degree plane spanner construction algorithms by implementing them carefully in \textsf{C}\texttt{++} using the popular \textsf{CGAL} library~\cite{cgal:eb-21b} and running them on large synthetic and real-world pointsets. The largest pointset contains $1.9$ million points approximately. For broader uses of these sophisticated algorithms, we share the \textsf{C}\texttt{++} implementations via \textsf{GitHub}\footnote{\url{https://github.com/ghoshanirban/BoundedDegreePlaneSpannersCppCode}}.  The comparisons are performed based on their runtimes and degree, stretch factor, and lightness of the generated spanners. We present a brief overview of the algorithms implemented and our experimental results in  Sections~\ref{chap:algorithms} and \ref{chap:experiments}, respectively.\\
(ii) In doing experiments with spanners, we find that stretch factor measurement turns out to be a severe bottleneck when $n$ is large. To address this, we have designed a new fast algorithm named \textsc{AppxStretchFactor} that can estimate the stretch factor of a given spanner (not necessarily plane). In our experiments, we find that it can estimate stretch factors with high accuracy for the class of geometric spanners dealt with in this work. It is considerably faster than the naive Dijkstra-based exact stretch factor measurement algorithm in practice. To our knowledge, no such practical algorithm exists in the literature. Further, it can be easily parallelized, making it very useful for estimating stretch factors of large spanners. See Section~\ref{chap:stretchfactor} for a description of this algorithm.

\section{Algorithms implemented}
\label{chap:algorithms}

Every algorithm designed to date for constructing bounded-degree plane geometric spanners relies on some variant of Delaunay triangulation. The rationale behind this is that such triangulations are geometric spanners~\cite{xia2013stretch, chew1989there, chew1986there, bonichon2012stretch} and are plane by definition. As a result, the family of plane spanner construction algorithms considered in this work has turned out to be a fascinating application of Delaunay triangulation. It is essential to know that  Delaunay triangulations have unbounded degrees and cannot be used as bounded-degree plane spanners.

In this section, we provide a brief description for each of the eleven algorithms considered in this work.  Appropriate abbreviations using the authors’ names and dates of publication are used for naming purposes. Since most of these algorithms are involved, so we urge the reader to refer to the original papers for a deeper understanding and correctness proofs.  For visualizing some of these algorithms, we recommend the interactive in-browser applet\footnote{\url{https://ghoshanirban.github.io/bounded-degree-plane-spanners/index.html}} developed by us; refer to~\cite{anderson2021interactive}. To observe variations in spanner construction between the algorithms see Appendix~\ref{chap:samples}.

In these algorithms, the surrounding of every input point is frequently divided into multiple cones (depending on the algorithm) for carefully selecting edges from the Delaunay triangulation used as the starting point. In our pseudocodes, the cone $i$ of point $u$, considered clockwise, is denoted by $C_i^u$. A triangulation $T$ of a pointset $P$ is said to be a $L_2$-Delaunay triangulation of $P$ if no point in $P$ lies inside the circumcircle of any triangle in $T$. Eight of the eleven algorithms use $L_2$-Delaunay triangulation as the starting point. The remaining three use $L_\infty$ or $TD$-Delaunay triangulations, as described later in this section. In the following, $n$ denotes the size of the input pointset.

\vspace*{-5pt}

\begin{itemize}[leftmargin=*]\itemsep0pt
	\item \texttt{BGS05}: \textbf{Bose, Gudmundsson, and Smid}~\cite{bose2005constructing}. This was the first algorithm  that can construct bounded-degree plane spanners using the classic $L_2$-Delaunay triangulation. 
	First, a Delaunay triangulation $DT$ of $P$ is constructed. Next, a degree-$3$ spanning subgraph $SG$ of $DT$ is computed that contains the convex hull of $P$ and is a (possibly degenerate) simple polygon with $P$ as its vertex set. The polygon is then transformed into a simple non-degenerate polygon $Q$. The vertices of $Q$ are processed in an order that is obtained from a breadth-first order of $DT$,  and then  Delaunay edges are carefully added to $Q$.  The resulting graph denoted $G'$, is a plane spanner for the vertices of $Q$. Refer to Algorithm~\ref{code:BSG2005} for a pseudocode of this algorithm. The authors show that their algorithm generates degree-$27$ plane spanners with a stretch factor of $1.998(\pi+1) \approx 8.3$ and runs in $O(n\log n)$ time.
	
	\vspace{-5pt}
	
	\begin{algorithm2e}
		{\small 	\SetAlgoLined
			Allocate $\Phi[1,\ldots,n]$\; 
			Make a copy of $DT$ and call it $H$\;
			Let \texttt{reserved} be a set of two consecutive vertices $v_1,v_2$ on the convex hull of $H$\;
			$\Phi_1 \gets v_1, \Phi_2 \gets v_2$\;
			
			\For{$i=1$ \KwTo $n-2$}{
				Let $u$ be a vertex of the outer face of $H \setminus \texttt{reserved}$ that is adjacent to at most two other vertices on the outer face\; 
				$\Phi_{u} \gets n-i+1$\;
				Remove $u$ and all incident edges from $H$\;
			}
			\Return $\Phi$\;}
		\caption{{\small \texttt{CanonicalOrdering}($DT$)}}
		\label{CanonicalOrdering}
	\end{algorithm2e}

\begin{algorithm2e}
	{\small \SetAlgoLined
	$\Phi[1,\ldots,n] \gets \texttt{CanonicalOrdering}(DT)$ (Algorithm~\ref{CanonicalOrdering})\;
	$SG \leftarrow \emptyset$\;
	Add edges between $v_1, v_2, v_3 \in \Phi$ to $SG$ and mark the vertices as \texttt{done}\;
	\For{$v_i \in \Phi\setminus \{v_1,v_2,v_3\}$ }{
		Let $u_1, ..., u_k$ be the vertices neighboring $v_i$ in $DT$ marked as \texttt{done}\;
		Remove edge $\{u_1,u_2\}$ from $SG$\;
		Add edges $\{v_i, u_1\}$ and $\{v_i, u_2\}$ to $SG$\;
		\If{$k>2$}{
			Remove edge $\{u_{k-1}, u_k\}$ from $SG$\;
			Add edge $\{v_i, u_k\}$ to $SG$\;
		}
	}
	
	\Return $SG$\;}
	\caption{{\small \texttt{SpanningGraph}($DT$})}
			\label{SpanningGraph}
\end{algorithm2e}

\vspace{-9pt}

\begin{algorithm2e}
{\small 	\SetAlgoLined
	$V \gets \emptyset, E \gets \emptyset$\;
	Let $s_1,v_1$ be two consecutive vertices on the convex hull of $SG$ in counterclockwise order\;
	$v_{prev} \gets s_1, v_i \gets v_1$\;
	Add $v_{prev}$ to $V$\;
	
	\Do{$v_{prev} \neq s_1$ \textup{\textbf{and}} $v_i \neq v_1$}{
		Add $v_i$ to $V$\;
		Add $\{v_i,v_{prev}\}$ to $E$\;
		Let $v_{next}$ be the neighbor of $v_i \in SG$ such that $v_{next}$ is the next neighbor clockwise from $v_{prev}$\;
		$v_{prev} \gets v_i, v_i \gets v_{next}$\;
	}
	$E=E \cup \{\{v_i,v_{prev}\}\} \cup DT \setminus SG$\;
	\Return $\left(V,E\right)$\;}
	\caption{{\small \texttt{TransformPolygon}($SG, DT$)}}
		\label{TransformPolygon}
\end{algorithm2e}

\vspace{-2pt}

\begin{algorithm2e}
	{\small \SetAlgoLined
		Let $V,E$ be the vertices and edges of $Q$, respectively\;
		Let $\rho[1,\ldots,n]$ be the breadth-first ordering of $V$ in $Q$, starting at any vertex in $V$\;
		$E' \gets SG$\;

		\ForEach{$u \in \rho$}{
			Let $s_1,s_2,...,s_m$ be the clockwise ordered neighbors of $u$ in $Q$\;
			$s_j,s_k \gets s_m$\;
			\If{$u \neq \rho_1$}{
				Set $s_j$ and $s_k$ to the first and last vertex in the ordered neighborhood of $u$ where $s_j,s_k \in E'$\;
			}
			Divide $\angle s_1 u s_j$ and $\angle s_k u s_m$ into an minimum number of cones with maximum angle $\pi/2$\;
			In each cone, add the shortest edge in $E$ incident upon $u$\ to $E'$ and  all edges  $\{s_\ell,s_{\ell+1}\}$ such that $ 1 \leq \ell < j$ \textbf{\textup{or}} $k \leq \ell < m$\;
		}
		
		\Return $E'$\;}
	\caption{{\small \texttt{PolygonSpanner}($Q,SG$)}}
	\label{PolygonSpanner}
\end{algorithm2e}


\begin{algorithm2e}
	{\small 	\SetAlgoLined
		$DT \gets \texttt{$L_2$-DelaunayTriangulation}(P)$\;
		$SG \gets \texttt{SpanningGraph}(DT)$  (Algorithm~\ref{SpanningGraph})\;
		$Q \gets \texttt{TransformPolygon}(SG, DT)$  (Algorithm~\ref{TransformPolygon})\;
		$G' \gets \texttt{PolygonSpanner}(Q, SG)$  (Algorithm~\ref{PolygonSpanner})\;
		
		\Return $G'$\;
		\caption{{\small \texttt{BGS05}($P$)}}
		\label{code:BSG2005}}
\end{algorithm2e}

\item \texttt{LW04}: \textbf{Li and Wang}~\cite{li2004efficient}. This algorithm is inspired from \texttt{BSG2005} but is a lot simpler and avoids the use of intermediate (possibly degenerate) polygons. The algorithm computes a reverse low degree ordering of the vertices of the $L_2$-Delaunay triangulation $DT$ constructed on $P$. Then it sequentially considers the vertices in this ordering, divides the surrounding of every such vertex into multiple cones, and then adds short edges from $DT$ to preserve planarity.   Refer to Algorithm~\ref{code:LW04} for a pseudocode of this algorithm. The authors have shown that this algorithm generates degree-$23$ plane spanners (when the input parameter $\alpha$ of this algorithm is set to $\pi/2$) having a stretch factor of $1.998\left(1 + \pi/\sqrt{2} \right)\approx 6.4$ and runs in $O(n\log n)$ time.

\begin{algorithm2e}
	
{\small 	Allocate $\Phi[1\ldots n]$\; 
	Make a copy of $DT$ and call it $H$\;
	\For{$i=1$ \KwTo $n$}{
		Let $u$  be a vertex in $H$ with minimal degree\;
		$\Phi[u] \gets n-i+1$\;
		Remove $u$ and all incident edges from $H$\;
	}
	
	\Return $\Phi$\;}
	\caption{{\small \texttt{ReverseLowDegreeOrdering}($DT$)}}
	\label{code:ReverseLowDegreeOrdering}
\end{algorithm2e}

\begin{algorithm2e}
	
	{\small 	$DT \gets \texttt{$L_2$-DelaunayTriangulation}(P)$\;
		$\Phi[1\ldots n] \gets \texttt{ReverseLowDegreeOrdering}(DT)$ (Algorithm~\ref{code:ReverseLowDegreeOrdering})\;
		$E \gets \emptyset$\;
		
		\ForEach{$u \in \Phi$}{
			Divide the area surrounding $u$ into sectors delineated by $u$'s already \texttt{processed} neighbors in $DT$\;
			Divide each sector into a minimum number of cones $C^0_u,C^1_u,\ldots$ with angle at most $\alpha$\;
			\ForEach{$C^i_u$}{
				Let $v_1, v_2, ..., v_m$ be the clockwise-ordered Delaunay neighbors of $u$ in $C^i_u$\;
				Add an edge between $u$ and the closest of such neighbors to $E$\;
				Add all edges $\{v_j,v_{j+1}\}$ such that $1 \leq j < m$ to $E$\;
			}
			Mark $u$ as \texttt{processed}\;
		}
		
		\Return $E$\;}
	\caption{{\small \texttt{LW04}($P,\ 0 < \alpha \leq \pi/2$})}
	\label{code:LW04}
\end{algorithm2e}

\item \texttt{BSX09}: \textbf{Bose, Smid, and Xu}~\cite{bose2009delaunay}. This algorithm is quite similar to \texttt{LW04} in design and also relies on reverse low-degree ordering of the vertices of the Delaunay triangulation. Refer to Algorithm~\ref{code:BSX09}. The authors have generalized their algorithm so that it can construct bounded-degree plane spanners from any triangulation of $P$, not necessarily just the $L_2$-Delaunay triangulation (although the $L_2$-Delaunay triangulation is of primary interest to us). When the $L_2$-Delaunay triangulation is used and the parameter $\alpha$  is set to $2\pi/3$, the algorithm generates degree-$17$ plane spanners having a stretch factor of $\sigma(2+2\sqrt{3}+\frac{3\pi}{2} + 2\pi \sin \frac{\pi}{12}) \approx 23.6$ in $O(n\log{n})$ time. After computing the triangulation and the reverse low-degree ordering, at every vertex $u$,  $\delta=\lceil 2\pi/\alpha \rceil$ Yao cones are initialized such that the closest unprocessed triangulation neighbor falls on a cone boundary and occupies both cones as the \textit{short} edge, which is added to the spanner. In the remaining cones, the closest unprocessed neighbor of $u$ in each cone is added. In all cones,  special edges between pairs of neighbors of $u$ are added to the spanner if both the neighbors are unprocessed.

\begin{algorithm2e}
	{\small 	\SetAlgoLined
		$DT \gets \texttt{$L_2$-DelaunayTriangulation}(P)$\;
		$\Phi[1\ldots n] \gets \texttt{ReverseLowDegreeOrdering}(DT)$ (use Algorithm~\ref{code:ReverseLowDegreeOrdering})\;
		$E \gets \emptyset$\;
		
		\ForEach{$u \in \Phi$}{
			\If{$u$ \textup{has \texttt{unprocessed} Delaunay neighbors}}{
				Let $v_{closest}$ be the closest \texttt{unprocessed} neighbor to $u$\;
				Add the edge $\{u,v_{closest}\}$ to $E$\;
				Divide the area surrounding $u$ into a minimum number of cones $C^0_u, C^1_u,\ldots$ with angle at most $\alpha$, such that $v_{closest}$ is on the boundary between the first and last cones\;
				
				\ForEach{\textup{$C^i_u$ except the first and last}}{
					
					\If{\textup{$u$ has \texttt{unprocessed} neighbors in $C^i_u$}}{
						Let $w$ be the closest \texttt{unprocessed} neighbor to $u$ in the cone\;
						Add edge $\{u,w\}$ to $E$\;
					}
					Let $v_0, v_1, ..., v_{m-1}$ be the clockwise-ordered neighbors of $u$\;
					Add all edges $\{v_j,v_{(j+1) \bmod m}\}$ to $E$ such that $0 \leq j < m$ and $v_j, v_{(j+1)\bmod m}$ are \texttt{unprocessed}\;
				}
			}
			Mark $u$ as \texttt{processed}\;
		}
		
		\Return $E$;}
	\caption{{\small \texttt{BSX09}($P,\ 0<\alpha\leq 2\pi / 3$)}}
	\label{code:BSX09}
\end{algorithm2e}

\item \texttt{BGHP10}: \textbf{Bonichon, Gavoille, Hanusse, and Perkovi{\'c}}~\cite{bonichon2010plane}.  
It was the first algorithm that deviated from the use of $L_2$-Delaunay triangulation; instead, it used $TD$-Delaunay triangulation to select  spanner edges, introduced by Chew back in $1989$~\cite{chew1989there}. For such triangulations, empty equilateral triangles are used for characterization instead of empty circles, as needed in the case of $L_2$-Delaunay triangulations. $TD$-Delaunay triangulations are plane $2$-spanners but may have an unbounded degree. \texttt{BGHP10} first extracts a degree-$9$ subgraph from the $TD$-Delaunay triangulation that has a stretch factor of $6$. Then using some local modifications, the degree is reduced from $9$ to $6$ but the stretch factor remains unchanged. Refer to Algorithm~\ref{code:BGHP10}. It uses internally Algorithms~\ref{irelevant} - \ref{charge}. In this algorithm, each edge incident to a node is charged to some cone of that node. The algorithm runs in $O(n\log n)$ time, as shown by the authors. 

\item \texttt{KPX10}: \textbf{Kanj,  Perkovi{\'c}, and Xia}~\cite{kanj2010spanners}.  
For every vertex $u$ in the $L_2$-Delaunay triangulation, its surrounding is divided into $k \geq 14$ cones. In every nonempty cone of $u$, the shortest Delaunay edge incident on $u$ is selected. After this, a few additional Delaunay edges are also selected using some  criteria based on cone sequences. See Algorithm~\ref{code:KPX10} for a complete description of this algorithm with the technical details. When $k$ is set to $14$, degree-$14$ plane spanners are generated having a stretch factor of $1.998 \left(1 + \frac{2\pi}{14 \cos (\pi/14)}\right) \approx 2.9$. Note that out of the $11$ algorithms we have implemented in this work, this algorithm gives the best theoretical guarantee on the stretch factor; see Table~\ref{tab:summary}. \texttt{KPX10} runs in $O(n\log n)$ time.

\begin{algorithm2e}
	{\small 	\SetAlgoLined
		$DT \gets \texttt{TD-DelaunayTriangulation}(P)$\;
		$E \gets \emptyset$\;
		
		\ForEach{\textup{nonempty cone $i$ of vertex }$u \in DT$ \textup{where} $i \in \{1,3,5\}$ }{
			Add edge $\{u,\texttt{closest}(u,i)\}$ to $E$\;
			$\texttt{charge}(u,i)\gets \texttt{charge}(u,i)+1$\;
			$\texttt{charge}(\texttt{closest}(u,i),i+3)) \gets \texttt{charge}(\texttt{closest}(u,i),i+3)+1$\;
			
			\If{\textup{\texttt{first}$(u,i) \neq$ \texttt{closest}($u,i$) $\land$ \texttt{i-relevant}(\texttt{first}$(u,i),u,i-1$)}}{
				Add edge $u,\texttt{first}(u,i)$ to $E$\;
				$\texttt{charge}(u,i-1) \gets \texttt{charge}(u,i-1)+1$\;
			}
			\If{\textup{\texttt{last}$(u,i) \neq$ \texttt{closest}($u,i$) $\land$ \texttt{i-relevant}(\texttt{last}$(u,i),u,i+1$)}}{
				
				Add edge $\{u, \texttt{last}(u,i)\}$ to $E$\;
				$\texttt{charge}(u,i+1)\gets \texttt{charge}(u,i+1)+1$\;
			}
		}
		
		\ForEach{\textup{cone $i$ of vertex $u \in DT$ \textup{where} $i \in \{0,2,4\}$, such that \texttt{i-distant}($u,i$) is true } }{
			$v_{next} \gets \texttt{first}(u,i+1)$\;
			$v_{prev} \gets \texttt{last}(u,i-1)$\;
			Add edge $\{v_{next},v_{prev}\}$ to $E$\;
			$\texttt{charge}(v_{next},i+1)\gets\texttt{charge}(v_{next},i+1)+1$\;
			$\texttt{charge}(v_{prev},i-1)\gets\texttt{charge}(v_{prev},i-1)+1$\;
			
			Let $v_{remove}$ be the vertex from ${v_{next},v_{prev}}$ where $\angle (\texttt{parent}(u,i), u, v_{remove})$ is maximized\;
			
			Remove edge $\{u,v_{remove}\}$ from $E$\;
			$\texttt{charge}(u,i)\gets \texttt{charge}(u,i)-1$\;
		}
		
		\ForEach{\textup{cone $i$ of vertex $u \in DT$ where $i \in \{0,1,\ldots,5\}$, such that \texttt{charge}($u,i$) $= 2$ $\land$ \texttt{charge}($u,i-1$) $= 1$ $\land$ \texttt{charge}($u,i+1$) $= 1$}}{
			\eIf{\textup{$u = \texttt{last}(\texttt{parent}(u,i),i)$}}{
				$v_{remove} \gets \texttt{last}(u,i-1)$\;
			}{
				$v_{remove} \gets \texttt{first}(u,i+1)$\;
			}
			Remove edge $\{u,v_{remove}\}$ from $E$\;
			$\texttt{charge}(u,i)\gets\texttt{charge}(u,i) -1$\;
		}
		
		\Return $E$\;}
	\caption{{\small \texttt{BGHP10}($P$)}}
	\label{code:BGHP10}
\end{algorithm2e}

\vspace{-3pt}
\begin{algorithm2e}
{\small 	\SetAlgoLined
	$w \gets \texttt{parent}(u,i)$\;
	\Return{\textup{$v \neq \texttt{closest}(u,i) \land v \in C^i_w$}}\;}
	\caption{{\small \texttt{i-relevant}($v,u,i$)}}
	\label{irelevant}
\end{algorithm2e}

\begin{algorithm2e}
{\small 	\SetAlgoLined
	$u \gets \texttt{parent}(w,i)$\;
	\Return{\textup{$\{w,u\} \notin E$ $\land$ \texttt{i-relevant}($\texttt{first}(w,i+1),u, i+1$) $\land$ \texttt{i-relevant}($\texttt{last}(w,i-1),u,i-1$)}}\;}
	\caption{{\small \texttt{i-distant}($w,i$)}}
\end{algorithm2e}

\begin{algorithm2e}
{\small 	\SetAlgoLined
	\Return{\textup{the closest vertex to $u$ in a even cone $i$, if it exists}}\;
}
	\caption{{\small \texttt{parent}($u,i$)}}
\end{algorithm2e}

\begin{algorithm2e}
{\small 	\SetAlgoLined
	\Return{\textup{the closest vertex to $u$ in a odd cone $i$, if it exists}}\;}
	\caption{{\small \texttt{closest}($u,i$)}}
\end{algorithm2e}

\begin{algorithm2e}
{\small 	\SetAlgoLined
	\Return{\textup{the first vertex in a odd cone $i$, if it exists}}\;}
	\caption{{\small \texttt{first}($u,i$)}}
\end{algorithm2e}

\begin{algorithm2e}
	{\small \SetAlgoLined
	
	\Return{\textup{the last vertex in a odd cone $i$, if it exists}}\;}
	\caption{{\small \texttt{last}($u,i$)}}
\end{algorithm2e}

\begin{algorithm2e}
{\small 	\SetAlgoLined
	\Return{\textup{the number of edges charged to cone $i$ of $u$}}\;}
	\caption{{\small \texttt{charge}($u,i$)}}
	\label{charge}
\end{algorithm2e}

\begin{algorithm2e}
{\small 	\SetAlgoLined
	$DT \gets \texttt{$L_2$-DelaunayTriangulation}(P)$\;
	
	\ForEach{\textup{vertex} $u \in DT$}{
		Partition the area surrounding $u$ into $k$ disjoint cones of angle $2\pi/k$\;
		In each nonempty cone, \texttt{select} the shortest edge in $DT$ incident to $u$\;
		\ForEach{\textup{maximal sequence of $\ell\geq 1$ consecutive empty cones}}{
			\eIf{$\ell>1$}{
				\texttt{select} the first $\lfloor \ell/2 \rfloor$ \texttt{unselected} incident $DT$ edges on $u$ clockwise from the sequence of empty cones and the  first  $\lceil \ell/2 \rceil$ \texttt{unselected} $DT$ edges incident on $u$ counterclockwise from the sequence of empty cones\;
			}{
				let $ux$ and $uy$ be the incident  $DT$ edges on $m$ clockwise and counterclockwise, respectively, from the empty cone\;
				
				if either $ux$ or $uy$ is selected then \texttt{select} the other edge (in case it has not been selected); otherwise \texttt{select} the shorter edge between $ux$ and $uy$ breaking ties arbitrarily;
				
			}
		}
	}
	
	\Return the $DT$ edges \texttt{selected} by both endpoints\;}
	\caption{{\small \texttt{KPX10}($P$, integer $k\geq 14$)}}
	\label{code:KPX10}
\end{algorithm2e}

\item \texttt{KX12}: \textbf{Kanj and Xia}~\cite{kanj2012improved}.  
This $O(n\log n)$-time algorithm takes a different approach in contrast with the previous ones, although it still uses the $L_2$-Delaunay triangulation $DT$ as the starting point. Every vertex $u$ in $DT$ selects at most $11$ of its incident edges in $DT$, and edges that are selected by both endpoints are kept. As  such, it is guaranteed that the degree of the resulting subgraph is at most $11$. The stretch factors of the generated spanners is shown to be at most $1.998 \left(\frac{2\sin (2\pi/5)\cos (\pi/5)}{2\sin (2\pi/5)\cos (\pi/5)-1}\right) \approx 5.7$. Refer to Algorithm~\ref{code:KX12}.

\item \texttt{BCC12-7}, \texttt{BCC12-6}: \textbf{Bose,  Carmi, and Chaitman-Yerushalmi}~\cite{bose2012bounded}.
The authors present two algorithms in their paper. Whereas previous algorithms used strategies involving iterating over the vertices one-by-one, this algorithm takes the approach of {iterating over the edges of the Delaunay triangulation in order of non-decreasing length} to query agreement among the vertices for bounding degrees. \texttt{BCC12-7}, the simpler of the two, produces $1.998 (1 + \sqrt{2})^2 \approx 11.6$-spanners with degree $7$. 
\texttt{BCC12-6}, on the other hand, constructs $11.998 \left(\frac{1}{1-\tan (\pi/7)(1 + 1/\cos (\pi/14))} \right)\approx 81.7$-spanners with degree $6$ but not all edges come from the $L_2$-Delaunay triangulation. Both these algorithms run in $O(n \log n)$ time. See Algorithm~\ref{alg:bcc}. The parameter $\Delta \in \{7,6\}$ is used to control the degree. Depending on $\Delta$, either Algorithm~\ref{alg:bcc_wedge6} or Algorithm~\ref{alg:bcc_wedge7} is invoked.

\begin{algorithm2e}
{\small 	\SetAlgoLined
	$DT \gets \texttt{$L_2$-DelaunayTriangulation}(P)$\;
	
	\ForEach{\textup{vertex} $u \in DT$}{
		In each \texttt{wide} sequence (a sequence of exactly three consecutive edges incident to a vertex whose overall angle is at least $4\pi/5$) around $u$, \texttt{select} the edges of the sequence\;
		Partition the remaining space surrounding $u$ not in a \texttt{wide} sequence into a minimum number of disjoint cones of maximum angle $\pi/5$\;
		In each nonempty cone, \texttt{select} the shortest edge incident to $u$\;
		In each empty cone, let $ux$ and $uy$ be the incident $DT$ edges on $u$ clockwise and counterclockwise, respectively, from the empty cone\;
		
		If either $ux$ or $uy$ is selected then \texttt{select} the other edge (in case it has not been selected); otherwise \texttt{select} the longer edge between $ux$ and $uy$ breaking ties arbitrarily;
		
	}
	
	\Return{\textup{all edges \texttt{selected} by both incident vertices}}\;}
	\caption{{\small \texttt{KX12}($P$)}}
	\label{code:KX12}
\end{algorithm2e}

\begin{algorithm2e}
{\small 	\SetAlgoLined
	$DT \gets \texttt{$L_2$-DelaunayTriangulation}(P)$\;
	
	$E, E^* \gets \emptyset$\;
	
	Initialize $k=\Delta+1$ cones surrounding each vertex $u$, oriented such that the shortest edge incident on $u$ falls on a boundary\;
	
	\ForEach{$\{u,v\} \in DT$\textup{ in order of non-decreasing length}}{
		\If{$\forall C^i_u$\textup{ containing }$\{u,v\}$, $C^i_u \cap E = \emptyset$\textup{ \textbf{and} }$\forall C^j_v$\textup{ containing }$\{u,v\}$, $C^j_v \cap E = \emptyset$}{
			Add edge $\{u,v\}$ to $E$\;
		}
	}
	
	\ForEach{$\{u,v\} \in E$}{
		$\texttt{Wedge}_\Delta(u,v)$\;
		$\texttt{Wedge}_\Delta(v,u)$\;
	}
	
	\Return $E \cup E^*$\;}
	\caption{{\small \texttt{BCC12}($P, \Delta \in \{6,7\}$})}
	\label{alg:bcc}
\end{algorithm2e}

\begin{algorithm2e}
	{\small \SetAlgoLined
		\ForEach{$C^z_u$\textup{ containing }$\{u,v_i\}$}{
			Let $\{u,v_j\}$ and $\{u,v_k\}$ be the first and last edges of $DT$ in the cone\;
			Add all edges $\{v_m,v_{m+1}\}$ to $E^*$ such that $j<m<i-1$ or $i<m<k-1$\;
			\If{$\{u,v_{i+1}\} \in C^z_u$\textup{ \textbf{and} }$v_{i+1} \neq v_k$\textup{ \textbf{and} }$\angle uv_iv_{i+1} > \pi/2$}{
				Add edge $\{v_i,v_{i+1}\}$ to $E^*$\;
			}
			\If{$\{u,v_{i-1}\} \in C^z_u$\textup{ \textbf{and} }$v_{i-1} \neq v_k$\textup{\textbf{ \textbf{and} }}$\angle uv_iv_{i-1} > \pi/2$}{
				Add edge $\{v_i,v_{i-1}\}$ to $E^*$\;
			}
		}
	}
	\caption{{\small $\texttt{Wedge}_7(u,v_i)$}}
	\label{alg:bcc_wedge7}
\end{algorithm2e}

\begin{algorithm2e}
	{\small \SetAlgoLined
		\ForEach{$C^z_u \textup{ containing } \{u,v_i\}$}{
			Let $Q = \{v_n: \{u,v_n\} \in C^z_u \cap DT\} = \{v_j,...,v_k\}$\;
			Let $Q' = \{v_n: \angle v_{n-1}v_nv_{n+1} < 6\pi/7, v_n \in Q \setminus \{v_j,v_i,v_k\}$\}\;
			Add all edges $\{v_n,v_{n+1}\}$ to $E^*$ such that $v_n,v_{n+1} \notin Q'$ and $n \in \left[j+1,i-2\right] \cup \left[i+1,k-2\right]$\;
			
			\tcc{W.l.o.g. the points of $Q'$ lie between $v_i$ and $v_k$ (the symmetric case is handled analogously)}
			\If{$\angle uv_iv_{i-1} > 4\pi/7$ \textup{ and } $i,i-1 \neq j$}{
				Add edge $\{v_i,v_{i-1}\}$ to $E^*$\;
			}
			Let $v_f$ be the first point in $Q'$\;
			Let $a = \min\{ n|n>f \textrm{ and } v_n \in Q \setminus Q' \}$\;
			\eIf{$f=i+1$}{
				\If{$\angle uv_iv_{i+1} \leq 4\pi/7 \textup{ and } a\neq k$}{
					Add edge $\{v_f,v_a\}$ to $E^*$\;
				}
				\If{$\angle uv_iv_{i+1} > 4\pi/7 \textup{ and } f+1\neq k$}{
					Add edge $\{v_i,v_{f+1}\}$ to $E^*$\;
				}
			}{
				Let $v_\ell$ be the last point in $Q'$\;
				Let $b = \max\{ n|n<\ell \textrm{ and } v_n \in Q \setminus Q' \}$\;
				\eIf{$\ell=k-1$}{
					Add edge $\{v_\ell,v_b\}$ to $E^*$\;
				}{
					Add edge $\{v_b,v_{\ell+1}\}$ to $E^*$\;
					\If{$v_{\ell-1} \in Q'$}{
						Add edge $\{v_\ell,v_{\ell-1}\}$ to $E^*$\;
					}
				}
			}
	}}
	\caption{{\small $\texttt{Wedge}_6(u,v_i)$}}
	\label{alg:bcc_wedge6}
\end{algorithm2e}

\vspace{-4pt}

\item \texttt{BKPX15}: \textbf{{Bonichon,  Kanj,  Perkovi{\'c}, and Xia}~\cite{bonichon2015there}}.  This algorithm uses the $L_\infty$-Delaunay triangulation and was the first degree-$4$ algorithm. For such triangulations, empty axis parallel squares are used for characterization instead of empty circles, as needed in the case of $L_2$-Delaunay triangulations. The $L_\infty$-distance between two points $u,w$ is defined as  $d_\infty(u,w) = \max(d_x(u,w),d_y(u,w))$. From the $L_\infty$-Delaunay triangulation, a directed $L_\infty$-distance-based Yao graph $\overrightarrow{Y_4^\infty}$ is constructed that is a plane $\sqrt{20+14\sqrt{2}}$-spanner. Then a degree-$8$ subgraph $H_8$ of $Y_4^\infty$ is constructed. Finally, some redundant edges are removed and new shortcut edges are added to obtain the final plane degree-$4$ spanner with a stretch factor of $\sqrt{20+14\sqrt{2}} (19+29\sqrt{2})\approx 156.8$. No runtime analysis is presented by the authors. Refer to Algorithm~\ref{code:BKPX15}. The algorithm divides the space around each point into $4$ cones, separated by the $x$ and $y$-axes after translating the point to the origin. Each cone has an associated \texttt{charge}, which can be $0, 1,$ or $2$. The algorithm labels certain edges as follows. Each edge will be an \texttt{anchor} or a \texttt{\texttt{non-anchor}} and \texttt{weak} or \texttt{strong}.
	Further, each edge may have an additional label of \texttt{start-of-odd-chain-anchor}.
 A \texttt{weak anchor chain} is a path $w_0,w_1,w_2,...w_k$ of maximal length consisting of \texttt{weak anchors} such that the cone of each edge (wrt the source vertex) alternates between some $i$ and $i+2$.
 \texttt{Canonical} edges are edges between consecutive vertices in the ordered neighborhood of a vertex $u$ in a common cone $i$.
 An edge $(u,v)$ is said to be \texttt{dual} if there are two or more edges of $\overrightarrow{Y^{\infty}_4}$ incident to cone $i$ of $u$ and cone $i+2$ of $v$.

\vspace*{-5pt}

\begin{algorithm2e}
	{\small \SetAlgoLined
		$\overrightarrow{Y^{\infty}_4} \gets \emptyset$\;
		
		\ForEach{$u \in DT$   }{
			\ForEach{$C^i_u$}{
				Let $v \in C^i_u$ be the vertex with the smallest $L_\infty$ distance\;
				Add $\left(u,v\right)$ to $\overrightarrow{Y^{\infty}_4}$\;
			}
		}
		\Return $\overrightarrow{Y^{\infty}_4}$\;}
	\caption{{\small \texttt{constructYaoInfinityGraph}($DT$})}
	\label{constructYaoInfinityGraph}
\end{algorithm2e}

\begin{algorithm2e}
	{\small 	\SetAlgoLined
		\ForEach{$\left(u,v\right) \in \overrightarrow{Y^{\infty}_4}$}{
			Let $i$ be the cone of $u$ containing $v$\;
			$v_{anchor} \gets v$\;
			
			\If{\textup{$\lnot$\texttt{isMutuallySingle}$(\overrightarrow{Y^{\infty}_4},u,v,i)$}\textup{ \textbf{and} $u$ has more than one $\overrightarrow{Y^{\infty}_4}$ edge in $C^i_u$}}{
				Let $\ell$ be the position of $v$ and
				$k$  the number of vertices in \texttt{fan}($DT,u,i$)\;
				\uIf{$\ell \geq 2 \textup{ \textbf{and}}\left(v_{\ell -1},v_\ell\right) \in \overrightarrow{Y^{\infty}_4}\textup{ \textbf{and}}\left(v_{\ell},v_{\ell-1}\right) \notin \overrightarrow{Y^{\infty}_4}$}{
					Let $v_{\ell'}$ such that $\ell' < \ell$ be the starting vertex of the maximal unidirectional \texttt{canonical} path ending at $v_\ell$\;
					$v_{anchor} \gets v_{\ell'}$\; 
				}
				
				\uElseIf{$\ell \leq k-1 \textup{ \textbf{and}}\left(v_{\ell +1},v_\ell\right) \in \overrightarrow{Y^{\infty}_4}\textup{ \textbf{and}}\left(v_{\ell},v_{\ell+1}\right) \notin \overrightarrow{Y^{\infty}_4}$}{
					Let $v_{\ell'}$ such that $\ell' > \ell$ be the starting vertex of the maximal unidirectional \texttt{canonical} path ending at $v_\ell$\;
					$v_{anchor} \gets v_{\ell'}$\; 
				}
			}
			Mark $(u,v_{anchor})$ as the \texttt{anchor} of $C^i_u$\;

		}
		$A \gets \emptyset$\;
		\ForEach{\textup{\texttt{anchor} $(u,v)$ in each $C^i_u$}}{
			\eIf{\textup{\texttt{anchor} of $C^{i+2}_v$ is $(v,u)$ \textbf{or} undefined}}{
				Mark $(u,v)$ as \texttt{strong} and add it to $A$\;
			}{
				Mark $(u,v)$ as \texttt{weak}\;
			}
		}
		\ForEach{\textup{\texttt{weak anchor} $(u,v)$ in each $C^i_u$}}{
			
			\If{\textup{$u$ begins the \texttt{weak anchor chain} $\left(w_0,w_1,...,w_k\right)$}}{
				\If{$k$ \textup{is odd}}{
					Mark $\left(w_0,w_1\right)$ as a \texttt{start-of-odd-chain-anchor}\;
				}
				\For{$\ell = k-1;\ \ell\geq 0;\ \ell=\ell-2$}{
					Add $\left(w_{\ell -1},w_\ell\right)$ to $A$\;
				}
			}
		}
		\Return $A$\;
	}
	
	\caption{{\small \texttt{selectAnchors}($\protect\overrightarrow{Y_4^\infty}, DT$)}}
	\label{selectAnchors}
\end{algorithm2e}

\begin{algorithm2e}
	{\small \SetAlgoLined
	\Return all neighboring vertices $\left(v_1,v_2,...,v_k\right)$ in $C^i_u$ in counterclockwise order\;}
	\caption{{\small \texttt{fan}($DT,u,i$)}}
\end{algorithm2e}

\vspace*{-8pt}
\begin{algorithm2e}
{\small 	\SetAlgoLined
	\Return $u$ has one $\overrightarrow{Y^{\infty}_4}$ edge in $C^i_u$ \textbf{and} $v$ has one $\overrightarrow{Y^{\infty}_4}$ edge in $C^{i+2}_v$\;}
	
	\caption{{\small \texttt{isMutuallySingle}($\protect\overrightarrow{Y^{\infty}_4},u,v,i$)}}
	\label{isMutuallySingle}
\end{algorithm2e}

\begin{algorithm2e}
{\small 	\SetAlgoLined
	\texttt{Charge} each anchor $\left(u,v\right) \in A$ to the cones of each vertex in which the edge lies\;
	$H_8 \gets A$\;
	\ForEach{\textup{vertex $u$ and cone $i$ of $u$}}{
		$\{v_1,...,v_k\} \gets \texttt{fan}(DT,u,i)$\;
		\If{$k \geq 2$}{
			Add all uni-directional \texttt{canonical} edges to $H_8$ except $\left(v_2,v_1\right)$ and $\left(v_{k-1},v_k\right)$\;
			
			\If{\textup{$\left(v_2,v_1\right)$ is a \texttt{non-anchor}, uni-directional edge such that $\left(v_2,v_1\right) \in \overrightarrow{Y_4^\infty} \land \left(v_1,v_2\right) \notin \overrightarrow{Y_4^\infty} \land \left(v_1,u\right)$ is a \texttt{dual} edge $\land$ not a \texttt{start-of-odd-chain} anchor chosen by $v_1$}}
			{
				Add  $\left(v_2,v_1\right)$ to $H_8$\;
			}
			\If {\textup{$\left(v_{k-1},v_k\right)$ is a \texttt{non-anchor}, uni-directional edge such that $\left(v_{k-1},v_k\right) \in \overrightarrow{Y_4^\infty} \land \left(v_k,v_{k-1}\right) \notin \overrightarrow{Y_4^\infty} \land \left(v_k,u\right)$ is a \texttt{dual} edge $\land$ not a \texttt{start-of-odd-chain} anchor chosen by $v_k$}}
			{
				Add  $\left(v_{k-1},v_k\right)$ to $H_8$\;
			}

			\ForEach{\textup{\texttt{canonical} edge $\left(v,w\right)$ added to $H_8$}}{
				$v_{charge} \gets v$\; \If{\textup{$\left(v,w\right)$ is a \texttt{non-anchor}}}{
					$v_{charge} \gets u$\;
				}
				\texttt{Charge} $\left(v,w\right)$ to the cone of $v$ containing $w$ and the cone of $w$ containing $v_{charge}$\;
			}
		}
	}
	\Return $H_8$\;}

	\caption{{\small \texttt{degree8Spanner}$(A,\protect\overrightarrow{Y^{\infty}_4},DT)$}}
\label{degree8Spanner}
\end{algorithm2e}


\begin{algorithm2e}
	{\small \SetAlgoLined
	
	$H_6 \gets H_8$\;
	
	\ForEach{\textup{uni-directional \texttt{non-anchor} $\left(u,v\right)$ in cone $i$ of $u$ in $H_8$ with \texttt{charge} $= 1$}}{
		\If{\textup{cone $i+1$ or $i-1$ of $v$ has \texttt{charge} $=2 \land \left(u,v\right)$ is \texttt{charged} to cone $i+1$ or $i-1$ of $v$}}{
			Let $j$ be the cone of $v$ where ($u$, $v$) is \texttt{charged}\;
			
			$v_{current}\gets u, v_{next}\gets v$, $D\gets \emptyset$\;
			
			\While{\textup{cone $j$ of $v_{next}$ has \texttt{charge} $=2 \land (v_{current},v_{n ext})$ is in cone $j$ of $v_{next}$}}{
				Add $(v_{current},v_{next})$ to $D$\;
				$v_{current}\gets v_{next}$\;
				Set $v_{next}$ to the target of the $\overrightarrow{Y^{\infty}_4}$ edge beginning in cone $j$ of $v_{current}$\;
				swap($i,j$)\;
			}
			
			Starting with the last edge in the path induced by $D$, remove every other edge from $H_6$\;
		}
	}
	
	\Return $H_6$\;}
	
	\caption{{\small \texttt{processDupEdgeChains}$(H_8,\protect\overrightarrow{Y^{\infty}_4})$}}
		\label{processDupEdgeChains}
\end{algorithm2e}

\begin{algorithm2e}
{\small 	\SetAlgoLined
	$H_4 \gets H_6$\;
	
	\ForEach{\textup{pair of \texttt{non-anchor} uni-directional \texttt{canonical} edges $(v_{r-1},v_r), (v_{r+1}, v_r)$ in cone $i$ of $u$}}{
		Remove $(v_{r-1},v_r)$ and $(v_{r+1}, v_{r})$ from $H_4$\;
		Add $(v_{r-1},v_{r+1})$ to $H_4$\;
		\texttt{Charge} this edge to the cones of each vertex in which the edge lies\;
	}
	
	\Return $H_4$\;}
	
	\caption{{\small \texttt{createShortcuts}$(H_6,\protect\overrightarrow{Y^{\infty}_4},DT)$}}
	\label{createShortcuts}
\end{algorithm2e}

\begin{algorithm2e}
	{\small \SetAlgoLined
		$DT \gets \texttt{$L_{\infty}$-DelaunayTriangulation}(P)$\;
		$\overrightarrow{Y^{\infty}_4} \gets \texttt{constructYaoInfinityGraph}(DT)$ (Algorithm~\ref{constructYaoInfinityGraph})\; 
		$A \gets \texttt{selectAnchors}(\overrightarrow{Y^{\infty}_4},DT)$ (Algorithm~\ref{selectAnchors})\;
		$H_8 \gets \texttt{degree8Spanner}(A,\overrightarrow{Y^{\infty}_4},DT)$ (Algorithm~\ref{degree8Spanner})\;
		$H_6 \gets \texttt{processDupEdgeChains}(H_8,\overrightarrow{Y^{\infty}_4})$ (Algorithm~\ref{processDupEdgeChains})\;
		$H_4 \gets \texttt{createShortcuts}(H_6,\overrightarrow{Y^{\infty}_4},DT)$ (Algorithm~\ref{createShortcuts})\;
		
		\Return $H_4$;}
	\caption{{\small \texttt{BKPX15}($P$)}}
	\label{code:BKPX15}
\end{algorithm2e}

\item \texttt{KPT17}: \textbf{Kanj,  Perkovi{\'c}, and T{\"u}rkoǧlu}~\cite{kanj2017degree}.  Akin to \texttt{BGHP10}, this algorithm uses the $TD$-Delaunay triangulation and $\Theta$-graph to introduce fresh techniques in spanner construction. Refer to Algorithm~\ref{code:KPT17} for a pseudocode of this algorithm. The authors show that their algorithm generates degree-$4$ spanners with a stretch factor of $20$ and runs in $O(n\log n)$ time. In the following, we define the notations used in the pseudocode. 
\begin{itemize}
	\item For each vertex, the shortest edge in each odd cone is called an anchor.
	\item Cones $1$ and $4$ are labelled as \texttt{blue} and the rest as \texttt{white}.
	\item The first and last edges incident upon a vertex $u$ in a cone $i$ are called the \texttt{boundary} edges of $u$ in $i$.
	\item The \texttt{canonical} path is made up of all \texttt{canonical} edges incident on $u$ in cone $i$, forming a path from one \texttt{boundary} edge in the cone to the other.
\end{itemize}

\item \texttt{BHS18}: {\textbf{Bose,  Hill, and Smid}~\cite{bose2018improved}}.
This algorithm produces a plane degree-$8$ spanner with stretch factor at most $1.998 \left( 1 + \frac{2\pi}{6 \cos (\pi/6) } \right) \approx 4.4$ using the $L_2$-Delaunay triangulation and $\Theta$-graph. However, the authors do not present any runtime analysis of their algorithm. In \texttt{BHS18}, the space around every point $p$ is divided into six cones and are oriented such that a boundary lies on the $x$-axis after translating $p$ to the origin. The algorithm starts with the $L_2$-Delaunay triangulation $DT$, then, in order of non-decreasing {bisector distance}, each edge is added to the spanner if the cones containing it are both empty. For each edge added here, certain \textit{canonical} edges will also be carefully added to the spanner. Refer to Algorithm~\ref{code:BHS18}. In the following, we define the notations used in this pseudocode. 

\vspace*{-5pt}

\begin{itemize}
	\item The \texttt{bisector-distance} $\left[pq\right]$ between $p$ and $q$ is the distance from $p$ to the orthogonal projection of $q$ onto the bisector of $C^p_i$ where $q \in C^p_i$.
	\item Let $\{q_0,q_1,...,q_{d-1}\}$ be the sequence of all neighbors of $p$ in $DT$ in consecutive clockwise order. The neighborhood $N_p$ with apex $p$ is the graph with the vertex set $\{p,q_0,q_1,...,q_{d-1}\}$ and the edge set $\{\{q_j,q_{j+1}\}\} \cup \{\{q_j,q_{j+1}\}\}, 0 \leq j \leq d-1$, with all values mod $d$.
 The edges $\{\{q_j,q_{j+1}\}\}$ are called \textit{canonical edges}.
	\item $N^p_i$ is the subgraph of $N_p$ induced by all the vertices of $N_p$ in $C^p_i$, including $p$.
	\item Let $Can^{\{p,r\}}_i$ be the subgraph of $DT$ consisting of the ordered subsequence of canonical edges $\{s,t\}$ of $N^p_i$ in clockwise order around apex $p$ such that $\left[ps\right] \geq \left[pr\right]$ and $\left[pt\right] \geq \left[pr\right]$.
\end{itemize}

\begin{algorithm2e}
{\small 	\SetAlgoLined
	$DT \leftarrow \texttt{TD-DelaunayTriangulation}(P)$\;
	$E, A \gets \emptyset$\;
	
	\ForEach{\textup{\texttt{white} anchor $\left(u,v\right)$ in increasing order of $d_\triangledown$ length}}
	{
		\If{\textup{$u$ and $v$ do not have a \texttt{white} anchor in a cone adjacent to $\left(u,v\right)$'s cone}}
		{
			Add $\left(u,v\right)$ to $A$\;
		}
	}
	Add all \texttt{blue} anchors to $A$\;
	\ForEach{\textup{\texttt{blue} anchor $u$}}{
		Let $s_1,s_2,...,s_m$ be the clockwise ordered neighbors of $u$ in $DT$\;
		Add all \texttt{canonical} edges $\left(s_\ell,s_{\ell+1}\right) \notin A$ to $E$ such that $1 \leq \ell < m$\;
	}
	\ForEach{\textup{pair of \texttt{canonical} edges $\left(u,v\right), \left(w,v\right) \in E$ in a \texttt{blue} cone}}{
		Remove $\left(u,v\right)$ and  $\left(w,v\right)$ from $E$\;
		Add a shortcut edge $\left(u,w\right)$ to E\;
	}
	\ForEach{\textup{\texttt{white} canonical edge $\left(u,v\right)$ on the \texttt{white} side of its anchor $a$}}{
		\If{$a \notin A$}{
			Add $\left(u,v\right)$ to $E$\;
		}
	}
	\ForEach{\textup{\texttt{white} anchor $\left(v,w\right)$ and its \texttt{boundary} edge $\left(u,w\right) \neq \left(v,w\right)$ on the \texttt{white} side}}{
		Let $u=s_1,s_2,...,s_m=v$ be the \texttt{canonical} path between $u$ and $v$\;
		\For{$i=0$ \KwTo $m$}{
			\If{$\left(s_{i+1},s_i\right)$ \textup{is \texttt{blue}}}{
				Let $j$ be the smallest index in $P_i=\{s_{i+1},...,s_m\}$ such that $s_j$ is in a \texttt{white} cone of $s_i$ and $P_i$ lies on the same side (or on) the straight line $s_{i}s_{j}$\;
				Add the shortcut $\left(s_{j},s_{i}\right)$ to $E$\;
				\If{$\left(s_{j},s_{j-1}\right) \in E$}{
					Remove $\left(s_{j},s_{j-1}\right)$ from $E$\;
				}
				$i \gets j$\;
			}
		}
	}
	
	\Return $E \cup A$;}
	\caption{\texttt{KPT17}($P$)}
	\label{code:KPT17}
\end{algorithm2e}

\begin{algorithm2e}
	{\small \SetAlgoLined
	
	$DT \gets \texttt{$L_2$-DelaunayTriangulation}(P)$\;
	Let $m$ be the number of edges in $DT$\;
	$L$ be the edges $\in DT$ sorted in non-decreasing order of \texttt{bisector-distance}\;
	
	$E_A \gets$ \texttt{addIncident}($L$), $E_{CAN} \gets \emptyset$\;
	\ForEach{$\{u,v\} \in E_A$}{
		$E_{CAN}\gets E_{CAN}\ \cup$ \texttt{addCanonical}($u,v$) $\cup$ \texttt{addCanonical}($v,u$)\;
	}
	\Return $E_A \cup E_{CAN}$\;}
	\caption{\texttt{BHS18}($P$)}
	\label{code:BHS18}
\end{algorithm2e}

\begin{algorithm2e}
	{\small \SetAlgoLined
	
	$E_A \gets \emptyset$\;
	\ForEach{$\{u,v\} \in L$}{
		Let $i$ be the cone of $u$ containing $v$\;
		\If{$\{u,w\} \notin E_A$ \textup{for all} $w\in N^u_i$ $\land$ $\{v,y\} \notin E_A$ \textup{for all} $y\in N^v_{i+3}$}{
			Add $\{u,v\}$ to $E_A$\;
		}
	}
	
	\Return{$E_A$}\;}
	\caption{\texttt{addIncident}($L$)}
\end{algorithm2e}

\begin{algorithm2e}
{\small 	\SetAlgoLined
	
	$E' \gets \emptyset$\;
	
	Let $i$ be the cone of $u$ containing $v$\;
	
	
	Let $e_{first}$ and $e_{last}$ be the first and last canonical edge in $Can^{\{u,v\}}_i$\;
	
	\If{$Can^{\{u,v\}}_i$ \textup{has at least} $3$ \textup{edges}}{
		\ForEach{$\{s,t\} \in Can^{\{u,v\}}_i \setminus \{e_{first},e_{last}\}$}{
			Add $\{s,t\}$ to $E'$\;
		}
	}
	
	\If{$v \in \{e_{first},e_{last}\}$ \textup{\textbf{and} there is more than one edge in} $Can^{\{u,v\}}_i$}{
		Add the edge of $Can^{\{u,v\}}_i$ incident to $v$ to $E'$\;
	}
	
	\ForEach{$\{y,z\} \in \{e_{first},e_{last}\}$}{
		\If{$\{y,z\} \in N^z_{i-1}$}{
			Add $\{y,z\}$ to $E'$\;
		}
		\If{$\{y,z\} \in N^z_{i-2}$}{
			\If{$N^z_{i-2} \cap E_A$ \textup{does not have an edge incident to} $z$}{
				Add $\{y,z\}$ to $E'$\;
			}
			\If{$N^z_{i-2} \cap E_A \setminus \{y,z\}$ \textup{has an edge incident to} $z$}{
				Let $\{w,y\}$ be the canonical edge of $z$ incident to $y$\;
				Add $\{w,y\}$ to $E'$\;
			}
		}
	}

	\Return{$E'$}\;}
	\caption{\texttt{addCanonical}($u,v$)}
\end{algorithm2e}
\end{itemize}

\section{Estimating stretch factors of large spanners}
\label{chap:stretchfactor}


Measuring exact stretch factors of large graphs is a tedious job, and so is for geometric spanners.
Although many algorithms exist in the literature for constructing geometric spanners, nothing is known about practical algorithms for
computing stretch factors of large geometric spanners. It is a severe bottleneck for conducting experiments with large spanners since the stretch factor is considered a fundamental quality of geometric spanners.

For any spanner (not necessarily geometric) on $n$ vertices, its exact stretch factor can be computed in $O(n^3\log n)$ time by running the folklore Dijkstra algorithm (implemented using a Fibonacci heap) from every vertex, and in $\Theta(n^3)$ time by running the classic Floyd-Warshall algorithm. Note that the Dijkstra-based algorithm runs $O(n^2\log n)$ time for plane spanners since the number of edges is $O(n)$. Both of these are very slow in practice. However, the latter has a quadratic space-complexity and is unusable when $n$ is large. Consequently, they are practically useless when $n$ is large. Stretch factor estimation of large geometric graphs appears to be a far cry despite theoretical studies on this problem; refer to \cite{agarwal2008computing,narasimhan2000approximating,cheng2012approximating,federickson1987fast,wulff2010computing}. We believe these algorithms are either involved from an algorithm engineering standpoint or rely on well-separated pair decomposition~\cite{callahan1995decomposition}, which may potentially slow down practical implementations due to the large number of well-separated pairs needed by those algorithms.   
This has motivated us to design a practical algorithm, named \textsc{AppxStretchFactor}, that gives a lower-bound on the actual stretch factor of any geometric spanner (not necessarily plane). However, we will consider the universe of plane geometric spanners as the input domain in this work. To our knowledge, we are not aware of any such algorithm in the literature. Refer to Algorithm~\ref{alg:approximatesf}. It takes as input an $n$-element pointset $P$ and a geometric graph $G$, constructed on $P$.

\begin{algorithm2e}
{\small 	\SetAlgoLined
	$DT \gets \texttt{$L_2$-DelaunayTriangulation}(P)$\;
	$t \gets 1$\;
	\ForEach{$p \in P$}{
		$h \gets 1$, $t_p \gets 1$\; 
		\While{\textup{true}}{
			Let $X$ denote the set of points which are exactly $h$ hops away from $p$ in $DT$\, found using a breadth-first traversal originating at $u$\;
			$t' \gets 1$\;
			\ForEach{$q \in X$}{
				$t' \gets \max{   \left( \frac{{|\pi_G(p,q)|}}{|pq|}, t' \right)   }$\;
			}
			\uIf{$t' > t_p$}{
				$h \gets h + 1$; $t_p \gets t'$\;
			}
			\uElse {
				\textbf{break}\;
			}
			
		}	
		$t \gets \max(t,t_p)$\;	
	}
	\Return $t$;}
	\caption{\textsc{AppxStretchFactor}($P, G$)}
	\label{alg:approximatesf}
\end{algorithm2e}

The underlying idea of our algorithm is as follows. We observe that most geometric spanners are well-constructed; meaning it is likely that far away points (having many hops in the shortest paths between them) have low detour ratios (ratio of the length of a shortest path to that of the Euclidean distance) between them and the worst-case detour is achieved by point pairs that are a few hops apart. Note that stretch factor of a graph is the maximum detour ratio over all vertex pairs. To capture \textit{closeness}, we use the $L_2$-Delaunay triangulation constructed on $P$ as the basis. For every point $p \in P$, we start a breadth-first traversal on the Delaunay triangulation $DT$. At every level, we compute the detour ratios in $G$ from $p$ to all the points in that level. If a worse detour ratio is found in the current level compared to the worst found in the previous level, we continue to the next level; otherwise, the process is terminated. For finding detour ratios in $G$, we use the folklore Dijkstra algorithm since computation of shortest paths are required. In our algorithm, $\pi_G(p,q)$ denotes a shortest path between the points $p,q \in P$ in $G$ and $|\pi_G(p,q)|$ its total length. The detour between $p,q$ in $G$ can be easily calculated as $|\pi_G(p,q)|/|pq|$. The current level is denoted by $h$. It is assumed that the neighbors of $p$ in $G$ are at level $1$. For efficiency reasons, we do not restart the Dijkstra at every level of the breadth-first traversal; instead, we save our progress from the previous level and continue after that. 

To our surprise, we find that for the class of spanners used in this work, \textsc{AppxStretchFactor}  returned exact stretch factors almost every time. The precision error was very low whenever it failed to compute the exact stretch factor. Further, our algorithm can be parallelized very easily by spawning parallel iterations of the \textbf{foreach} loop. Apart from the $L_2$-Delaunay triangulation (which can be constructed very fast in practice), it does not use any advanced geometric structure, making it fast in practice. We present our experimental observations for this algorithm in Section~\ref{subsec:appx}. 

%
%

\section{Experiments}
\label{chap:experiments}

We have implemented the algorithms in \textsf{GNU C}\texttt{++}$17$ using the \textsf{CGAL} library~\cite{cgal:eb-21b}. The machine used for experiments is equipped with a  \textsf{AMD Ryzen 5 1600 (3.2 GHz)} processor and $24$ GB of main memory, and runs \textsf{Ubuntu Linux 20.04 LTS}.  The \texttt{g++} compiler was invoked with \texttt{-O3} flag to achieve fast real-world speed. 
From \textsf{CGAL}, the \texttt{Exact\_predicates\_inexact\_constructions\_kernel} is used for accuracy and speed.  We have tried our best to tune our codes to run fast.

All the eleven algorithms considered in this work use one of the following three kinds of Delaunay triangulation as the starting point: $L_2$, $TD$, and $L_\infty$. For constructing  $L_2$ and $L_\infty$-Delaunay triangulations, the \texttt{CGAL::Delaunay\_triangulation\_2} and \texttt{CGAL::Segment\_Delaunay\_graph\_Linf\_2} implementations have been used, respectively. As of now, a $TD$-Delaunay triangulation implementation is not available in the \textsf{CGAL}. It was pointed out by Chew in~\cite{chew1989there} that such triangulations can be constructed in $O(n\log n)$ time. However, no precise implementable algorithm was presented.  But luckily, it is shown  in~\cite{bonichon2010connections} by Bonichon~\etal~that $TD$-Delaunay triangulation of a pointset is same as its $\frac{1}{2}$-$\Theta$ graph. We leveraged this result and used the $O(n\log n)$ time  \texttt{CGAL::Construct\_theta\_graph\_2} implementation for constructing the $TD$-Delaunay triangulations. For faster speed, the input pointsets are always sorted using \texttt{CGAL::spatial\_sort}  before constructing Delaunay triangulations on them. We share the implementations via  \textsf{GitHub}\footnote{\url{https://github.com/ghoshanirban/BoundedDegreePlaneSpannersCppCode}}. 

In our experiments, we have used both synthetic and real-world pointsets, as described next. 

\subsection{Synthetic pointsets}

\begin{figure}[H]
	\centering
	\begin{minipage}{\linewidth}
		\begin{subfigure}[b]{0.27\linewidth}
			\centering
			\resizebox{0.65\linewidth}{!}{\includegraphics{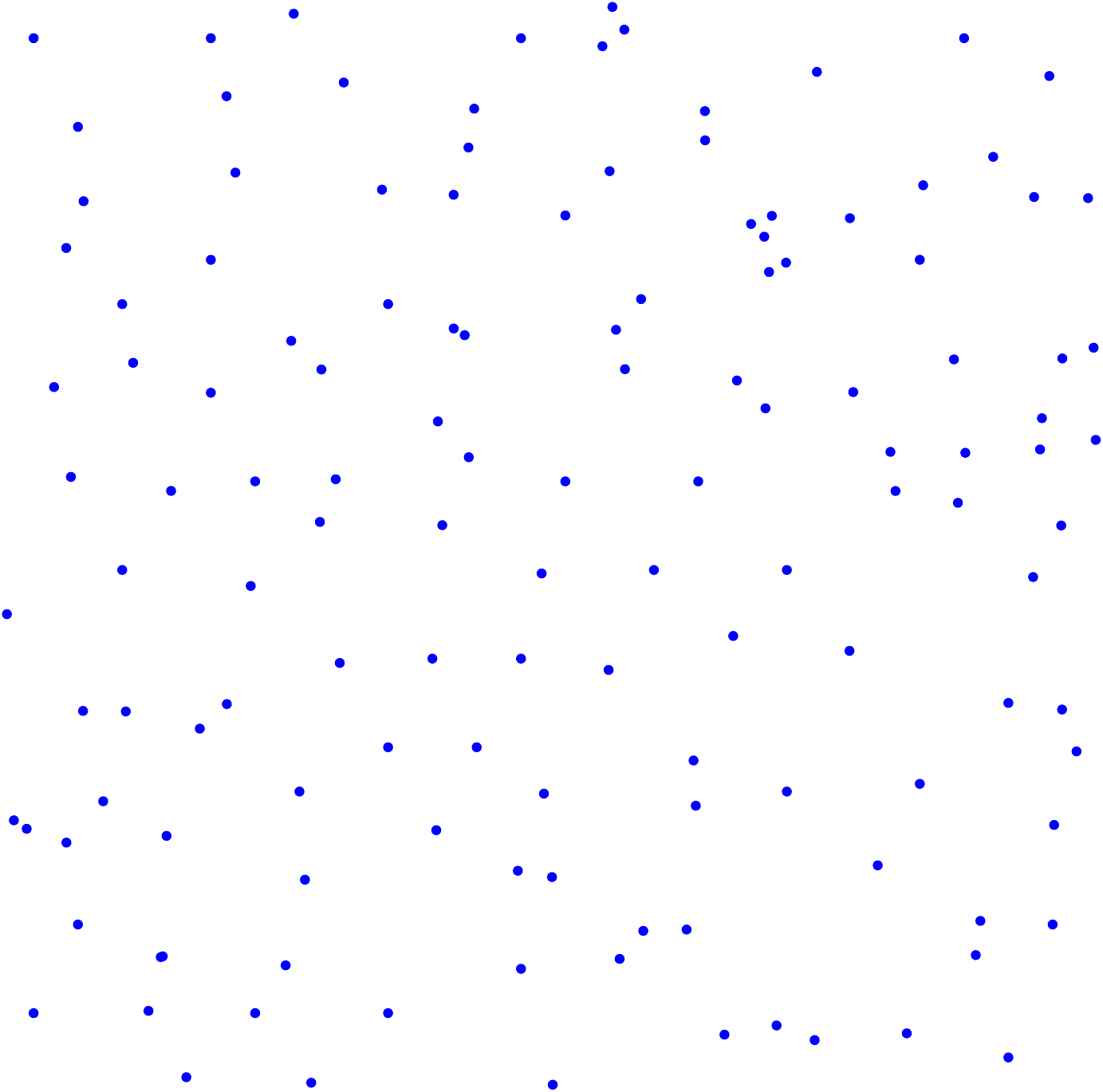}}
			\caption{\texttt{uni-square}}
		\end{subfigure}
		\hfill
		\begin{subfigure}[b]{0.22\linewidth}
			\centering
			\resizebox{0.77\linewidth}{!}{\includegraphics{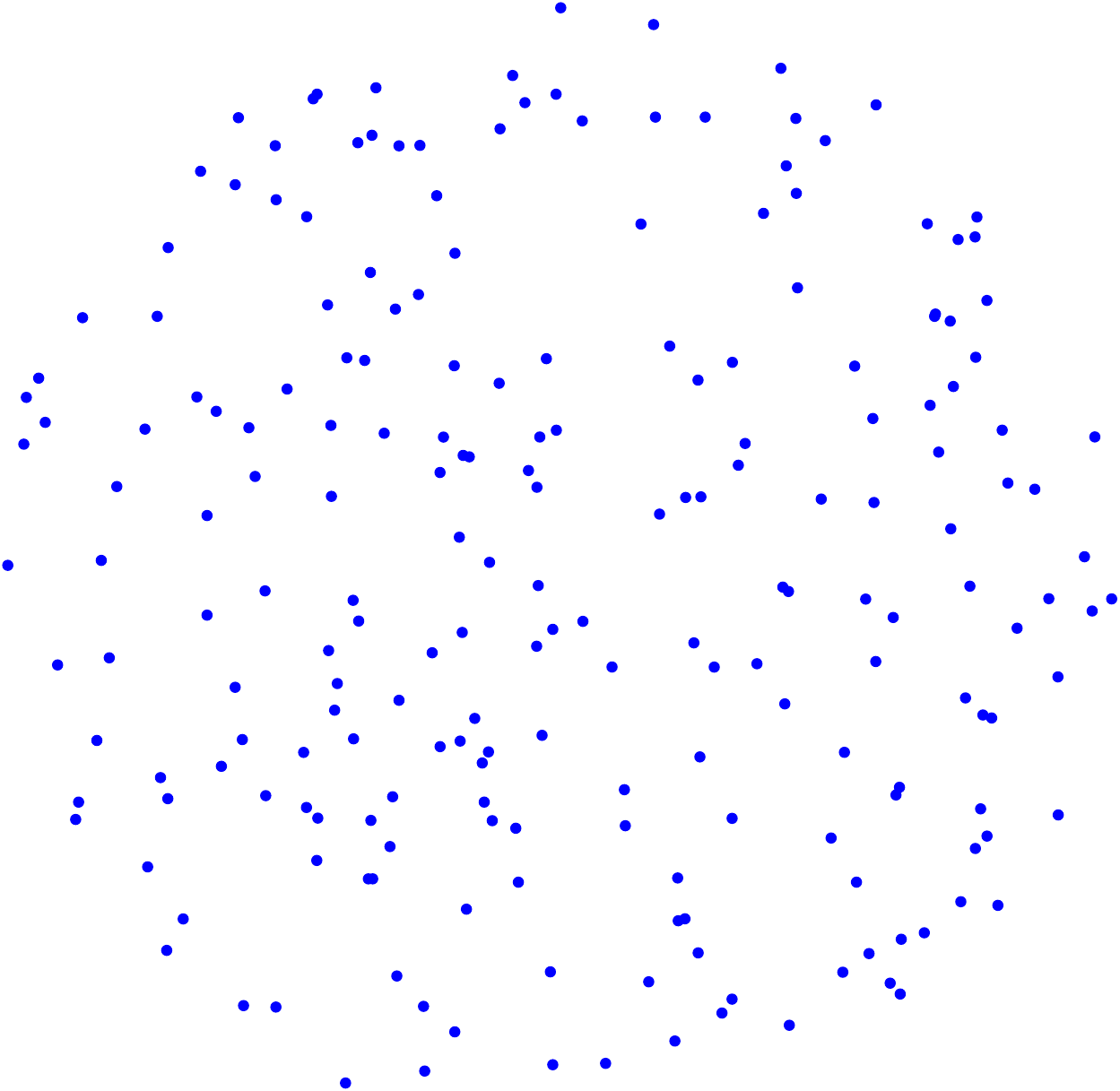}}
			\caption{\texttt{uni-disk}}
		\end{subfigure}
		\hfill
		\begin{subfigure}[b]{0.28\linewidth}
			\centering
			\resizebox{0.75\linewidth}{!}{\includegraphics{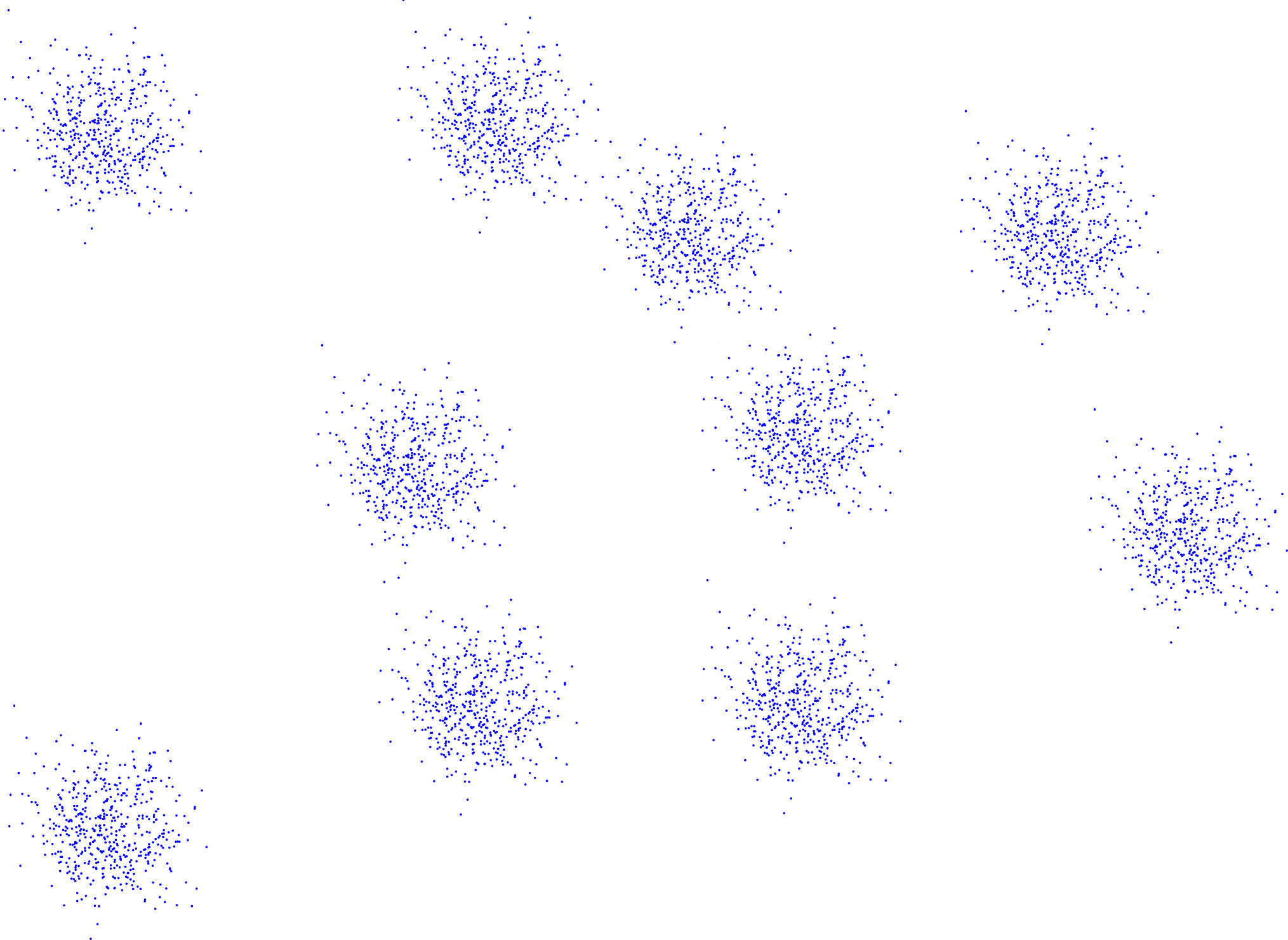}}
			\caption{\texttt{normal-clustered}}
		\end{subfigure}
		\hfill
		\begin{subfigure}[b]{0.20\linewidth}
			\centering
			\resizebox{0.75\linewidth}{!}{\includegraphics{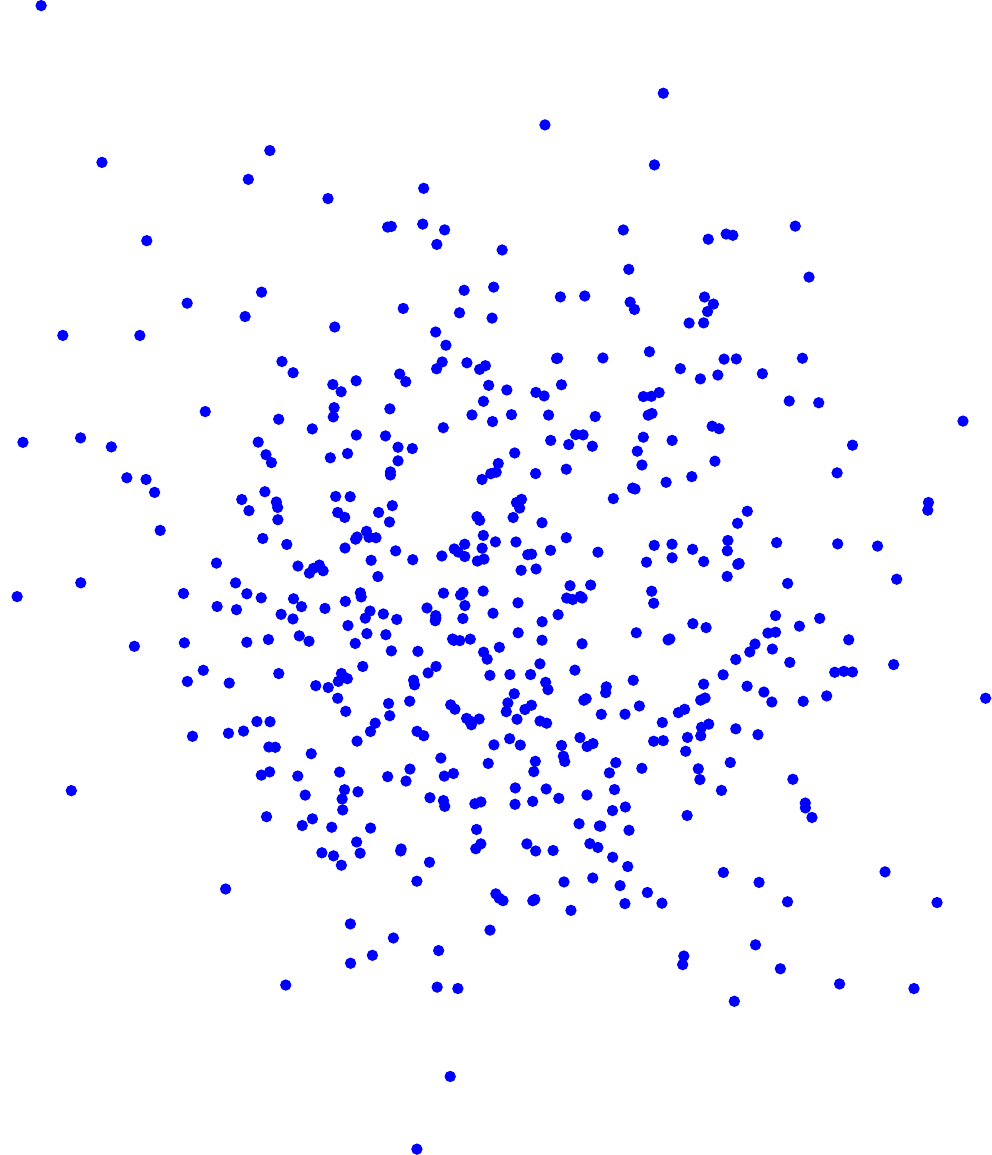}}
			\caption{\texttt{normal}}
		\end{subfigure}
	\end{minipage}
	
	\vspace{1em}
	
	\begin{minipage}{\linewidth}
		\begin{subfigure}[b]{0.27\linewidth}
			\centering
			\resizebox{0.6\linewidth}{!}{\includegraphics{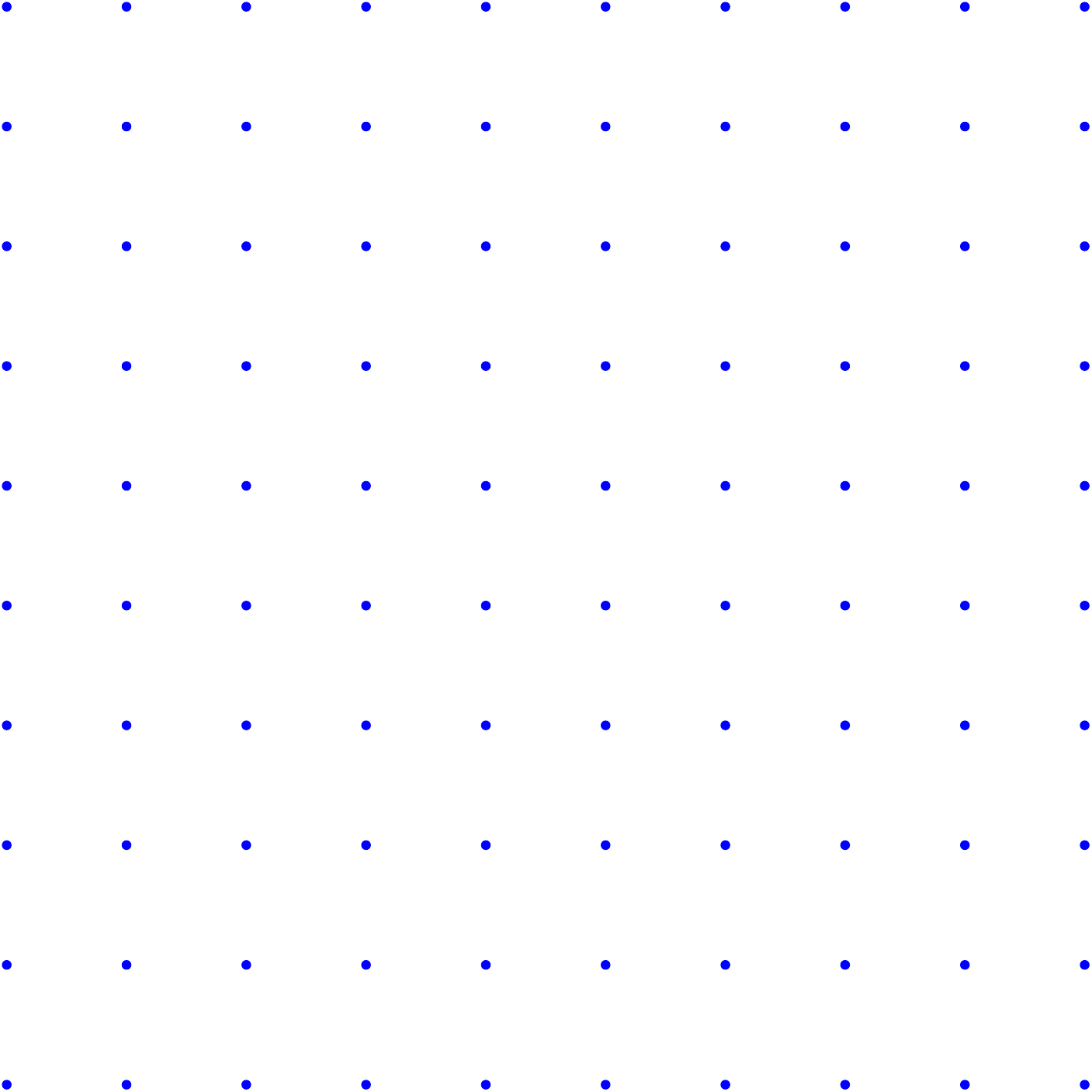}}
			\caption{\texttt{grid-contiguous}}
		\end{subfigure}
		\hfill
		\begin{subfigure}[b]{0.22\linewidth}
			\centering
			\resizebox{0.75\linewidth}{!}{\includegraphics{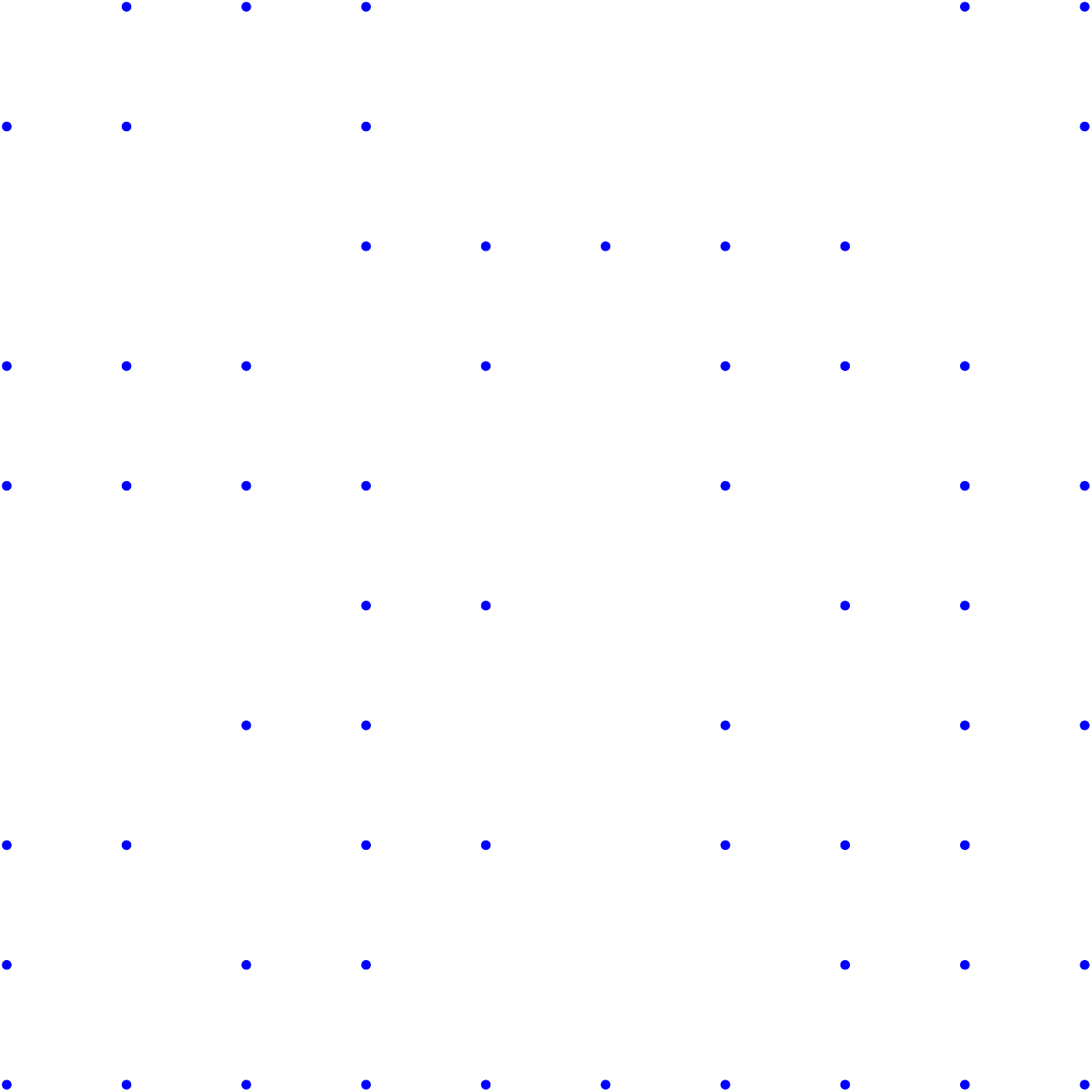}}
			\caption{\texttt{grid-random}}
		\end{subfigure}
		\hfill
		\begin{subfigure}[b]{0.28\linewidth}
			\centering
			\resizebox{0.67\linewidth}{!}{\includegraphics{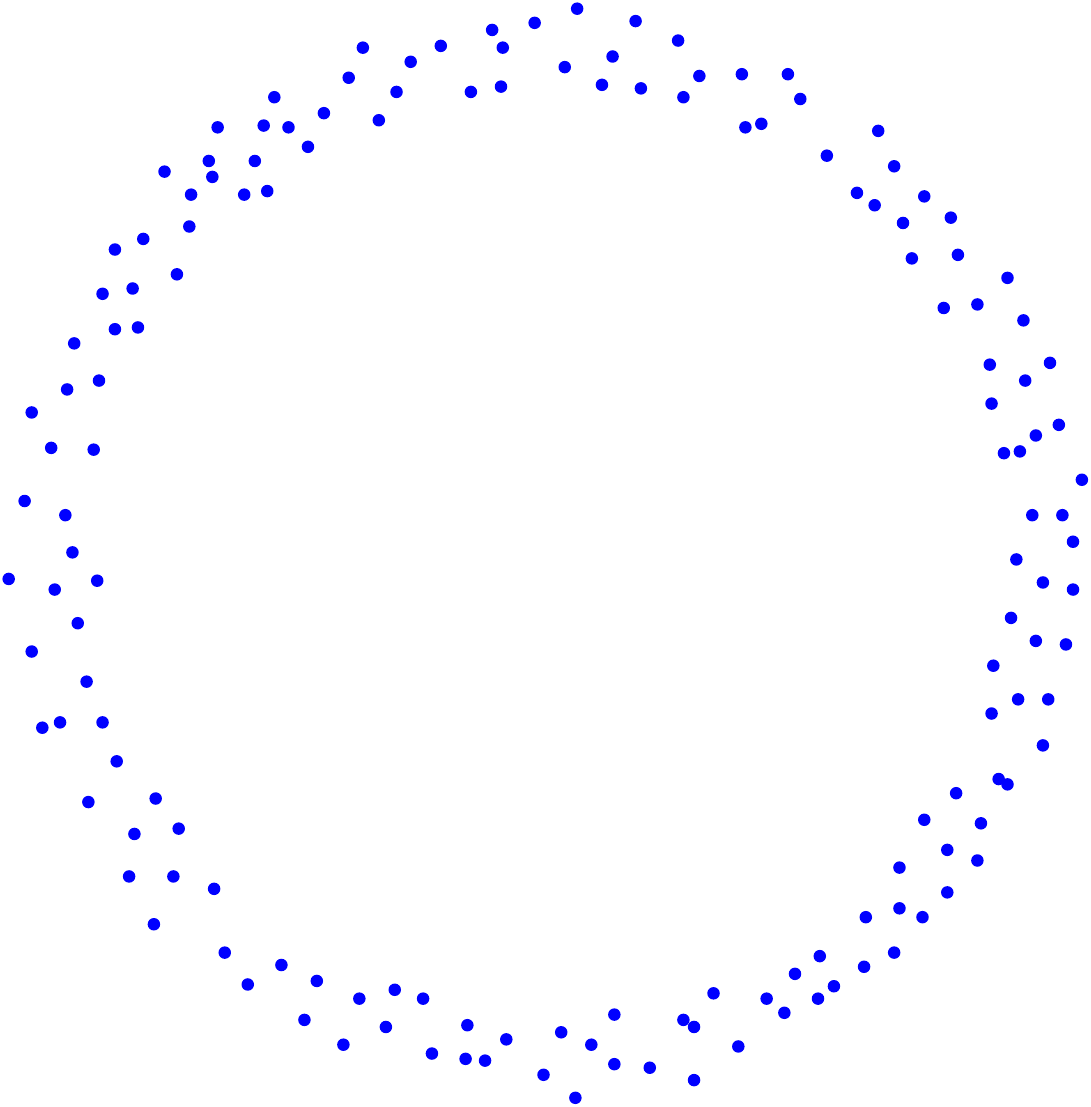}}
			\caption{\texttt{annulus}}
		\end{subfigure}
		\hfill
		\begin{subfigure}[b]{0.20\linewidth}
			\centering
			\resizebox{0.85\linewidth}{!}{\includegraphics{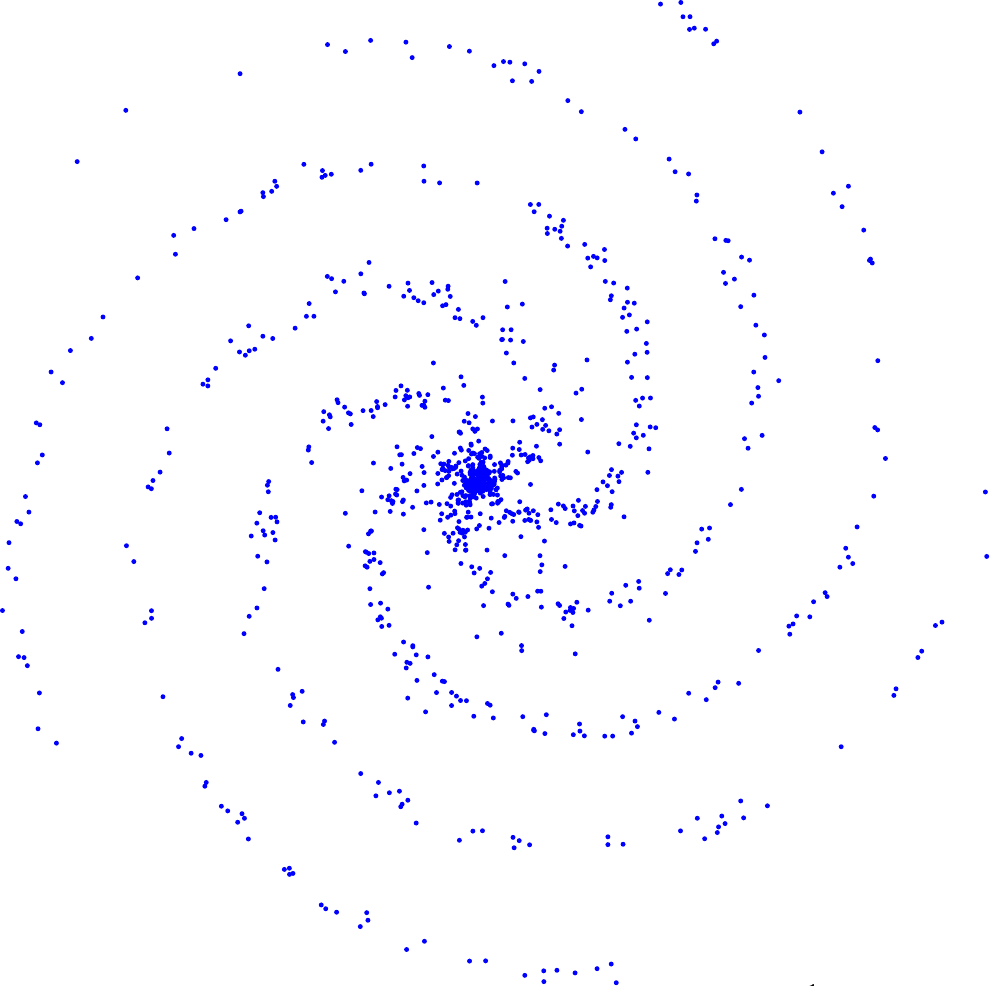}}
			\caption{\texttt{galaxy}}
		\end{subfigure}
	\end{minipage}
	
	\caption{The eight  distributions used to generate synthetic pointsets for our experiments.}
	\label{fig:dist}
\end{figure}

We have used the following eight distributions to generate synthetic pointsets for our experiments. The selection of these distributions are inspired by the ones used in \cite{narasimhan2001geometric, ghosh2019unit,bentley1990k,FriederichUDC} for geometric experiments. See Figure~\ref{fig:dist} to visualize these eight distributions. 

\begin{enumerate}[label=(\alph*)]\itemsep0pt

	\item \texttt{uni-square}. Points are generated uniformly inside a square of side length of $1000$ using the \texttt{CGAL::Random\_points\_in\_square\_2} generator.

	\item  \texttt{uni-disk}. Points are generated uniformly inside a disc of radius $1000$ using the \\ \texttt{CGAL::Random\_points\_in\_disc\_2} generator.

	\item  \texttt{normal-clustered}. A set of $10$ normally distributed clusters placed randomly in the plane. Each cluster contains $n/10$ normally distributed points (mean and standard-deviation are set to $2.0$). We have used \texttt{std::normal\_distribution<double>} to generate the point coordinates.

	\item  \texttt{normal}. This is same as \texttt{normal-clustered} except that only one cluster is used.

	\item  \texttt{grid-contiguous}. Points are generated contiguously on a $\lceil\sqrt{n}\rceil \times \lceil \sqrt{n} \rceil$ square grid using the  \texttt{CGAL::points\_on\_square\_grid\_2} generator.

	\item  \texttt{grid-random}. Points are generated on a $\lceil 0.7n \rceil \times \lceil 0.7n \rceil$ unit square grid. The value $0.7$ is chosen arbitrarily to obtain well-separated non-contiguous grid points. The coordinates of the generated points are integers and are generated independently using  \texttt{std::uniform\_int\_distribution}.

	\item  \texttt{annulus}. Points are generated inside an annulus whose outer radius is set to $1000$ and the inner-radius is set to $800$. The coordinates are generated using \texttt{std::uniform\_real\_distribution}.

	\item \texttt{galaxy}. Points are generated in the shape of a spiral galaxy having outer five arms; see \cite{itinerantgames_2014}.
\end{enumerate}

For seeding the random number generators from \textsf{C}\texttt{++}, we have used the Mersenne twister engine \texttt{std::mt19937}. Since some of the  algorithms assume that no two points must have the same value $x$ or $y$-coordinates, the generated pointsets have been perturbed using the  \texttt{CGAL::perturb\_points\_2} function with $0.0001, 0.0001$ as the two required parameters.

\subsection{Real-world pointsets}

The following real-world pointsets are obtained from various publicly available sources.  
We have removed duplicate points (wherever present) from the pointsets. The main reason behind the use of such pointsets is that they do not follow the popular synthetic distributions. Hence,  experimenting with them is beneficial to see how the algorithms perform on them.

\begin{itemize}\itemsep0pt
\item \texttt{burma}~\cite{tsp}. An $33,708$-element pointset representing cities in Burma.

	\item \texttt{birch3}~\cite{bus2018practical,ghosh2019unit}. An $99,801$-element pointset representing random clusters at random locations.
	
	\item \texttt{monalisa}~\cite{tsp,ghosh2019unit}: A $100,000$-city TSP instance representing a continuous-line drawing of the Mona Lisa. 
	
	\item  \texttt{KDDCU2D}~\cite{bus2018practical,ghosh2019unit}.  An $104,297$-element pointset representing the first two dimensions of a protein data-set.
	
	\item \texttt{usa}~\cite{tsp,ghosh2019unit}.  A $115,475$-city TSP instance representing (nearly) all towns, villages, and cities in the United States.
	
	\item \texttt{europe}~\cite{bus2018practical,ghosh2019unit}.  An $168,896$-element pointset representing differential coordinates of the map of Europe.
	
	\item \texttt{wiki}\footnote{\url{ https://github.com/placemarkt/wiki_coordinates}}.  An $317,695$-element pointset of coordinates found in English-language Wikipedia articles.
	
	\item \texttt{vlsi}~\cite{tsp}.  An $744,710$-element pointset representing a Very Large Scale Integration chip.
	
	\item \texttt{china}~\cite{bus2018practical,ghosh2019unit}.  An $808,693$-element pointset representing cities in China.
	
	\item \texttt{uber}\footnote{\url{https://www.kaggle.com/fivethirtyeight/uber-pickups-in-new-york-city}}.  An $1,381,253$-element pointset representing Uber pickup locations in New York City.
	
	\item \texttt{world}~\cite{tsp,ghosh2019unit}.  An $1,904,711$-element pointset representing  all locations in the world that are registered as populated cities or towns, as well as several research bases in Antarctica.

\end{itemize}

\subsection{Efficacy of \textsc{AppxStretchFactor}}
\label{subsec:appx}

We have seen in Section~\ref{chap:stretchfactor}, it is quite challenging to measure stretch factor of large spanners. This motivated us to design and use the \textsc{AppxStretchFactor} algorithm  in our experiments for estimating stretch factors of the generated spanners. In the following, we compare \textsc{AppxStretchFactor} with the Dijkstra's algorithm (run from every vertex) and show that for the eight distributions it is not only much faster than Dijkstra but can also estimate stretch factors of plane spanners with high accuracy. 

The main reason behind the fast practical performance of \textsc{AppxStretchFactor} is early terminations of the breadth-first traversals (one traversal per vertex), which in turn makes Dijkstra run fast to find the shortest paths to the vertices in all the levels. We have noticed in our experiments that the pair that achieves the stretch factor for a bounded-degree plane spanner are typically a few hops away and pairwise stretch factors (ratio of detour between two vertices to that of their Euclidean distance) drop with the increase in hops. Consequently, the breadth-first traversals terminate very early most of the time.

\begin{figure}[ht!]
	\begin{center}
		\begin{subfigure}{0.35\textwidth}
			\centering
			\includegraphics[scale=1]{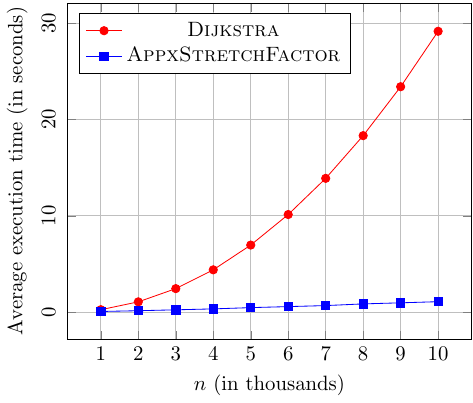} 
			\caption{\texttt{uni-square}}
		\end{subfigure}
		\hspace{8mm}
		\begin{subfigure}{0.35\textwidth}
			\centering
			\includegraphics[scale=1]{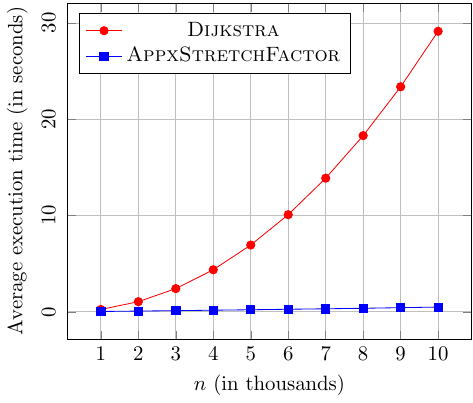} 
			\caption{\texttt{uni-disk}}
		\end{subfigure}
		\vspace{2mm}
		
		\begin{subfigure}{0.35\textwidth}
			\centering	\includegraphics[scale=1]{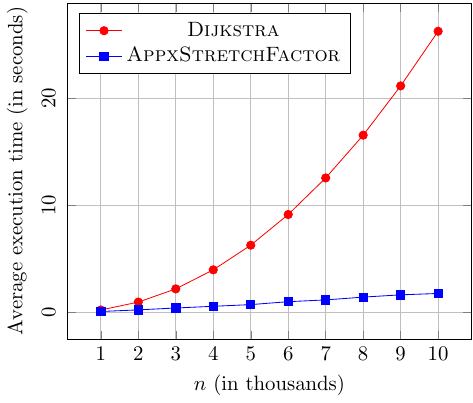} 
			\caption{\texttt{normal-clustered}}
		\end{subfigure}
		\hspace{8mm}
		\begin{subfigure}{0.35\textwidth}
			\centering
			\includegraphics[scale=1]{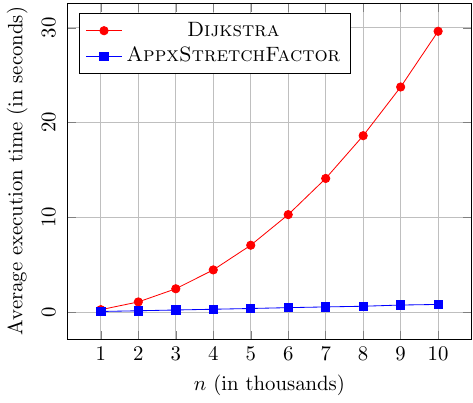} 
			\caption{\texttt{normal}}
		\end{subfigure}
		
		\vspace{2mm}
		
		\begin{subfigure}{0.35\textwidth}
			\centering
			\includegraphics[scale=1]{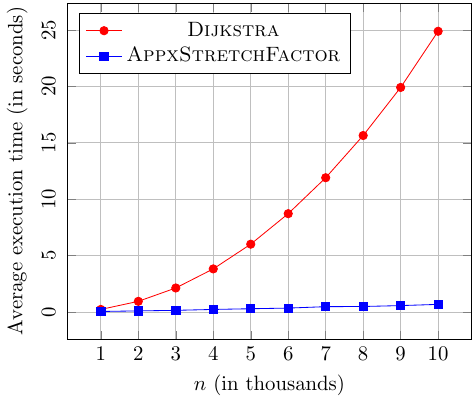} 
			\caption{\texttt{grid-contiguous}}
		\end{subfigure}
		\hspace{8mm}
		\begin{subfigure}{0.35\textwidth}
			\centering
			\includegraphics[scale=1]{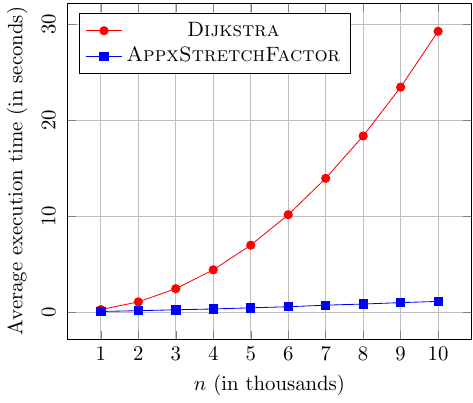} 
			\caption{\texttt{grid-random}}
		\end{subfigure}
		
		\vspace{2mm}
		
		\begin{subfigure}{0.35\textwidth}
			\centering
			\includegraphics[scale=1]{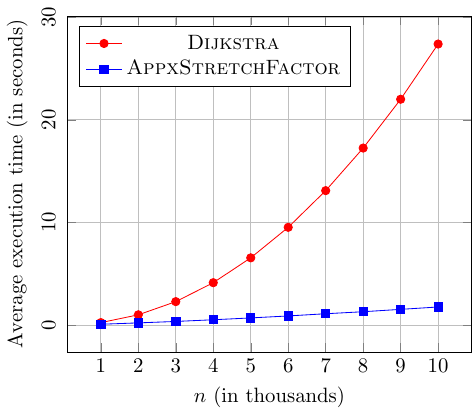} 
			\caption{\texttt{annulus}}
		\end{subfigure}
		\hspace{8mm}
		\begin{subfigure}{0.35\textwidth}
			\centering
			\includegraphics[scale=1]{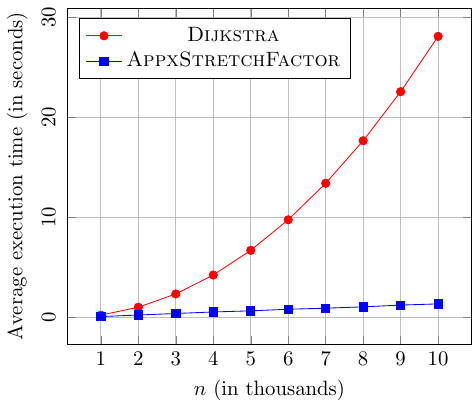} 
			\caption{\texttt{galaxy}}
		\end{subfigure}
		\caption{Runtime comparison: \text{Dijkstra} (run from every vertex) vs \textsc{AppxStretchFactor}. For every value of $n$, we have used $11 \cdot 5 = 55$ spanner samples since there are $11$ algorithms and $5$ pointsets were generated for that value of $n$ using the same distribution.}
		\label{fig:sfcomp}
	\end{center}
\end{figure}

The total number of pointsets used in this comparison experiment is $11 \cdot 8 \cdot 10 \cdot 5 = 4400$ since there are $11$ algorithms, $8$ distributions, $10$ distinct values of $n~ (1K, 2K, . . . , 10K)$, and $5$ samples were used for every value of $n$. Out of these many, the number of times  \textsc{AppxStretchFactor} has failed to return the exact stretch factor is just $8$. So, the observed failure rate is $\approx 0.18\%$. Interestingly, in the cases where \textsc{AppxStretchFactor} failed to compute the exact stretch factor, the largest observed  error percentage between the exact stretch factor (found using Dijkstra) and the stretch factor returned by it is just $\approx 0.15$. This gives us the confidence that our algorithm can be safely used to estimate stretch factor of large spanners. 
Refer to the Figure~\ref{fig:sfcomp}. As evident from these graphs, \textsc{AppxStretchFactor} is substantially faster than Dijkstra everywhere. 
Henceforth, we  use \textsc{AppxStretchFactor} (Algorithm~\ref{alg:approximatesf}) to estimate the stretch factors of the spanners in our experiments.


%
%
%
%
%
%
%
%
%
%
%

%
%
%
%

\subsection{Experimental comparison of the algorithms}
\label{chap:results}


We compare the $11$ implemented algorithms based on their runtimes and degree,  stretch factor, and lightness\footnote{The \emph{lightness} of a geometric graph $G$ on a pointset $P$ is defined as  ratio of the weight of $G$ to that of a Euclidean minimum spanning tree on $P$.} of the generated spanners. 

\begin{figure}[H]
	\centering 
	\includegraphics[scale=1]{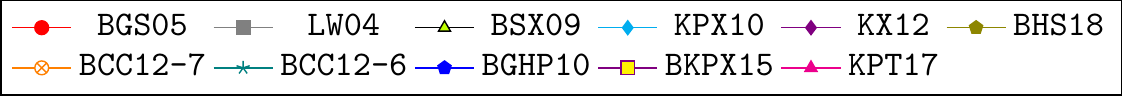}
	\caption{The plot legends.}
		\label{fig:legends}
\end{figure}

In the interest of space, we avoid legend tables everywhere in our plots. Since the legends are used uniformly everywhere, we present them here for an easy reference; see Fig.~\ref{fig:legends}. 

For synthetic pointsets, we vary $n$  from $10K$ to $100K$. For every value of $n$, we have used five random samples to measure runtimes and the above  characteristics of the spanners. In the case of real-world pointsets, we run every one of them five times and report the average time taken. 

	\begin{figure}[H]
	\centering 
	\includegraphics[scale=0.7]{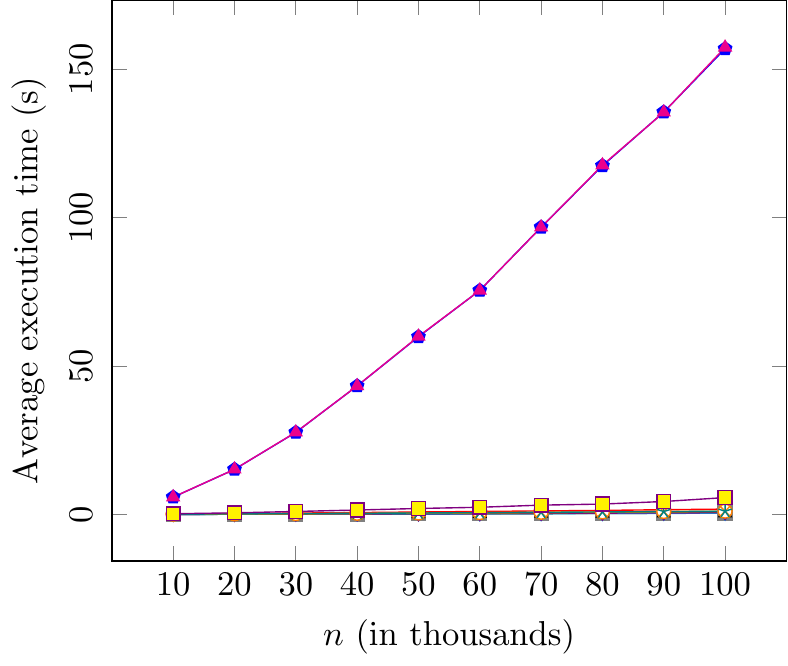}
	\caption{Points are generated using the \texttt{uni-square} distribution. For this particular experiment, $n$ is in the range $10K, 20K,\ldots,100K$. The plots for \texttt{BGHP10} and \texttt{KPT17} have overlapped in this figure.}
	\label{fig:slowDemo}
\end{figure}

In our experiments, we find that  \texttt{BGHP10} and \texttt{KPT17} are considerably slower than the other algorithms considered in this work. The reason behind this is slow construction of $TD$-Delaunay triangulations. Refer to Fig.~\ref{fig:legends} for an illustration. When $n=100K$, both take more than $150$ seconds to finish. In contrast, other nine algorithms take less than $10$ seconds.
Since real-world speed is an important factor for spanner construction algorithms, we do not consider them further in our runtime comparisons. 

\begin{figure}
\begin{center}
	
	\begin{subfigure}{0.35\textwidth}
		\centering
	\includegraphics[scale=0.7]{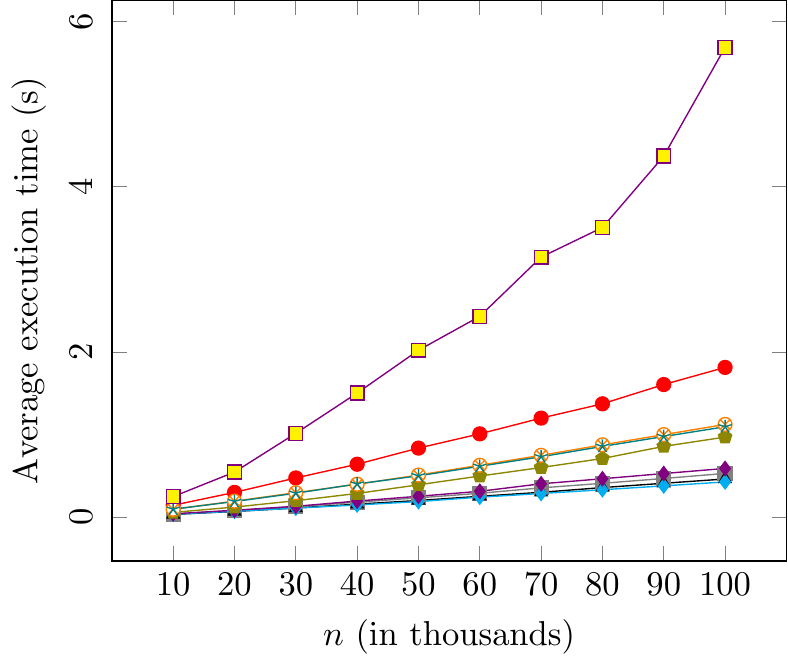}
			\vspace{-8pt}
				\caption{\texttt{uni-square}}
			\end{subfigure}
	\hspace{8mm}
		\begin{subfigure}{0.35\textwidth}
			\centering
		\includegraphics[scale=0.7]{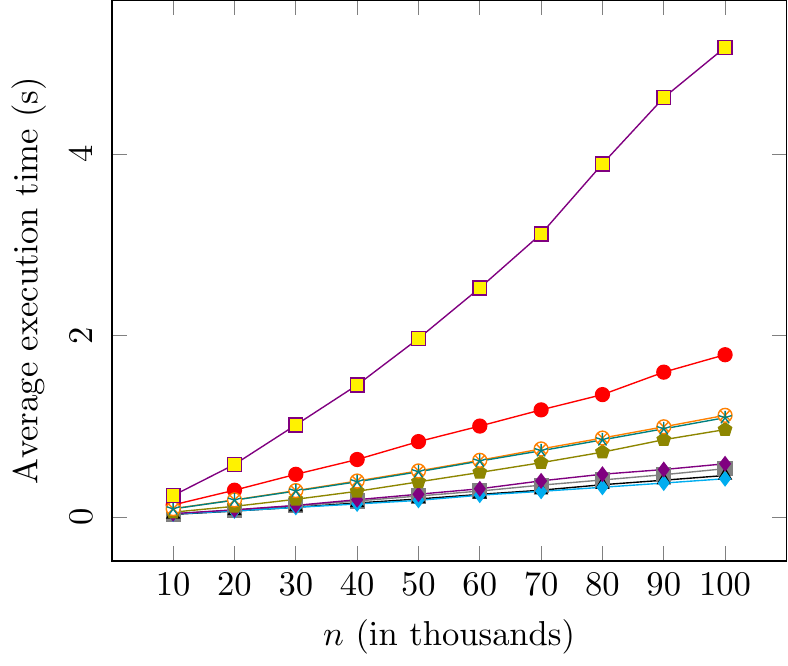}
			\vspace{-8pt}
		\caption{\texttt{uni-disk}}
		\end{subfigure}

		\vspace{5mm}

	\begin{subfigure}{0.35\textwidth}
		\centering
		\includegraphics[scale=0.7]{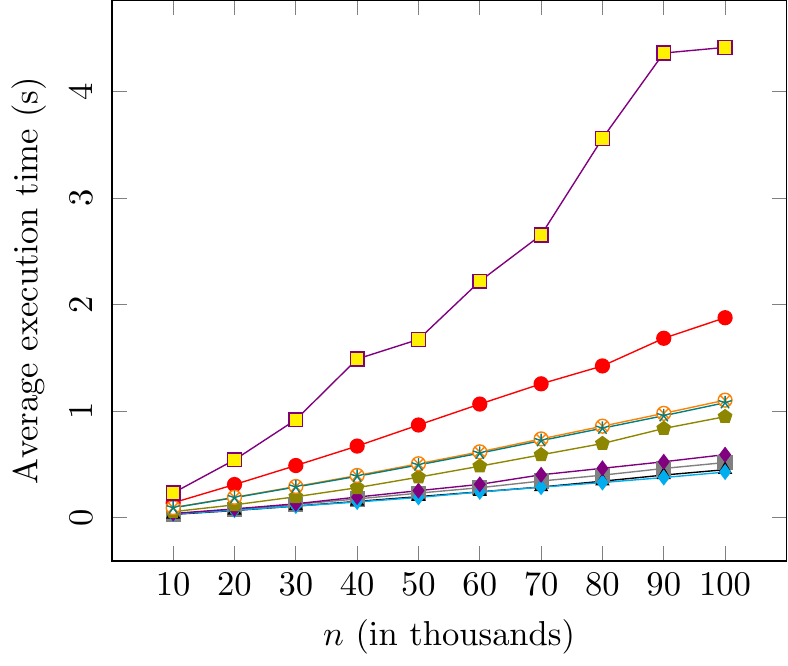}
			\vspace{-8pt}
					\caption{\texttt{normal-clustered}}
				\end{subfigure}
	\hspace{8mm}
	\begin{subfigure}{0.35\textwidth}
		\centering
		\includegraphics[scale=0.7]{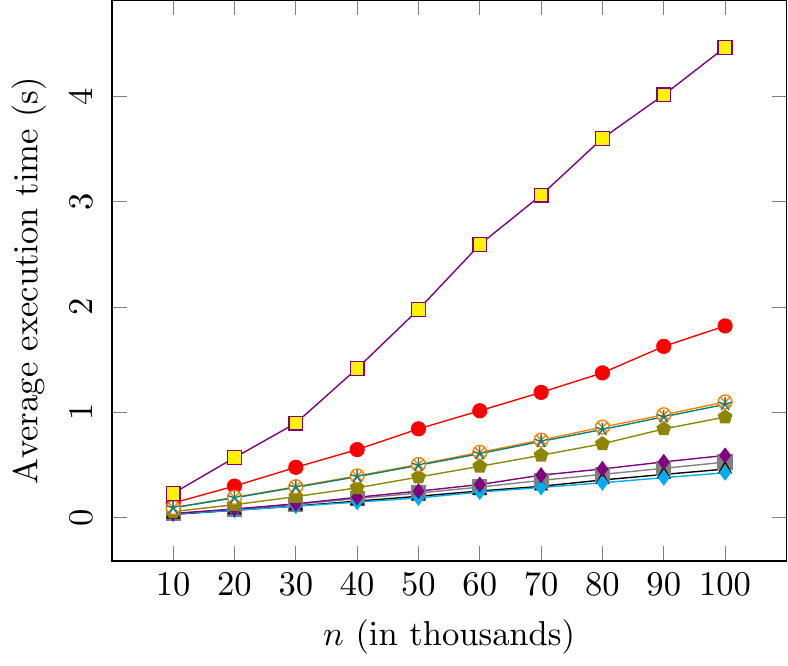}
			\vspace{-8pt}
					\caption{\texttt{normal}}
				\end{subfigure}
			
					\vspace{5mm}
					
	\begin{subfigure}{0.35\textwidth}
		\centering
	\includegraphics[scale=0.7]{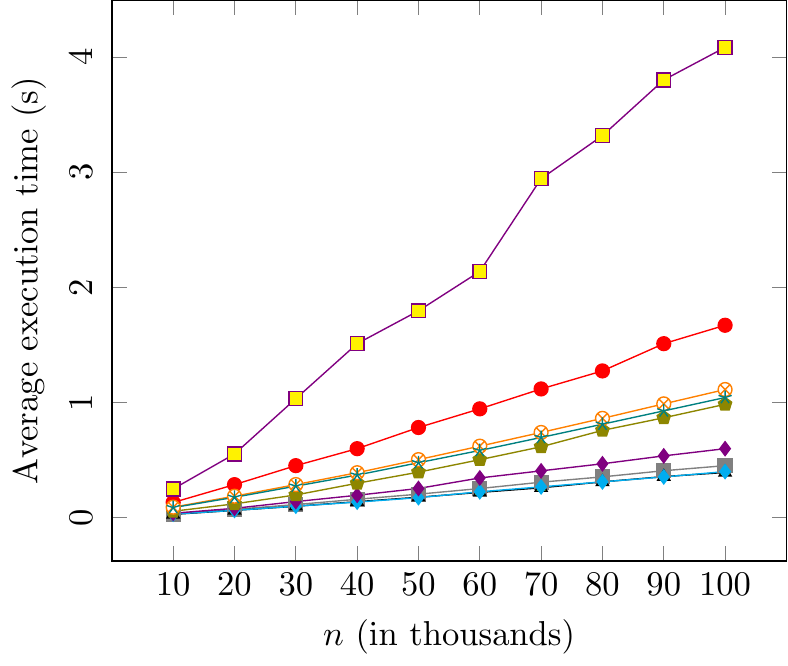}
			\vspace{-8pt}
				\caption{\texttt{grid-contiguous}}
			\end{subfigure}
	\hspace{8mm}
	\begin{subfigure}{0.35\textwidth}
		\centering
	 \includegraphics[scale=0.7]{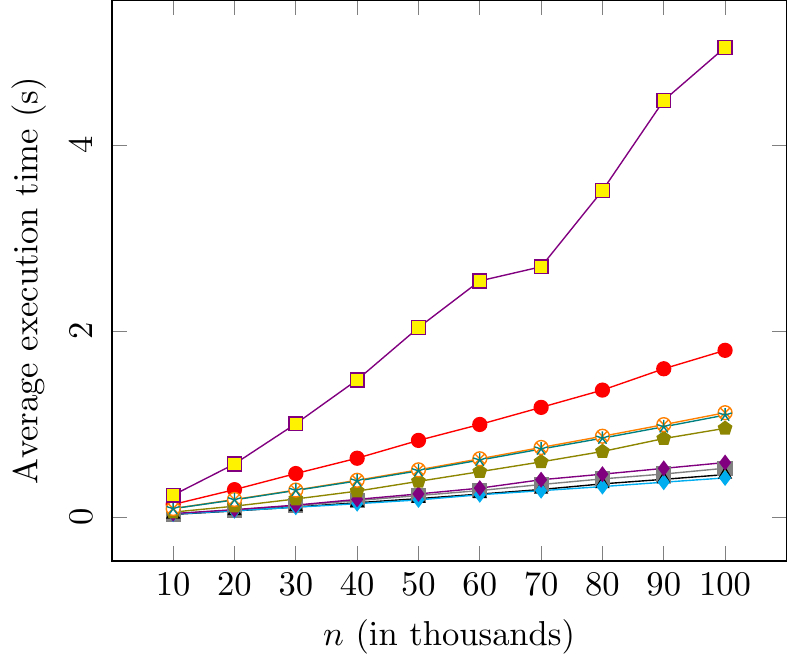}
			\vspace{-8pt}
	 			\caption{\texttt{grid-random}}
	 		\end{subfigure}

					\vspace{5mm}

\begin{subfigure}{0.35\textwidth}
	\centering
	\includegraphics[scale=0.7]{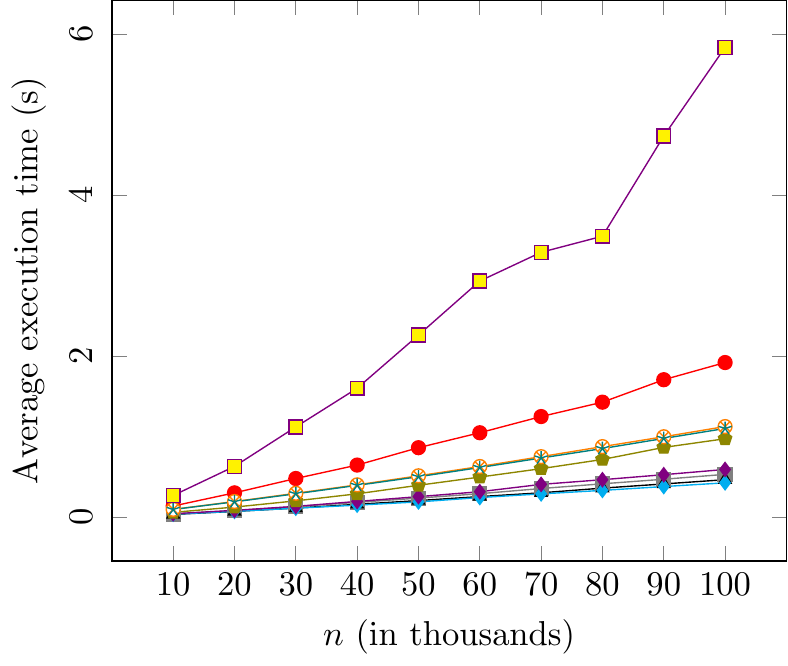}
			\vspace{-8pt}
				\caption{\texttt{annulus}}
			\end{subfigure}
	\hspace{8mm}
\begin{subfigure}{0.35\textwidth}
	\centering
	\includegraphics[scale=0.7]{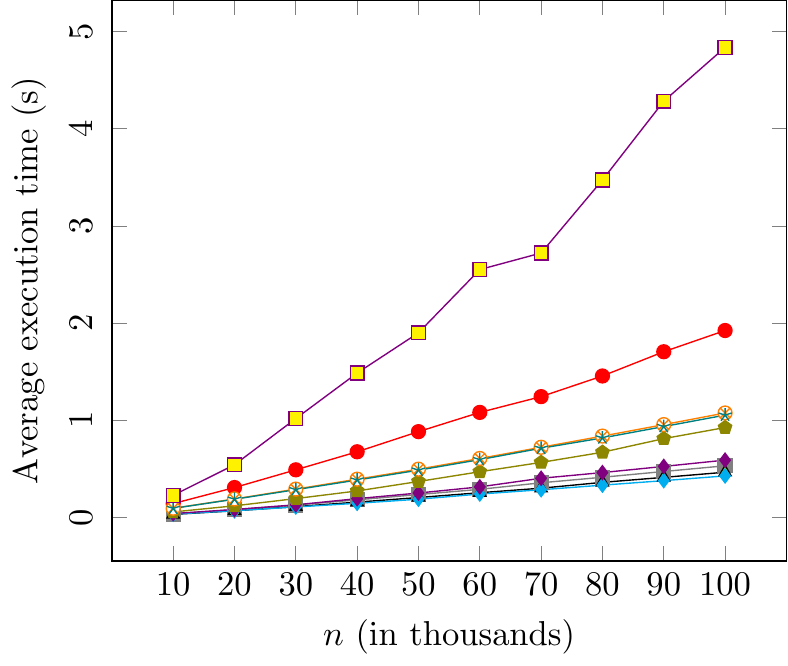}
			\vspace{-8pt}
				\caption{\texttt{galaxy}}
			\end{subfigure}

	\caption{Runtime comparisons of the nine algorithms (\texttt{BGHP10} and \texttt{KPT17} are excluded). }
	\label{fig:runtimePlots}
\end{center}
\end{figure}

\begin{figure}
	\begin{center}
		
		\begin{subfigure}{0.49\textwidth}
			\centering
	\includegraphics[scale=0.7]{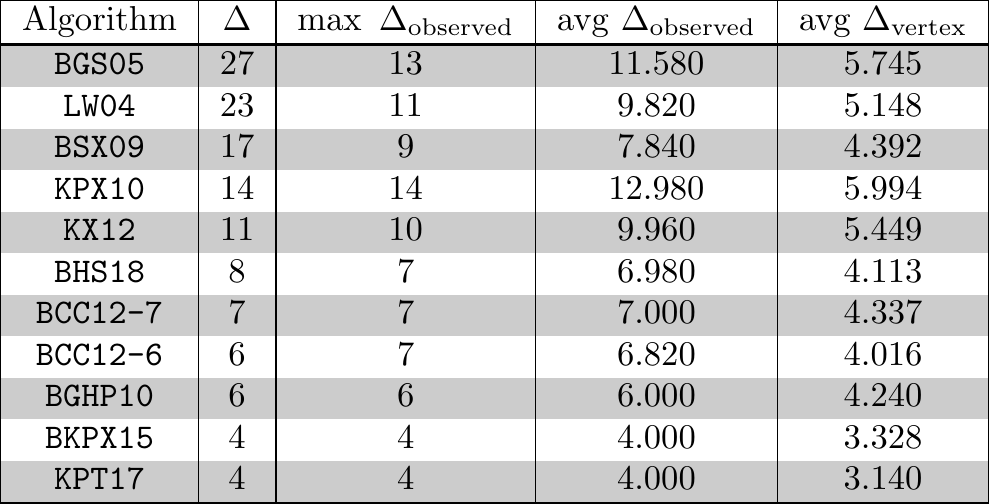}
			\caption{\texttt{uni-square}}
		\end{subfigure}
\begin{subfigure}{0.49\textwidth}
			\centering
	\includegraphics[scale=0.7]{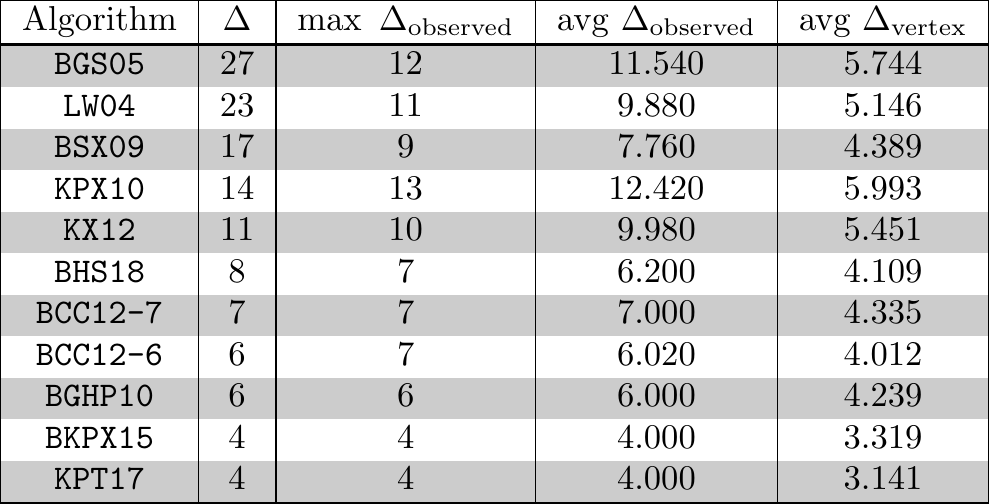}
			\caption{\texttt{uni-disk}}
		\end{subfigure}
		
		\vspace{5mm}

		\begin{subfigure}{0.49\textwidth}
\centering
	\includegraphics[scale=0.7]{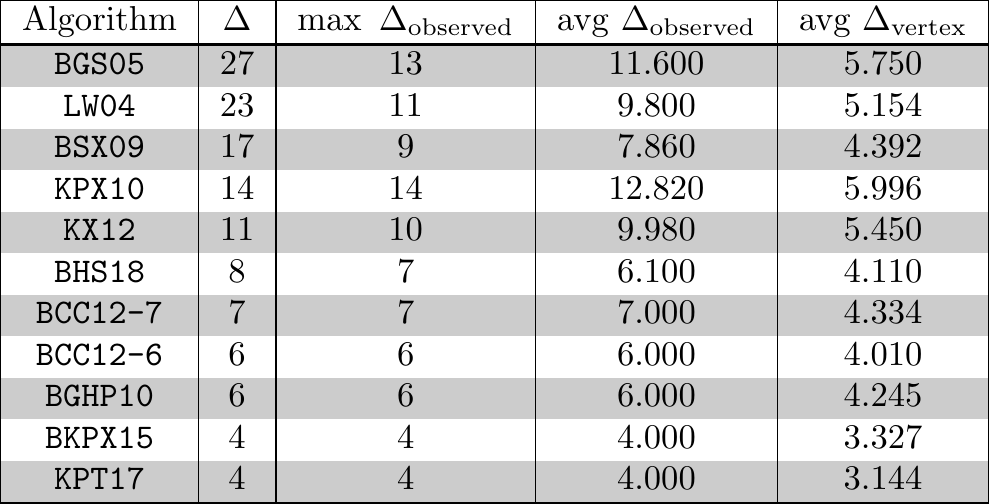}
			\caption{\texttt{normal-clustered}}
		\end{subfigure}
		\begin{subfigure}{0.49\textwidth}
			\centering
	\includegraphics[scale=0.7]{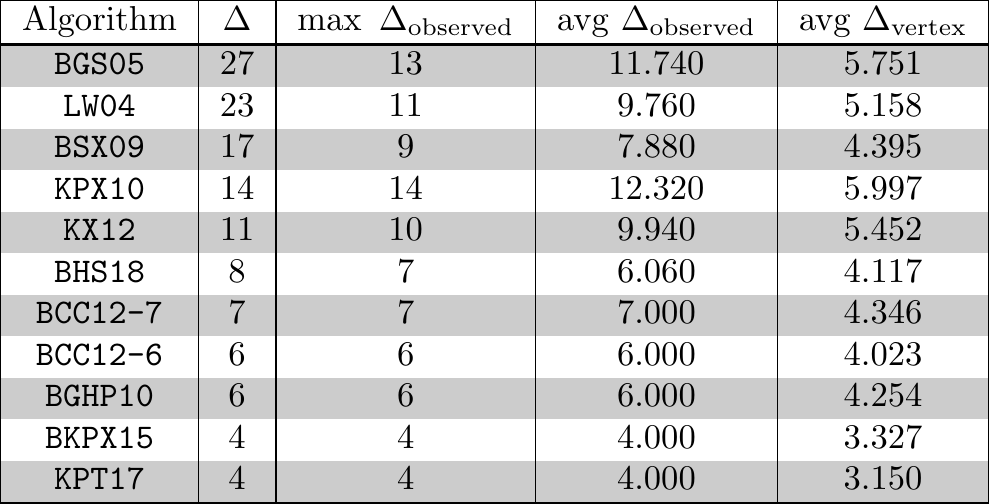}
			\caption{\texttt{normal}}
		\end{subfigure}
		
		\vspace{5mm}

		\begin{subfigure}{0.49\textwidth}
			\centering
	\includegraphics[scale=0.7]{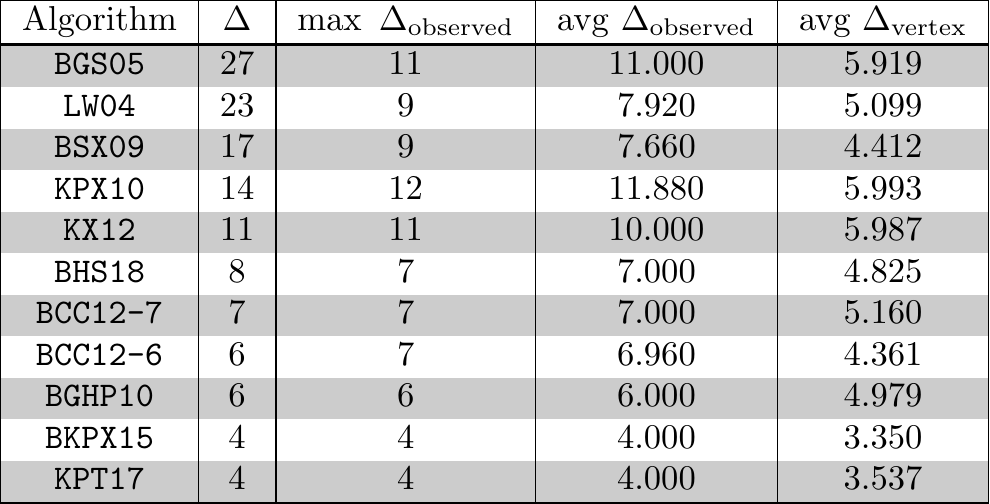}
			\caption{\texttt{grid-contiguous}}
		\end{subfigure}
			\begin{subfigure}{0.49\textwidth}
			\centering
	\includegraphics[scale=0.7]{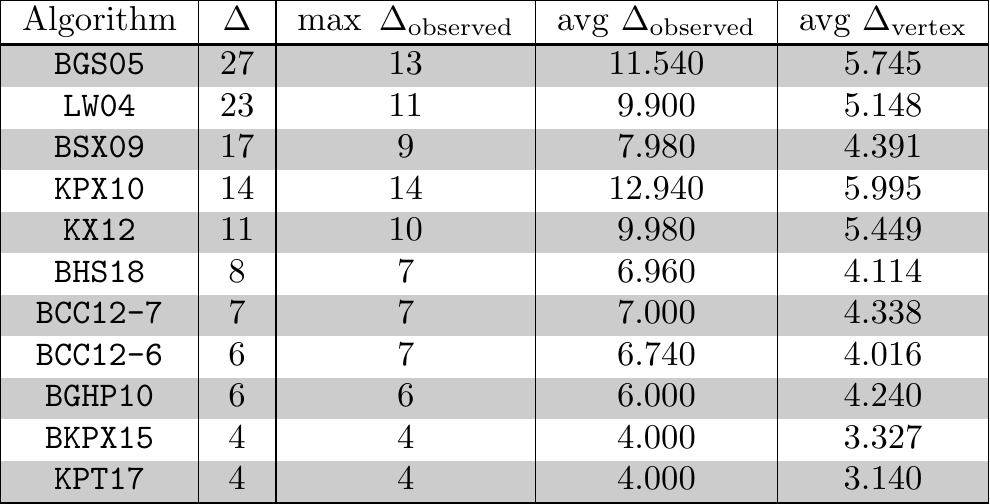}
			\caption{\texttt{grid-random}}
		\end{subfigure}
		
		\vspace{5mm}
		\begin{subfigure}{0.49\textwidth}
			\centering
	\includegraphics[scale=0.7]{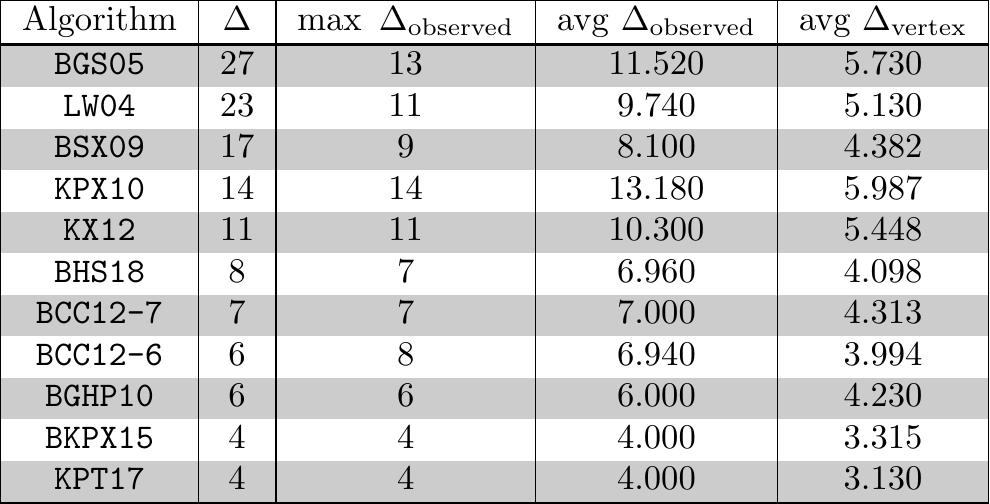}
			\caption{\texttt{annulus}}
		\end{subfigure}
		\begin{subfigure}{0.49\textwidth}
			\centering
	\includegraphics[scale=0.7]{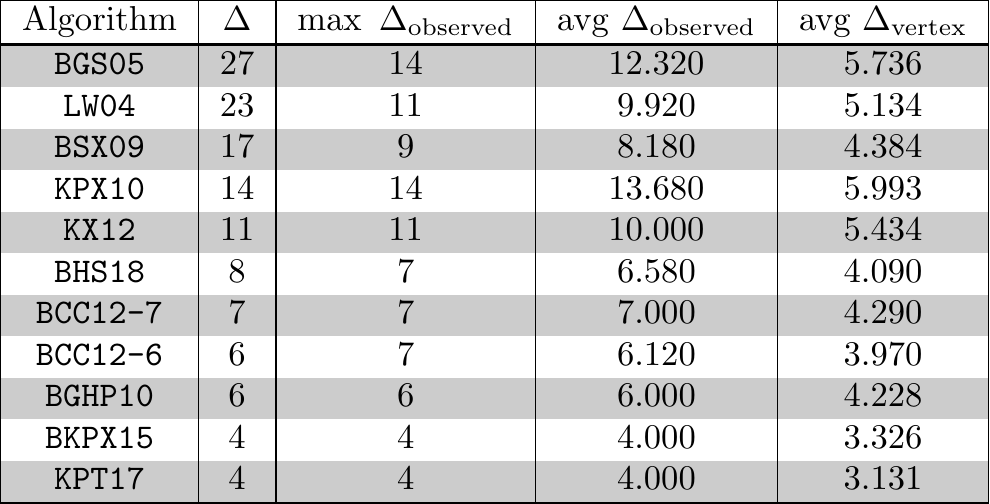}
			\caption{\texttt{galaxy}}
		\end{subfigure}
		
		\caption{Degree comparisons of the spanners generated by the eleven algorithms.}
		\label{fig:degree}
	\end{center}
\end{figure}

\begin{figure}
	\begin{center}
		
		\begin{subfigure}{0.45\textwidth}
			\centering
			\includegraphics[scale=0.7]{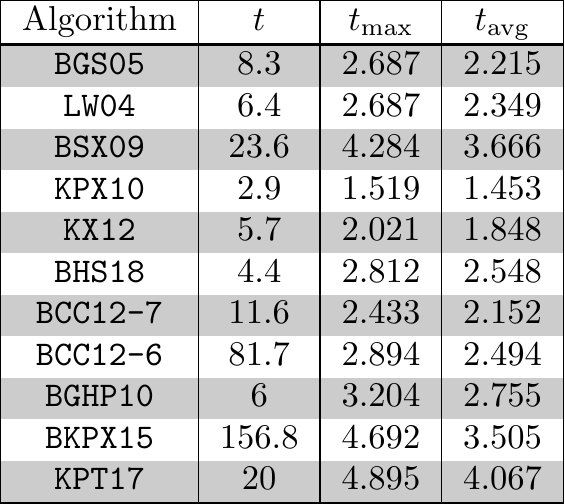}~~~\includegraphics[scale=0.7]{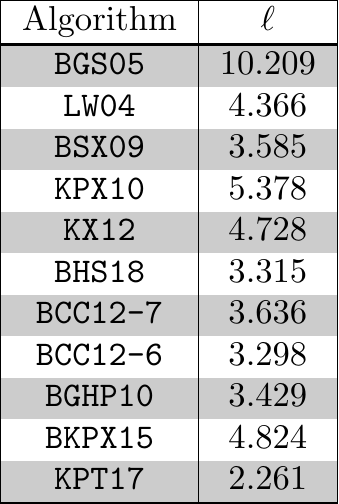}
			\caption{\texttt{uni-square}}
		\end{subfigure}
		\begin{subfigure}{0.45\textwidth}
			\centering
			\includegraphics[scale=0.7]{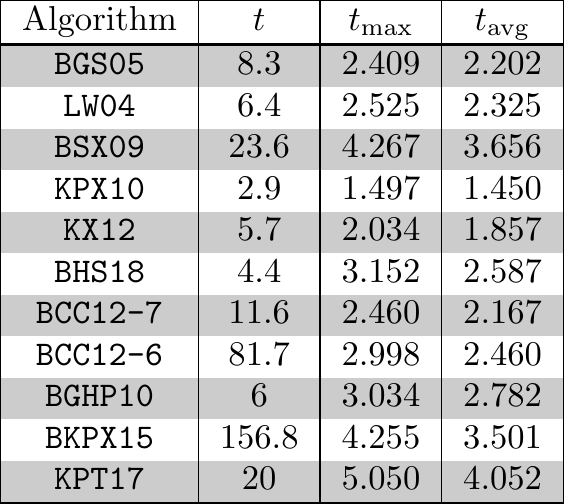}~~~\includegraphics[scale=0.7]{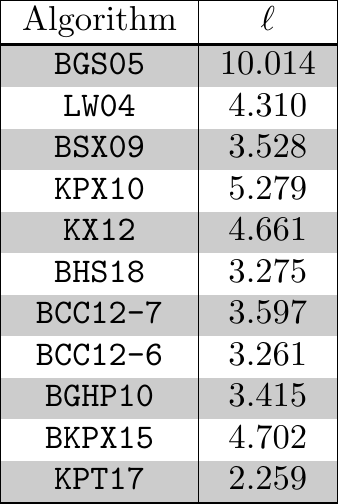}
			\caption{\texttt{uni-disk}}
		\end{subfigure}
		
		\vspace{5mm}

		\begin{subfigure}{0.45\textwidth}
			\centering
			\includegraphics[scale=0.7]{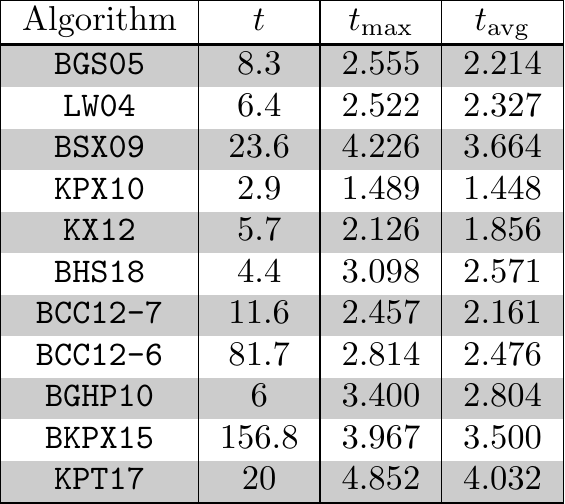}~~~\includegraphics[scale=0.7]{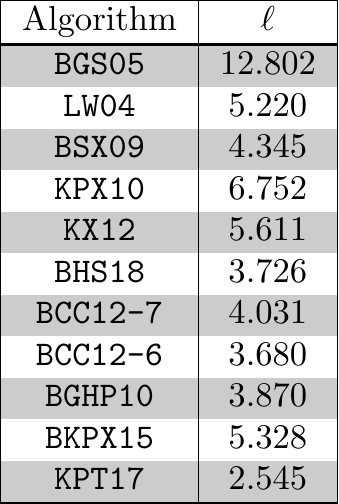}
			\caption{\texttt{normal-clustered}}
		\end{subfigure}
		\begin{subfigure}{0.45\textwidth}
			\centering
			\includegraphics[scale=0.7]{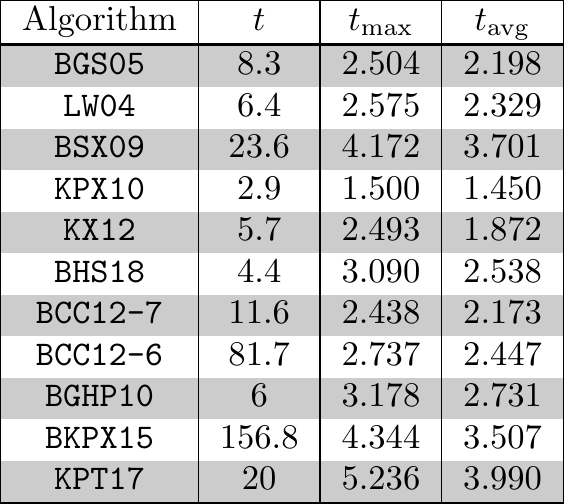}~~~\includegraphics[scale=0.7]{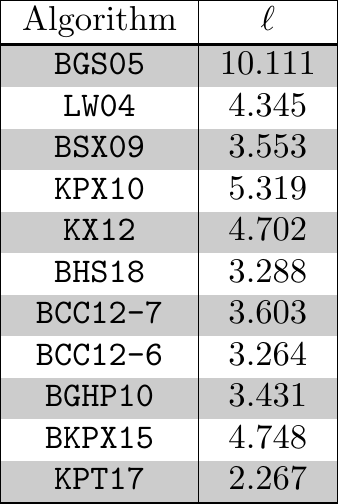}
			\caption{\texttt{normal}}
		\end{subfigure}
		
		\vspace{5mm}

		\begin{subfigure}{0.45\textwidth}
			\centering
			\includegraphics[scale=0.7]{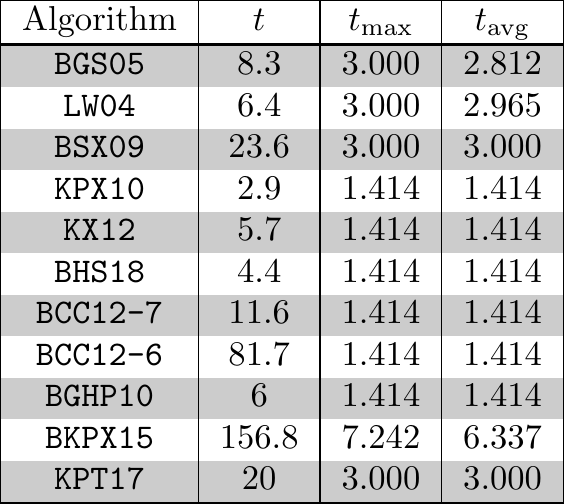}~~~\includegraphics[scale=0.7]{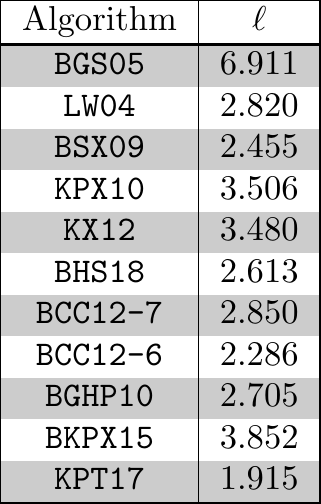}
			\caption{\texttt{grid-contiguous}}
		\end{subfigure}
		\begin{subfigure}{0.45\textwidth}
			\centering
			\includegraphics[scale=0.7]{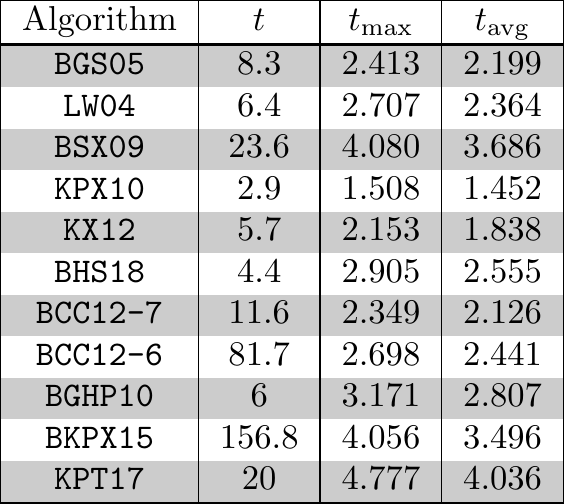}~~~\includegraphics[scale=0.7]{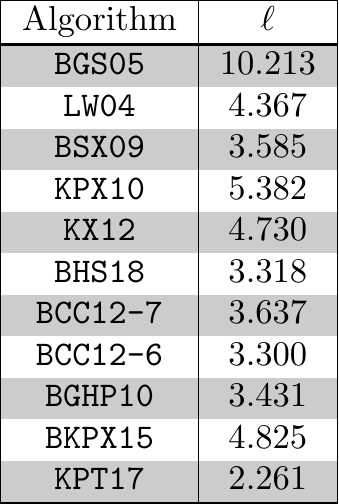}
			\caption{\texttt{grid-random}}
		\end{subfigure}
		
		\vspace{5mm}
		\begin{subfigure}{0.45\textwidth}
			\centering
			\includegraphics[scale=0.7]{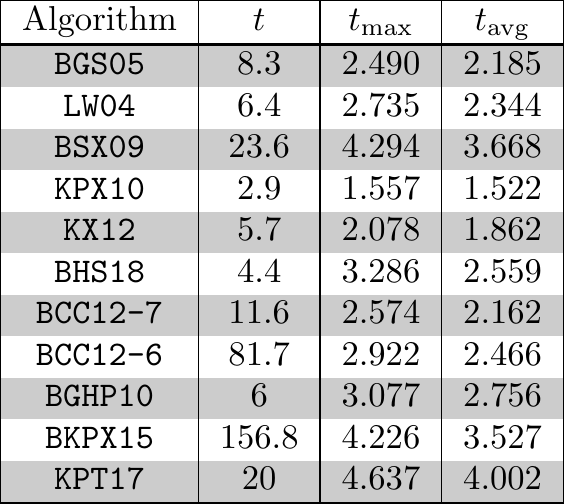}~~~\includegraphics[scale=0.7]{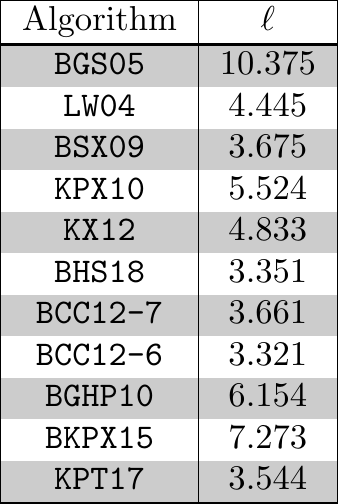}
			\caption{\texttt{annulus}}
		\end{subfigure}
		\begin{subfigure}{0.45\textwidth}
			\centering
			\includegraphics[scale=0.7]{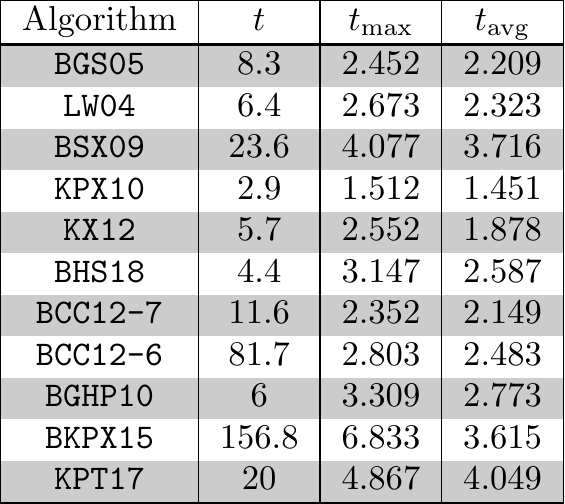}~~~\includegraphics[scale=0.7]{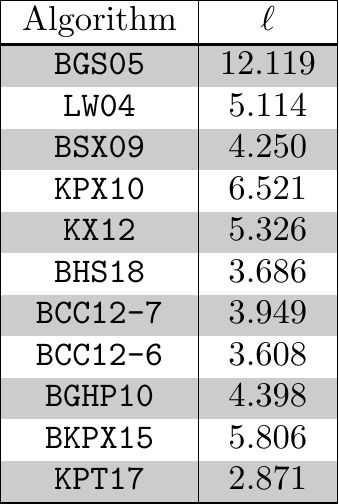}
			\caption{\texttt{galaxy}}
		\end{subfigure}
		
		\caption{Stretch-factor and lightness comparisons of the spanners generated by the eleven algorithms.}
		\label{fig:sf-lightness}
	\end{center}
\end{figure}

\begin{enumerate}

	\item \textit{Runtime}. Fast execution speed is highly desired for spanner construction on large pointsets.  
	We present the runtimes for all eight distributions  in the  Fig.~\ref{fig:runtimePlots}. As explained above, we have excluded \texttt{BGHP10} and \texttt{KPT17} from these plots since they are considerably slower than the other nine algorithms. Interestingly, we find that the relative performance of these algorithms is independent of the point distributions. 
    For all the eight distributions, we find that \texttt{BKPX15} is much slower than the others. This is mainly due to the time taken to construct $L_\infty$-Delauanay triangulations. Among the ones that use $L_2$-Delauanay triangulations, \texttt{BGS05} is the slowest due to the overhead of creation of temporary geometric graphs needed to control the degree and stretch-factor of the ouput spanners. Refer to Section~\ref{chap:algorithms} to see more details on this algorithm. The fastest algorithms are \texttt{KPX10}, \texttt{BSX09}, \texttt{LW04}, and \texttt{KX12}. The main reason behind their speedy performance is fast construction of $L_2$-Delaunay triangulations and lightweight processing of the triangulations for spanner construction. The \texttt{BHS18}, \texttt{BCC12-7}, and \texttt{BCC12-6} algorithms came out quite close the above four algorithms.  Note that these three algorithms also use $L_2$-Delaunay triangulation as the starting point. The same observations hold for the real-world pointsets used in our experiments. See the table presented in Fig.~\ref{fig:realworld-time} for the runtimes in seconds.

	\item \textit{Degree}. Refer to Fig.~\ref{fig:degree}. In the tables, $\Delta$ denotes the theoretical degree upper bound,  as claimed by the authors of these algorithms, max $\Delta_{\text{observed}}$ denotes the maximum degree observed in our experiments, avg $\Delta_{\text{observed}}$ denotes the observed average degree, and avg $\Delta_{\text{vertex}}$ denotes the observed average degree per vertex. 
	In our experiments, we find that spanners generated by \texttt{BGS05}, \texttt{LW04}, and \texttt{BSX09} have degrees much less than the degree upper-bounds derived by the authors. While it cannot be denied that there could be special examples where these upper-bounds are actually achieved,  the maximum degrees achieved in our experiments are $14$, $11$, and $9$, respectively. Note that the theoretical degree upper bounds are $27$, $23$, and $17$, respectively. For the remaining eight algorithms, the claimed degree upper-bounds are achieved in our experiments thereby showing the analyses obtained by the authors of those algorithms are tight. However, the degree  bound claimed by the authors of \texttt{BCC12-6} appears incorrect. We present an example in the Appendix (Section~\ref{chap:bcc_bad}) where the degree of the spanner generated by this algorithm exceeds $6$ (in fact, it is $7$ in this example). For every algorithm, we find that the average degree of the generated spanners is not far away from the maximum observed degrees. It shows that the algorithms are consistent in constructing the spanners. The average degree per vertex is another way to judge the quality of the spanners. In this regard, we find that it is always between $6$ and $3$ everywhere and is quite reasonable for practical purposes. This shows that all these algorithms are very careful when it comes to the selection of spanner edges. The lowest values are achieved by \texttt{BKPX15} and \texttt{KPT17}. For the real-world pointsets, we find similar performance from the algorithms when it comes to the degree and degree per vertex of the spanners. This is quite surprising since these real-world pointsets do not follow specific distributions. Refer to Figs.~\ref{fig:realworld-degree} and \ref{fig:realworld-degpervertex} for more details. Note that \texttt{BSG05} has produced a degree-$15$ spanner for the \texttt{vlsi} pointset. In contrast, for the synthetic pointsets the highest degree we could observe is $14$.
	
	\item \textit{Stretch factor}.  Refer to Fig.~\ref{fig:sf-lightness}. In the tables, $t$ denotes the theoretical stretch factor, as derived by the authors of these algorithms; $t_\text{max}$ denotes the maximum stretch factor observed in our experiments, and $t_\text{avg}$ denotes the average observed stretch factor. Among the eleven algorithms, \texttt{KPX10} has the lowest guaranteed stretch factor - it is $2.9$.  The stretch factors of the spanners generated by \texttt{KPX10} are always less than $1.6$, thereby making it the best among the eleven algorithms in terms of stretch factor.  In this regard, \texttt{BKPX15} turned out to be the worst; the largest stretch factor we have observed is  $7.242$, although it is substantially less than the theoretical stretch factor upper bound of $156.8$. Its competitor \texttt{KPX17} that can also generate degree-$4$ plane spanners has a lower observed maximum stretch factor - it is $5.236$ (theoretical upper bound is $20$ for this algorithm).
	Overall, we find that the stretch factors of the generated spanners are much less than the claimed theoretical upper bounds. This shows that the generated spanners are well-constructed in practice. With the exception of \texttt{BKPX15}, we find that the average stretch factors are quite close to the maximum stretch factors. Now let us turn our attention to the  real-world pointsets. Refer to Fig.~\ref{fig:realworld-sf}. Once again \texttt{KPX10} produced lowest stretch factor spanners.  The stretch factors seem quite reasonable everywhere except the two cases of \texttt{vlsi} and \texttt{uber} pointsets when fed to \texttt{BKPX15}. The produced spanners have stretch factors  of $11.535$ and $27.929$, respectively. The later is interesting since the  lower bound example  constructed by the authors in \cite{bonichon2015there} for the worst-case stretch factor of the spanners produced by \texttt{BKPX15} has a stretch factor of $7+7\sqrt{2} \approx 16.899$. The \texttt{uber} pointset beats this lower bound.
	
	\item \textit{Lightness}.  Since a minimum spanning tree is the cheapest (in terms of the sum of the total length of the edges) way to connect $n$ points, lightness can be used to judge the quality of spanners. This metric is beneficial when spanners are used for constructing computer or transportation networks. Refer to Fig.~\ref{fig:sf-lightness}. Lightness is denoted by $\ell$. With a few exceptions, we find that lightness somewhat correlates with degree. This is because using a lower number of carefully placed spanner edges usually leads to lower lightness. The spanners generated by \texttt{BGS05} are always found to have the highest lightness. This is expected because of their high degrees. Although, the difference in degree of the spanners generate by \texttt{BGXS05} and \texttt{LW04} is marginal (around $2$), the difference between their lightness is substantial (approximately $6$ for some cases). On the other hand, the degree-$4$ spanners generated by \texttt{KPT17} have the lowest lightnesses (less than $2.9$ everywhere). Interestingly, although \texttt{BKPX15} generates degree-$4$ spanners, their lightness is found to be approximately twice that of the ones generated by \texttt{KPT17}. In fact, their lightness turned out to be one of the highest. This shows that \texttt{KPT17} is more careful when it comes to placing long edges. The lightnesses of the spanners generated for real-world pointsets follow a similar trend, and we did not observe anything special. Refer to Fig.~\ref{fig:realworld-lightness} for more details.
	
\end{enumerate}

\textbf{Remark.} In our experiments, we find that the spanners' degree, stretch factor, and lightness remained somewhat constant with the increase in $n$. Hence, we do not present plots for them.

\section{Conclusions}
\label{chap:conclusion}

Since there are various ways (speed, degree, stretch factor, lightness) to judge the eleven algorithms, it is hard to declare the winner(s). So, based on our experimental observations, we come to the following conclusions (are our recommendations as well):
\begin{itemize}\itemsep0pt
	\item If speedy performance is the main concern, we recommend using \texttt{KPX10}, \texttt{BSX09}, \texttt{LW04}, or \texttt{KX12}.
	\item When it comes to minimization of degree, we recommend using \texttt{BCC12-7} or \texttt{BHS18} since they produce spanners of reasonable degrees in practice. If degree-$4$ spanners are desired, we recommend using \texttt{BKPX15} since \texttt{KPT17} is much slower in practice.
	\item In terms of stretch factor, we find the \texttt{KPX10} as the clear winner. This is particularly important in the study of geometric spanners since not much is known about  fast construction of low stretch factor spanners ($t \approx 1.6$) in the plane having at most $3n$ edges. However, the spanners produced by it have higher degrees compared to the ones produced by some of the other algorithms such as \texttt{BCC12} and \texttt{BHS18}. 
	
	\item In our experiments, \texttt{KPT17} produced spanners with the lowest lightnesses. But in practice, we found it to be very slow compared to the other algorithms except for \texttt{BGHP10} (which is as slow as \texttt{KPT17}).  If degree-$4$ spanners is not a requirement, we recommend using \texttt{BHS18} or \texttt{BCC12-7} since they produced spanners of reasonable lightness (less than $4$ most of the times).
\end{itemize}

\textbf{Acknowledgment.} We sincerely thank Nicolas Bonichon (one of the authors of \texttt{BKPX15}) for sharing the applet code for the algorithm \texttt{BKPX15}~\cite{bonichon2015there}. The code has helped us to understand the algorithm clearly and create a \textsf{CGAL} implementation of the algorithm.

\clearpage

\begin{figure}
	
	\centering
	\small

	\rowcolors{1}{ffffff}{cccccc}
	
	{
		\tiny
		
		\begin{tabular}{|c|c|c|c|c|c|c|c|c|c|c|c|c|}
			\hline
			
			Pointset & $n$ & \texttt{BGS05} & \texttt{LW04} & \texttt{BSX09} & \texttt{KPX10} & \texttt{KX12} & \texttt{BHS18} & \texttt{BCC12-7} & \texttt{BCC12-6} & \texttt{BGHP10} & \texttt{BKPX15} & \texttt{KPT17} \\
			
			\hline

			\hline
			$\texttt{burma}$ & $33708\mathrm{}$ &$2.888$ &$0.711$ &$0.627$ &$0.599$ &$0.832$ &$1.101$ &$1.569$ &$1.549$ &$173.528$ &$5.885$ &$173.447$ \\
			
			$\texttt{birch3}$ & $99801\mathrm{}$ &$9.700$ &$2.459$ &$2.139$ &$2.117$ &$3.001$ &$4.458$ &$5.188$ &$5.079$ &$640.619$ &$20.003$ &$641.839$ \\
			
			$\texttt{mona-lisa}$ & $100000$  &$8.527$ &$2.450$ &$2.127$ &$2.236$ &$3.189$ &$4.960$ &$5.863$ &$5.588$ &$704.980$ &$27.988$ &$706.000$ \\
			
			$\texttt{KDDCU2D}$ & $104297$  &$10.024$ &$2.579$ &$2.252$ &$2.250$ &$3.146$ &$4.837$ &$5.569$ &$5.460$ &$811.921$ &$21.424$ &$812.665$ \\
			
			$\texttt{usa}$ & $115475$  &$10.575$ &$2.878$ &$2.497$ &$2.521$ &$3.559$ &$5.600$ &$6.426$ &$6.258$ &$1033.99$ &$35.769$ &$1035.64$ \\
			
			$\texttt{europe}$ & $168896$  &$15.516$ &$4.520$ &$3.985$ &$3.968$ &$5.685$ &$8.386$ &$9.402$ &$9.230$ &$1494.69$ &$39.133$ &$1497.09$ \\
			
			$\texttt{wiki}$ & $317695$   &$35.374$ &$9.041$ &$7.687$ &$8.004$ &$11.432$ &$17.274$ &$18.105$ &$17.794$ &$3507.06$ &$98.975$ &$3510.20$ \\
			
			$\texttt{vsli}$ & $744710$   &$81.461$ &$21.655$ &$19.457$ &$19.693$ &$28.165$ &$46.630$ &$47.297$ &$46.183$ &$13296.6$ &$509.795$ &$13312.2$ \\
			
			$\texttt{china}$ & $808693$  &$91.243$ &$24.078$ &$21.642$ &$21.775$ &$30.793$ &$52.637$ &$54.367$ &$53.029$ &$17028.4$ &$451.383$ &$17057.3$ \\
			
			$\texttt{uber}$ & $1381253$  &$170.493$ &$42.199$ &$38.227$ &$38.268$ &$54.486$ &$89.577$ &$94.444$ &$92.359$ &$29584.6$ &$1689.042$ &$29633.1$ \\
			
			$\texttt{world}$ & $1904711$  &$259.147$ &$57.657$ &$52.078$ &$52.337$ &$74.570$ &$125.078$ &$128.316$ &$125.787$ &$58707.7$ &$1358.70$ &$58780.3$ \\
			
			\hline
		\end{tabular}
	}
	
		\caption{Average execution time (in seconds).}
		\label{fig:realworld-time}
\end{figure}

\begin{figure}
\centering
	
	\rowcolors{1}{ffffff}{cccccc}

	{\tiny \begin{tabular}{|c|c|c|c|c|c|c|c|c|c|c|c|c|c|}
			\hline
			
			Pointset & $n$ & \texttt{BGS05} & \texttt{LW04} & \texttt{BSX09} & \texttt{KPX10} & \texttt{KX12} & \texttt{BHS18} & \texttt{BCC12-7} & \texttt{BCC12-6} & \texttt{BGHP10} & \texttt{BKPX15} & \texttt{KPT17} \\
			
			\hline

			\hline
			$\texttt{burma}$ & $33708\mathrm{}$ &$11$ &$10$ &$8$ &$14$ &$11$ &$6$ &$7$ &$6$ &$6$ &$4$ &$4$ \\
			
			$\texttt{birch3}$ & $99801\mathrm{}$ &$13$ &$10$ &$8$ &$13$ &$11$ &$6$ &$7$ &$6$ &$6$ &$4$ &$4$ \\
			
			$\texttt{mona-lisa}$ & $100000$  &$11$ &$8$ &$8$ &$12$ &$10$ &$7$ &$7$ &$7$ &$6$ &$4$ &$4$ \\
			
			$\texttt{KDDCU2D}$ & $104297$  &$13$ &$10$ &$8$ &$14$ &$10$ &$6$ &$7$ &$7$ &$6$ &$4$ &$4$ \\
			
			$\texttt{usa}$ & $115475$  &$12$ &$10$ &$8$ &$14$ &$11$ &$7$ &$7$ &$7$ &$6$ &$4$ &$4$ \\
			
			$\texttt{europe}$ & $168896$  &$12$ &$10$ &$8$ &$13$ &$10$ &$6$ &$7$ &$6$ &$6$ &$4$ &$4$ \\
			
			$\texttt{wiki}$ & $317695$   &$14$ &$10$ &$9$ &$14$ &$11$ &$7$ &$7$ &$7$ &$6$ &$4$ &$4$ \\
			
			$\texttt{vsli}$ & $744710$   &$15$ &$11$ &$9$ &$14$ &$10$ &$7$ &$7$ &$7$ &$6$ &$4$ &$4$ \\
			
			$\texttt{china}$ & $808693$  &$13$ &$11$ &$9$ &$14$ &$11$ &$7$ &$7$ &$6$ &$6$ &$4$ &$4$ \\
			
			$\texttt{uber}$ & $1381253$  &$13$ &$11$ &$9$ &$14$ &$10$ &$7$ &$7$ &$6$ &$6$ &$4$ &$4$ \\
			
			$\texttt{world}$ & $1904711$  &$14$ &$11$ &$9$ &$14$ &$11$ &$7$ &$7$ &$7$ &$6$ &$4$ &$4$ \\
			
			\hline
		\end{tabular}
	}
	
		\caption{Degree of the spanners.}
	\label{fig:realworld-degree}
\end{figure}

\begin{figure}
\centering
	
	\rowcolors{1}{ffffff}{cccccc}

	{\tiny 
		\begin{tabular}{|c|c|c|c|c|c|c|c|c|c|c|c|c|}
			\hline
			
			Pointset & $n$ & \texttt{BGS05} & \texttt{LW04} & \texttt{BSX09} & \texttt{KPX10} & \texttt{KX12} & \texttt{BHS18} & \texttt{BCC12-7} & \texttt{BCC12-6} & \texttt{BGHP10} & \texttt{BKPX15} & \texttt{KPT17} \\
			
			\hline

			\hline
			$\texttt{burma}$ & $33708\mathrm{}$ &$5.761$ &$5.192$ &$4.454$ &$5.994$ &$5.512$ &$4.166$ &$4.292$ &$4.068$ &$4.347$ &$3.346$ &$3.187$ \\
			
			$\texttt{birch3}$ & $99801\mathrm{}$ &$5.745$ &$5.149$ &$4.417$ &$5.995$ &$5.429$ &$4.081$ &$4.287$ &$3.970$ &$4.222$ &$3.327$ &$3.130$ \\
			
			$\texttt{mona-lisa}$ & $100000$  &$5.938$ &$5.596$ &$4.617$ &$5.996$ &$5.981$ &$5.259$ &$5.741$ &$5.165$ &$5.434$ &$3.572$ &$3.613$ \\
			
			$\texttt{KDDCU2D}$ & $104297$  &$5.720$ &$5.127$ &$4.399$ &$5.985$ &$5.390$ &$4.047$ &$4.294$ &$4.015$ &$4.216$ &$3.325$ &$3.122$ \\
			
			$\texttt{usa}$ & $115475$  &$5.761$ &$5.208$ &$4.447$ &$5.993$ &$5.529$ &$4.248$ &$4.493$ &$4.132$ &$4.398$ &$3.366$ &$3.211$ \\
			
			$\texttt{europe}$ & $168896$  &$5.745$ &$5.160$ &$4.427$ &$5.997$ &$5.438$ &$4.090$ &$4.310$ &$3.991$ &$4.234$ &$3.325$ &$3.135$ \\
			
			$\texttt{wiki}$ & $317695$   &$5.679$ &$5.061$ &$4.379$ &$5.987$ &$5.333$ &$3.931$ &$4.016$ &$3.732$ &$4.070$ &$3.308$ &$3.016$ \\
			
			$\texttt{vsli}$ & $744710$   &$5.749$ &$5.152$ &$4.416$ &$5.994$ &$5.438$ &$4.096$ &$4.334$ &$4.007$ &$4.277$ &$3.316$ &$3.176$ \\
			
			$\texttt{china}$ & $808693$  &$5.770$ &$5.209$ &$4.437$ &$5.996$ &$5.519$ &$4.245$ &$4.506$ &$4.154$ &$4.368$ &$3.344$ &$3.208$ \\
			
			$\texttt{uber}$ & $1381253$  &$5.742$ &$5.147$ &$4.394$ &$5.996$ &$5.437$ &$4.086$ &$4.288$ &$3.968$ &$4.232$ &$3.326$ &$3.130$ \\
			
			$\texttt{world}$ & $1904711$  &$5.748$ &$5.171$ &$4.438$ &$5.991$ &$5.489$ &$4.151$ &$4.371$ &$4.020$ &$4.318$ &$3.344$ &$3.168$ \\
			
			\hline
		\end{tabular}
	}

	\caption{Average degree per vertex.}
	\label{fig:realworld-degpervertex}
\end{figure}

\begin{figure}

	\centering

	\rowcolors{1}{ffffff}{cccccc}

	{\tiny \begin{tabular}{|c|c|c|c|c|c|c|c|c|c|c|c|c|}
			\hline
			
			Pointset & $n$ & \texttt{BGS05} & \texttt{LW04} & \texttt{BSX09} & \texttt{KPX10} & \texttt{KX12} & \texttt{BHS18} & \texttt{BCC12-7} & \texttt{BCC12-6} & \texttt{BGHP10} & \texttt{BKPX15} & \texttt{KPT17} \\
			
			\hline

			\hline
			$\texttt{burma}$ & $33708\mathrm{}$ &$2.414$ &$2.414$ &$3.681$ &$1.482$ &$1.738$ &$2.856$ &$2.156$ &$2.162$ &$3.161$ &$4.404$ &$4.409$ \\
			
			$\texttt{birch3}$ & $99801\mathrm{}$ &$2.233$ &$2.234$ &$3.520$ &$1.481$ &$1.923$ &$2.719$ &$2.318$ &$2.460$ &$2.933$ &$3.624$ &$4.102$ \\
			
			$\texttt{mona-lisa}$ & $100000$  &$2.523$ &$2.237$ &$3.373$ &$1.413$ &$1.609$ &$2.872$ &$1.778$ &$2.278$ &$2.872$ &$4.190$ &$3.768$ \\
			
			$\texttt{KDDCU2D}$ & $104297$  &$2.211$ &$2.435$ &$3.953$ &$1.492$ &$2.068$ &$2.937$ &$2.174$ &$2.603$ &$2.937$ &$4.299$ &$4.218$ \\
			
			$\texttt{usa}$ & $115475$  &$2.300$ &$2.351$ &$3.564$ &$1.480$ &$2.038$ &$2.765$ &$2.241$ &$2.576$ &$3.430$ &$3.740$ &$4.455$ \\
			
			$\texttt{europe}$ & $168896$  &$2.245$ &$2.343$ &$4.072$ &$1.459$ &$1.840$ &$2.745$ &$2.310$ &$2.659$ &$2.981$ &$4.081$ &$4.121$ \\
			
			$\texttt{wiki}$ & $317695$   &$2.408$ &$2.421$ &$3.926$ &$1.458$ &$1.978$ &$2.757$ &$2.350$ &$2.610$ &$3.222$ &$4.353$ &$4.017$ \\
			
			$\texttt{vsli}$ & $744710$   &$2.468$ &$2.999$ &$3.650$ &$1.471$ &$1.970$ &$2.942$ &$2.355$ &$2.263$ &$3.521$ &$11.535$ &$5.472$ \\
			
			$\texttt{china}$ & $808693$  &$2.478$ &$2.421$ &$4.082$ &$1.511$ &$2.055$ &$2.731$ &$2.237$ &$2.711$ &$2.981$ &$4.061$ &$4.506$ \\
			
			$\texttt{uber}$ & $1381253$  &$2.535$ &$2.418$ &$3.987$ &$1.485$ &$2.204$ &$2.902$ &$2.407$ &$2.816$ &$3.073$ &$27.929$ &$4.966$ \\
			
			$\texttt{world}$ & $1904711$  &$2.989$ &$2.961$ &$4.228$ &$1.522$ &$1.997$ &$3.056$ &$2.357$ &$2.657$ &$3.545$ &$6.140$ &$5.422$ \\
			
			\hline
	\end{tabular}}
	\caption{Stretch factor of the spanners.}
	\label{fig:realworld-sf}
\end{figure}

\begin{figure}
	\centering

	\rowcolors{1}{ffffff}{cccccc}

	{\tiny 
		
		\begin{tabular}{|c|c|c|c|c|c|c|c|c|c|c|c|c|}
			\hline
			
			Pointset & $n$ & \texttt{BGS05} & \texttt{LW04} & \texttt{BSX09} & \texttt{KPX10} & \texttt{KX12} & \texttt{BHS18} & \texttt{BCC12-7} & \texttt{BCC12-6} & \texttt{BGHP10} & \texttt{BKPX15} & \texttt{KPT17} \\
			
			\hline

			\hline
			$\texttt{burma}$ & $33708\mathrm{}$ &$10.755$ &$4.538$ &$3.768$ &$5.672$ &$4.922$ &$3.374$ &$3.609$ &$3.365$ &$3.620$ &$5.048$ &$2.345$ \\
			
			$\texttt{birch3}$ & $99801\mathrm{}$ &$11.008$ &$4.660$ &$3.845$ &$5.805$ &$4.989$ &$3.453$ &$3.770$ &$3.434$ &$3.699$ &$5.124$ &$2.426$ \\
			
			$\texttt{mona-lisa}$ & $100000$  &$7.070$ &$3.259$ &$2.656$ &$3.574$ &$3.542$ &$3.012$ &$3.326$ &$2.934$ &$3.147$ &$3.934$ &$1.994$ \\
			
			$\texttt{KDDCU2D}$ & $104297$  &$10.576$ &$4.491$ &$3.719$ &$5.605$ &$4.830$ &$3.316$ &$3.670$ &$3.355$ &$3.511$ &$4.954$ &$2.311$ \\
			
			$\texttt{usa}$ & $115475$  &$10.264$ &$4.427$ &$3.663$ &$5.431$ &$4.753$ &$3.336$ &$3.642$ &$3.305$ &$3.602$ &$4.935$ &$2.336$ \\
			
			$\texttt{europe}$ & $168896$  &$10.136$ &$4.365$ &$3.593$ &$5.395$ &$4.736$ &$3.274$ &$3.588$ &$3.233$ &$3.428$ &$4.746$ &$2.254$ \\
			
			$\texttt{wiki}$ & $317695$   &$12.137$ &$5.087$ &$4.227$ &$6.555$ &$5.385$ &$3.607$ &$3.877$ &$3.525$ &$3.860$ &$5.477$ &$2.481$ \\
			
			$\texttt{vsli}$ & $744710$   &$11.344$ &$4.850$ &$4.024$ &$5.989$ &$5.110$ &$3.521$ &$3.899$ &$3.539$ &$3.864$ &$5.130$ &$2.513$ \\
			
			$\texttt{china}$ & $808693$  &$9.918$ &$4.304$ &$3.531$ &$5.232$ &$4.605$ &$3.286$ &$3.594$ &$3.247$ &$3.445$ &$4.719$ &$2.263$ \\
			
			$\texttt{uber}$ & $1381253$  &$11.225$ &$4.497$ &$4.291$ &$5.900$ &$5.424$ &$2.797$ &$2.843$ &$2.888$ &$3.015$ &$4.584$ &$1.861$ \\
			
			$\texttt{world}$ & $1904711$  &$11.145$ &$4.744$ &$3.923$ &$5.917$ &$5.003$ &$3.476$ &$3.777$ &$3.432$ &$3.967$ &$5.272$ &$2.541$ \\
			
			\hline
		\end{tabular}
	}
		\caption{Lightness of the spanners.}
		\label{fig:realworld-lightness}
\end{figure}

\newpage
\clearpage
\bibliographystyle{spmpsci.bst}
\bibliography{ref}

\begin{thebibliography}{10}

\bibitem{tsp}
\url{www.math.uwaterloo.ca/tsp/}.

\bibitem{agarwal2008computing}
Pankaj~K Agarwal, Rolf Klein, Christian Knauer, Stefan Langerman, Pat Morin,
  Micha Sharir, and Michael Soss.
\newblock Computing the detour and spanning ratio of paths, trees, and cycles
  in 2d and 3d.
\newblock {\em Discrete \& Computational Geometry}, 39(1):17--37, 2008.

\bibitem{anderson2021interactive}
Fred Anderson, Anirban Ghosh, Matthew Graham, Lucas Mougeot, and David
  Wisnosky.
\newblock An interactive tool for experimenting with bounded-degree plane
  geometric spanners (media exposition).
\newblock In {\em 37th International Symposium on Computational Geometry (SoCG
  2021)}. Schloss Dagstuhl-Leibniz-Zentrum f{\"u}r Informatik, 2021.

\bibitem{bakhshesh2020degree}
Davood Bakhshesh and Mohammad Farshi.
\newblock A degree 3 plane 5.19-spanner for points in convex position.
\newblock In {\em CCCG}, pages 226--232, 2020.

\bibitem{bentley1990k}
Jon~Louis Bentley.
\newblock K-d trees for semidynamic point sets.
\newblock In {\em Proceedings of the Sixth Annual Symposium on Computational
  Geometry}, pages 187--197, 1990.

\bibitem{biniaz2020plane}
Ahmad Biniaz.
\newblock Plane hop spanners for unit disk graphs: {S}impler and better.
\newblock {\em Comput. Geom.}, 89:101622, 2020.

\bibitem{biniaz2017towards}
Ahmad Biniaz, Prosenjit Bose, Jean-Lou De~Carufel, Cyril Gavoille, Anil
  Maheshwari, and Michiel Smid.
\newblock Towards plane spanners of degree 3.
\newblock {\em Journal of Computational Geometry}, 8(1):11--31, 2017.

\bibitem{bonichon2010connections}
Nicolas Bonichon, Cyril Gavoille, Nicolas Hanusse, and David Ilcinkas.
\newblock Connections between theta-graphs, {D}elaunay triangulations, and
  orthogonal surfaces.
\newblock In {\em International Workshop on Graph-Theoretic Concepts in
  Computer Science}, pages 266--278. Springer, 2010.

\bibitem{bonichon2010plane}
Nicolas Bonichon, Cyril Gavoille, Nicolas Hanusse, and Ljubomir Perkovi{\'c}.
\newblock Plane spanners of maximum degree six.
\newblock In {\em International Colloquium on Automata, Languages, and
  Programming}, pages 19--30. Springer, 2010.

\bibitem{bonichon2012stretch}
Nicolas Bonichon, Cyril Gavoille, Nicolas Hanusse, and Ljubomir Perkovi{\'c}.
\newblock The stretch factor of ${L}_1$-and ${L}_\infty$-{D}elaunay
  triangulations.
\newblock In {\em European Symposium on Algorithms}, pages 205--216. Springer,
  2012.

\bibitem{bonichon2015there}
Nicolas Bonichon, Iyad Kanj, Ljubomir Perkovi{\'c}, and Ge~Xia.
\newblock There are plane spanners of degree 4 and moderate stretch factor.
\newblock {\em Discrete \& Computational Geometry}, 53(3):514--546, 2015.

\bibitem{bose2012bounded}
Prosenjit Bose, Paz Carmi, and Lilach Chaitman-Yerushalmi.
\newblock On bounded degree plane strong geometric spanners.
\newblock {\em Journal of Discrete Algorithms}, 15:16--31, 2012.

\bibitem{bose2005constructing}
Prosenjit Bose, Joachim Gudmundsson, and Michiel Smid.
\newblock Constructing plane spanners of bounded degree and low weight.
\newblock {\em Algorithmica}, 42(3-4):249--264, 2005.

\bibitem{bose2018improved}
Prosenjit Bose, Darryl Hill, and Michiel Smid.
\newblock Improved spanning ratio for low degree plane spanners.
\newblock {\em Algorithmica}, 80(3):935--976, 2018.

\bibitem{bose2013plane}
Prosenjit Bose and Michiel Smid.
\newblock On plane geometric spanners: A survey and open problems.
\newblock {\em Computational Geometry}, 46(7):818--830, 2013.

\bibitem{bose2009delaunay}
Prosenjit Bose, Michiel Smid, and Daming Xu.
\newblock {D}elaunay and diamond triangulations contain spanners of bounded
  degree.
\newblock {\em International Journal of Computational Geometry \&
  Applications}, 19(02):119--140, 2009.

\bibitem{bus2018practical}
Norbert Bus, Nabil~H Mustafa, and Saurabh Ray.
\newblock Practical and efficient algorithms for the geometric hitting set
  problem.
\newblock {\em Discrete Applied Mathematics}, 240:25--32, 2018.

\bibitem{callahan1995decomposition}
Paul~B Callahan and S~Rao Kosaraju.
\newblock A decomposition of multidimensional point sets with applications to
  k-nearest-neighbors and n-body potential fields.
\newblock {\em Journal of the ACM (JACM)}, 42(1):67--90, 1995.

\bibitem{catusse2010planar}
Nicolas Catusse, Victor Chepoi, and Yann Vax{\`e}s.
\newblock Planar hop spanners for unit disk graphs.
\newblock In {\em International Symposium on Algorithms and Experiments for
  Sensor Systems, Wireless Networks and Distributed Robotics}, pages 16--30.
  Springer, 2010.

\bibitem{cheng2012approximating}
Siu-Wing Cheng, Christian Knauer, Stefan Langerman, and Michiel Smid.
\newblock Approximating the average stretch factor of geometric graphs.
\newblock {\em Journal of Computational Geometry}, 3(1):132--153, 2012.

\bibitem{chew1986there}
L~Paul Chew.
\newblock There is a planar graph almost as good as the complete graph.
\newblock In {\em Proceedings of the Second Annual Symposium on Computational
  Geometry}, 1986.

\bibitem{chew1989there}
L~Paul Chew.
\newblock There are planar graphs almost as good as the complete graph.
\newblock {\em Journal of Computer and System Sciences}, 39(2):205--219, 1989.

\bibitem{das1996constructing}
Gautam Das and Paul~J Heffernan.
\newblock Constructing degree-3 spanners with other sparseness properties.
\newblock {\em International Journal of Foundations of Computer Science},
  7(02):121--135, 1996.

\bibitem{dumitrescu2016lattice}
Adrian Dumitrescu and Anirban Ghosh.
\newblock Lattice spanners of low degree.
\newblock {\em Discrete Mathematics, Algorithms and Applications},
  8(03):1650051, 2016.

\bibitem{dumitrescu2016lower}
Adrian Dumitrescu and Anirban Ghosh.
\newblock Lower bounds on the dilation of plane spanners.
\newblock {\em International Journal of Computational Geometry \&
  Applications}, 26(02):89--110, 2016.

\bibitem{dumitrescu2020sparse}
Adrian Dumitrescu, Anirban Ghosh, and Csaba~D T{\'o}th.
\newblock Sparse hop spanners for unit disk graphs.
\newblock In {\em 31st International Symposium on Algorithms and Computation
  (ISAAC 2020)}. Schloss Dagstuhl-Leibniz-Zentrum f{\"u}r Informatik, 2020.

\bibitem{farshi2009experimental}
Mohammad Farshi and Joachim Gudmundsson.
\newblock Experimental study of geometric $t$-spanners.
\newblock {\em Journal of Experimental Algorithmics (JEA)}, 14:3, 2009.

\bibitem{federickson1987fast}
Greg~N Federickson.
\newblock Fast algorithms for shortest paths in planar graphs, with
  applications.
\newblock {\em SIAM Journal on Computing}, 16(6):1004--1022, 1987.

\bibitem{FriederichUDC}
Rachel Friederich, Matthew Graham, Anirban Ghosh, Brian Hicks, and Ronald
  Shevchenko.
\newblock Experiments with unit disk cover algorithms for covering massive
  pointsets, 2022.
\newblock URL: \url{https://arxiv.org/abs/2205.01716}.

\bibitem{ghosh2019unit}
Anirban Ghosh, Brian Hicks, and Ronald Shevchenko.
\newblock Unit disk cover for massive point sets.
\newblock In {\em International Symposium on Experimental Algorithms}, pages
  142--157. Springer, 2019.

\bibitem{itinerantgames_2014}
Itinerantgames.
\newblock A 2d procedural galaxy with c++, Mar 2014.
\newblock URL:
  \url{https://itinerantgames.tumblr.com/post/78592276402/a-2d-procedural-galaxy-with-c}.

\bibitem{kanj2017degree}
Iyad Kanj, Ljubomir Perkovic, and Duru T{\"u}rkoǧlu.
\newblock Degree four plane spanners: Simpler and better.
\newblock {\em Journal of Computational Geometry}, 8(2):3--31, 2017.

\bibitem{kanj2010spanners}
Iyad~A Kanj, Ljubomir Perkovi{\'c}, and Ge~Xia.
\newblock On spanners and lightweight spanners of geometric graphs.
\newblock {\em SIAM Journal on Computing}, 39(6):2132--2161, 2010.

\bibitem{kanj2012improved}
Iyad~A Kanj and Ge~Xia.
\newblock Improved local algorithms for spanner construction.
\newblock {\em Theoretical Computer Science}, 453:54--64, 2012.

\bibitem{klein2015most}
Rolf Klein, Martin Kutz, and Rainer Penninger.
\newblock Most finite point sets in the plane have dilation $>1$.
\newblock {\em Discrete \& Computational Geometry}, 53(1):80--106, 2015.

\bibitem{li2004efficient}
Xiang-Yang Li and Yu~Wang.
\newblock Efficient construction of low weighted bounded degree planar spanner.
\newblock {\em International Journal of Computational Geometry \&
  Applications}, 14(01n02):69--84, 2004.

\bibitem{mulzer2004minimum}
Wolfgang Mulzer.
\newblock Minimum dilation triangulations for the regular $n$-gon.
\newblock {\em Master's thesis, Freie Universit{\"a}t Berlin, Germany}, 2004.

\bibitem{narasimhan2000approximating}
Giri Narasimhan and Michiel Smid.
\newblock Approximating the stretch factor of {E}uclidean graphs.
\newblock {\em SIAM Journal on Computing}, 30(3):978--989, 2000.

\bibitem{narasimhan2007geometric}
Giri Narasimhan and Michiel Smid.
\newblock {\em Geometric spanner networks}.
\newblock Cambridge University Press, 2007.

\bibitem{narasimhan2001geometric}
Giri Narasimhan and Martin Zachariasen.
\newblock Geometric minimum spanning trees via well-separated pair
  decompositions.
\newblock {\em Journal of Experimental Algorithmics (JEA)}, 6:6--es, 2001.

\bibitem{cgal:eb-21b}
{The CGAL Project}.
\newblock {\em {CGAL} User and Reference Manual}.
\newblock {CGAL Editorial Board}, {5.3} edition, 2021.
\newblock URL: \url{https://doc.cgal.org/5.3/Manual/packages.html}.

\bibitem{toth2017handbook}
Csaba~D Toth, Joseph O'Rourke, and Jacob~E Goodman.
\newblock {\em Handbook of Discrete and Computational Geometry}.
\newblock Chapman and Hall/CRC, 2017.

\bibitem{wulff2010computing}
Christian Wulff-Nilsen.
\newblock Computing the maximum detour of a plane geometric graph in
  subquadratic time.
\newblock {\em Journal of Computational Geometry}, 1(1):101--122, 2010.

\bibitem{xia2013stretch}
Ge~Xia.
\newblock The stretch factor of the {D}elaunay triangulation is less than
  1.998.
\newblock {\em SIAM Journal on Computing}, 42(4):1620--1659, 2013.

\end{thebibliography}

\clearpage


\section{Appendix}

\subsection{Sample outputs}
\label{chap:samples}

\begin{figure}[h]\begin{minipage}[h]{0.46\linewidth}
		\centering
		\resizebox{\linewidth}{!}{\includegraphics{Demo-RPIS-150.pdf}}
		\caption{A $150$-element pointset, drawn randomly from a square. }
		\label{fig:n150}
	\end{minipage}
	\hfill
	\begin{minipage}[h]{0.46\linewidth}
		\resizebox{\linewidth}{!}{\includegraphics{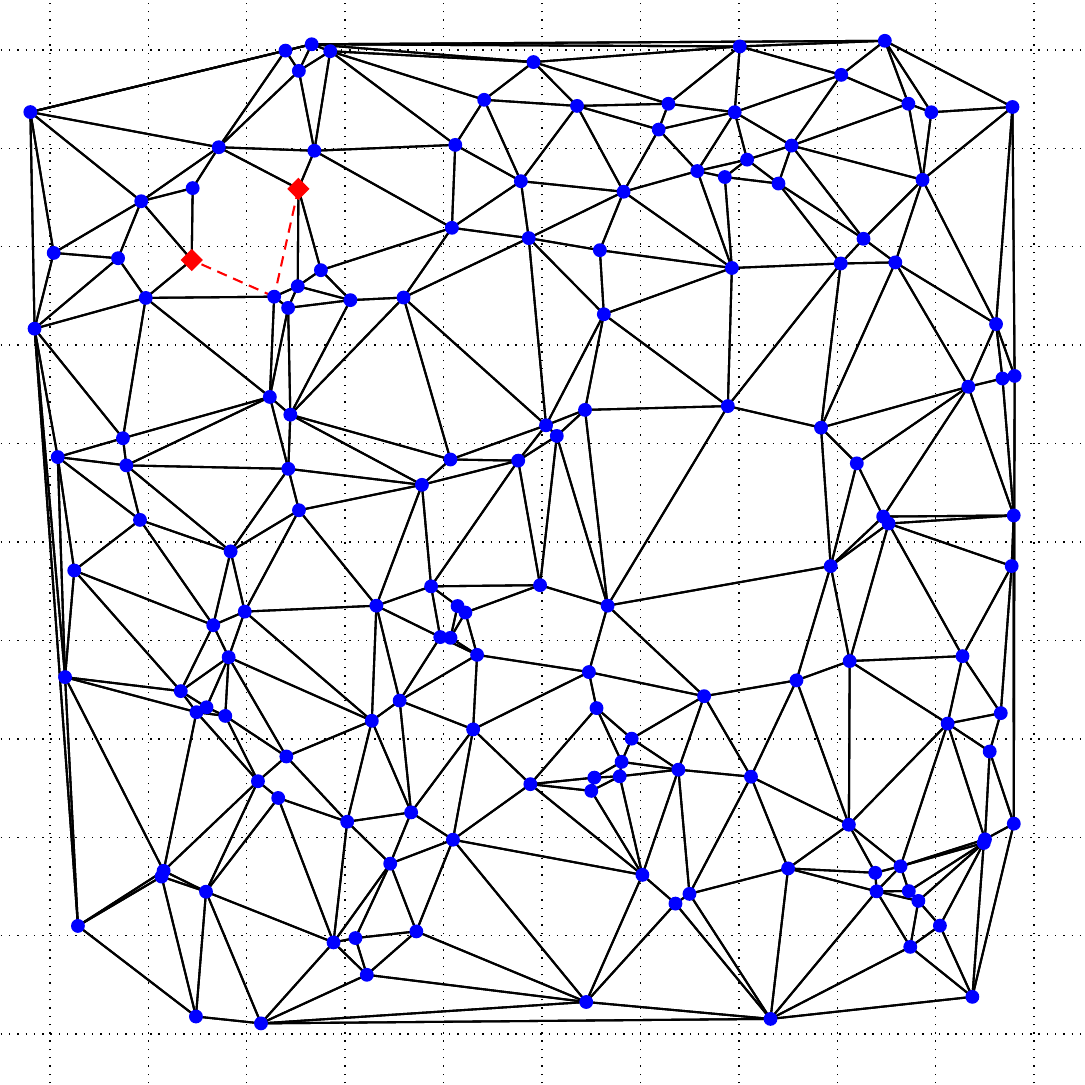}}
		\caption{The spanner generated by \texttt{BGS05} on the pointset shown in Fig.~\ref{fig:n150};  degree: $8$, stretch factor: $1.565763$}
		\label{fig:demo-BGS05}
	\end{minipage}

	\begin{minipage}[h]{0.46\linewidth}
		\centering
		\resizebox{\linewidth}{!}{\includegraphics{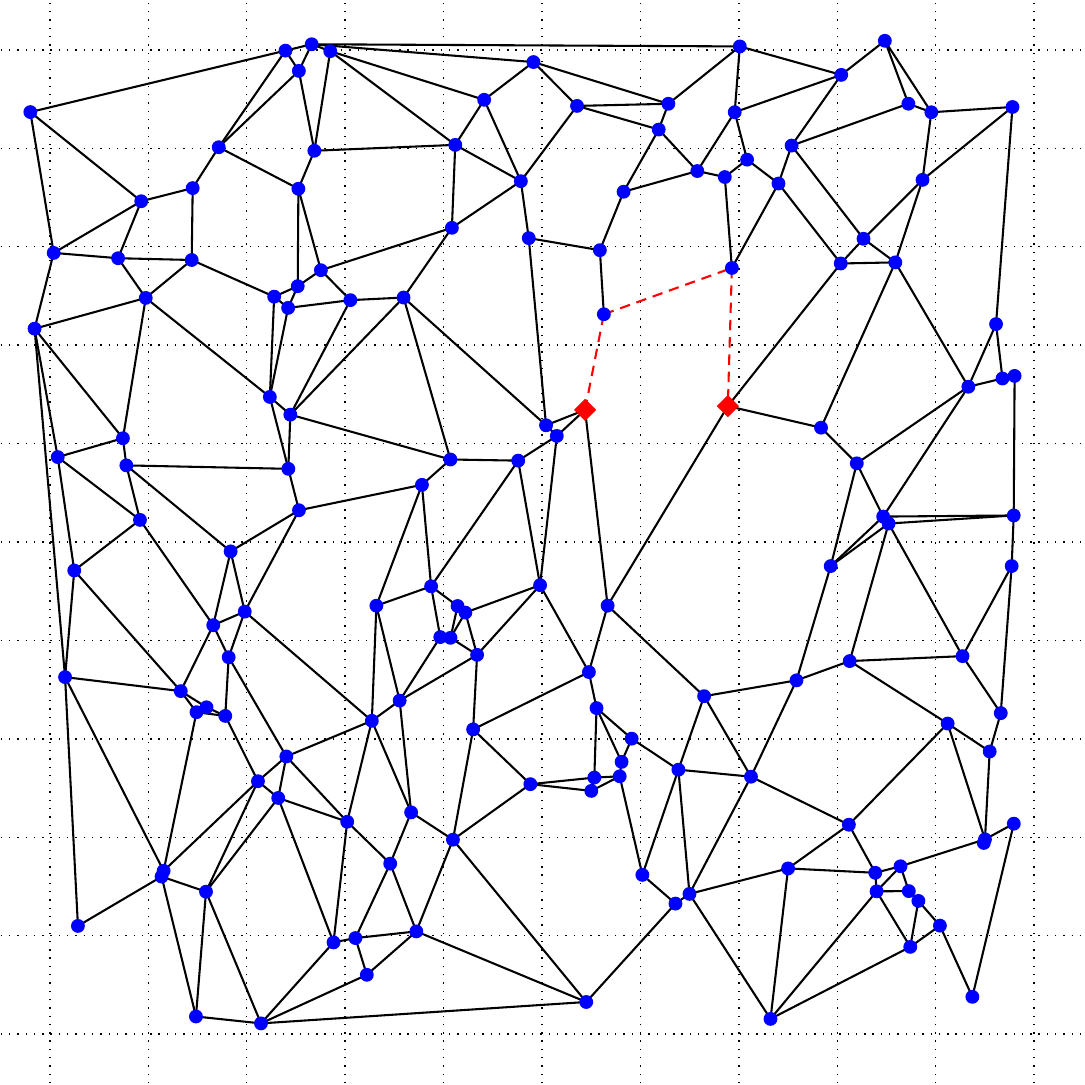}}
		\caption{The spanner generated by \texttt{LW04} on the pointset shown in Fig.~\ref{fig:n150};  degree: $6$, stretch factor: $2.602559$}
		\label{fig:demo-LW04}
	\end{minipage}
	\hfill
	\begin{minipage}[h]{0.46\linewidth}
		\centering
		\resizebox{\linewidth}{!}{\includegraphics{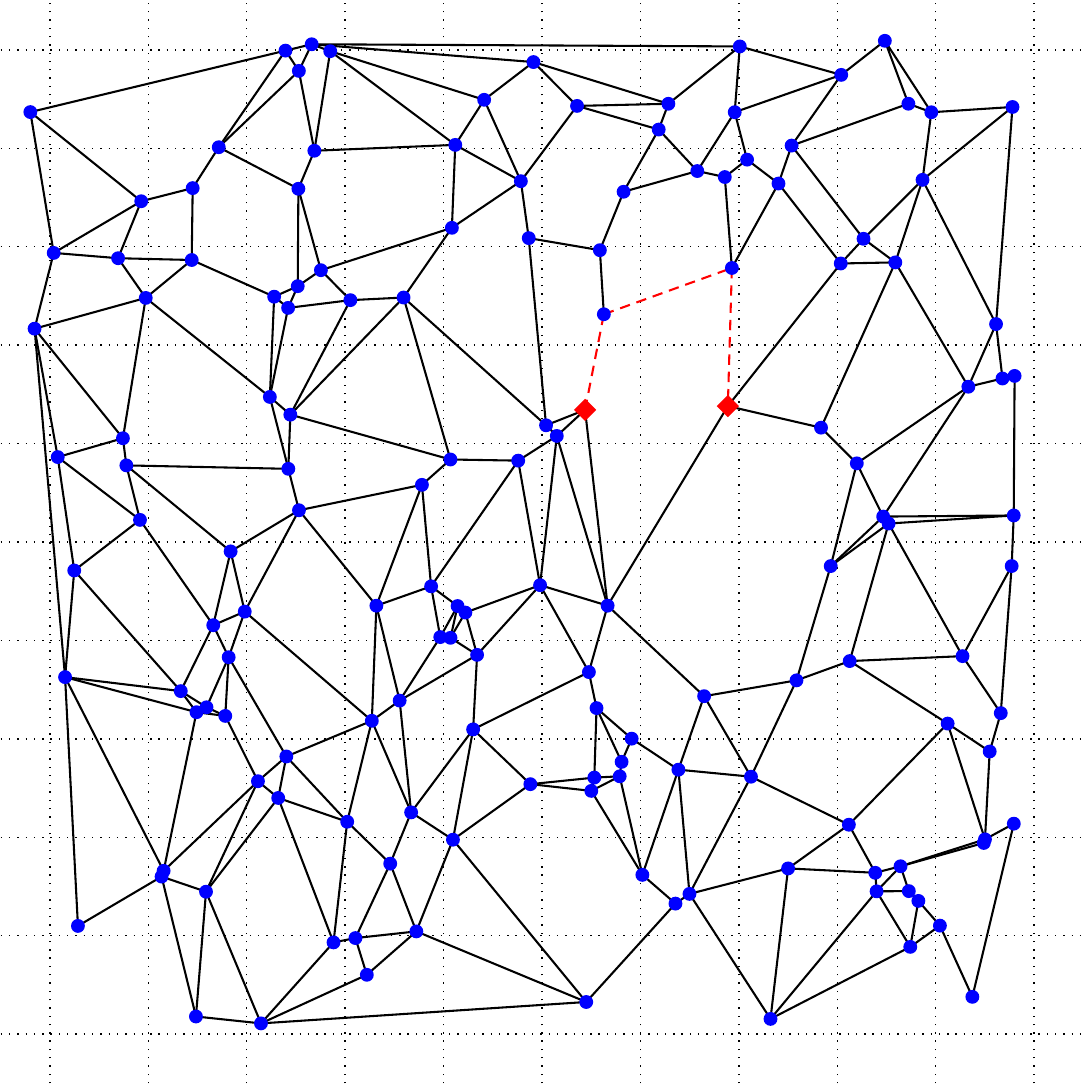}}
		\caption{The spanner generated by \texttt{BSX09} on the pointset shown in Fig.~\ref{fig:n150};  degree: $6$, stretch factor: $2.602559$}
		\label{fig:demo-BSX09}
	\end{minipage}
	\vspace{-10cm}
\end{figure}

\clearpage

\begin{figure}[h]
	\begin{minipage}[h]{0.46\linewidth}
		\centering
		\resizebox{\linewidth}{!}{\includegraphics{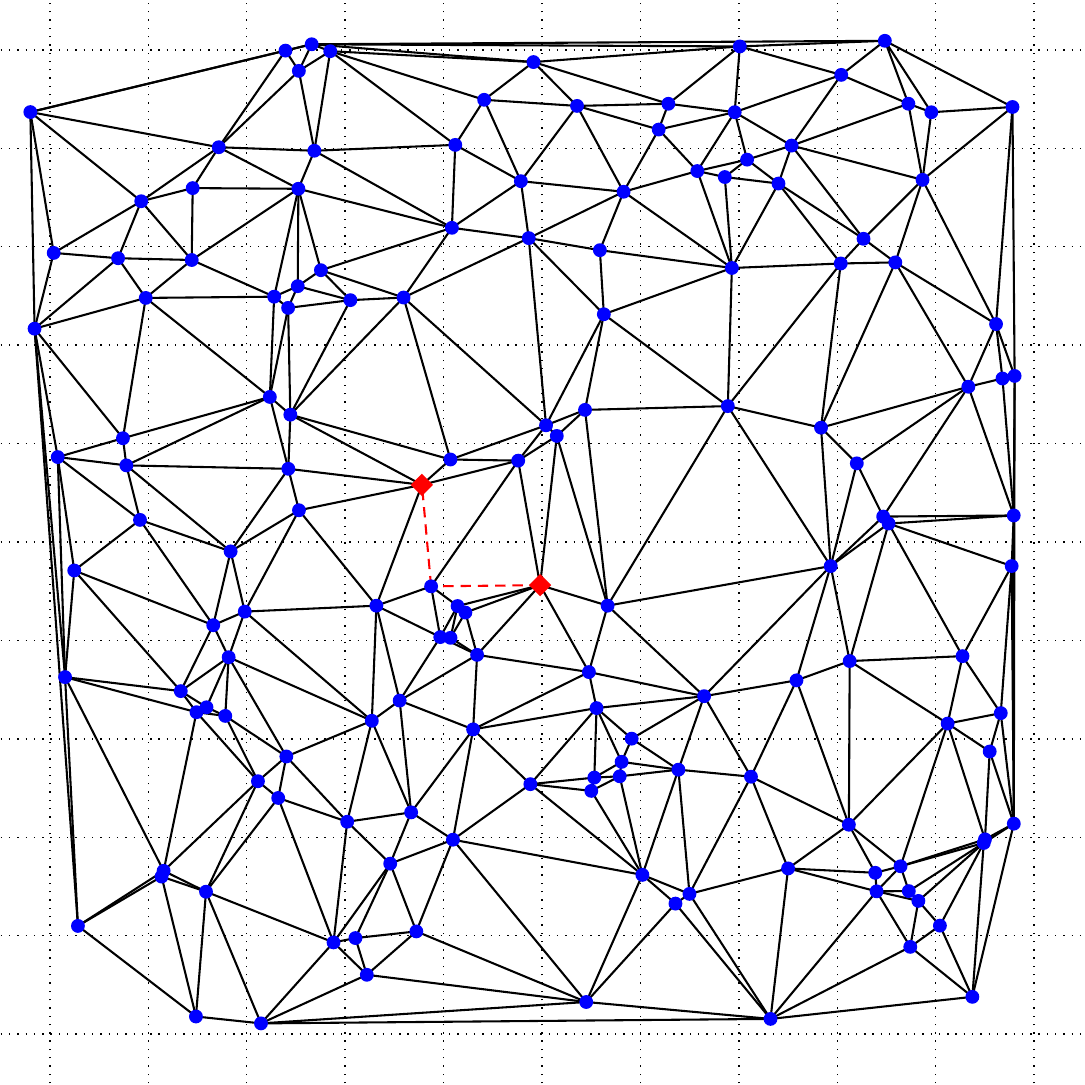}}
		\caption{The spanner generated by \texttt{KPX10} on the pointset shown in Fig.~\ref{fig:n150};  degree: $9$, stretch factor: $1.360771$}
		\label{fig:demo-KPX10}
	\end{minipage}
	\hfill
	\begin{minipage}[h]{0.46\linewidth}
		\centering
		\resizebox{\linewidth}{!}{\includegraphics{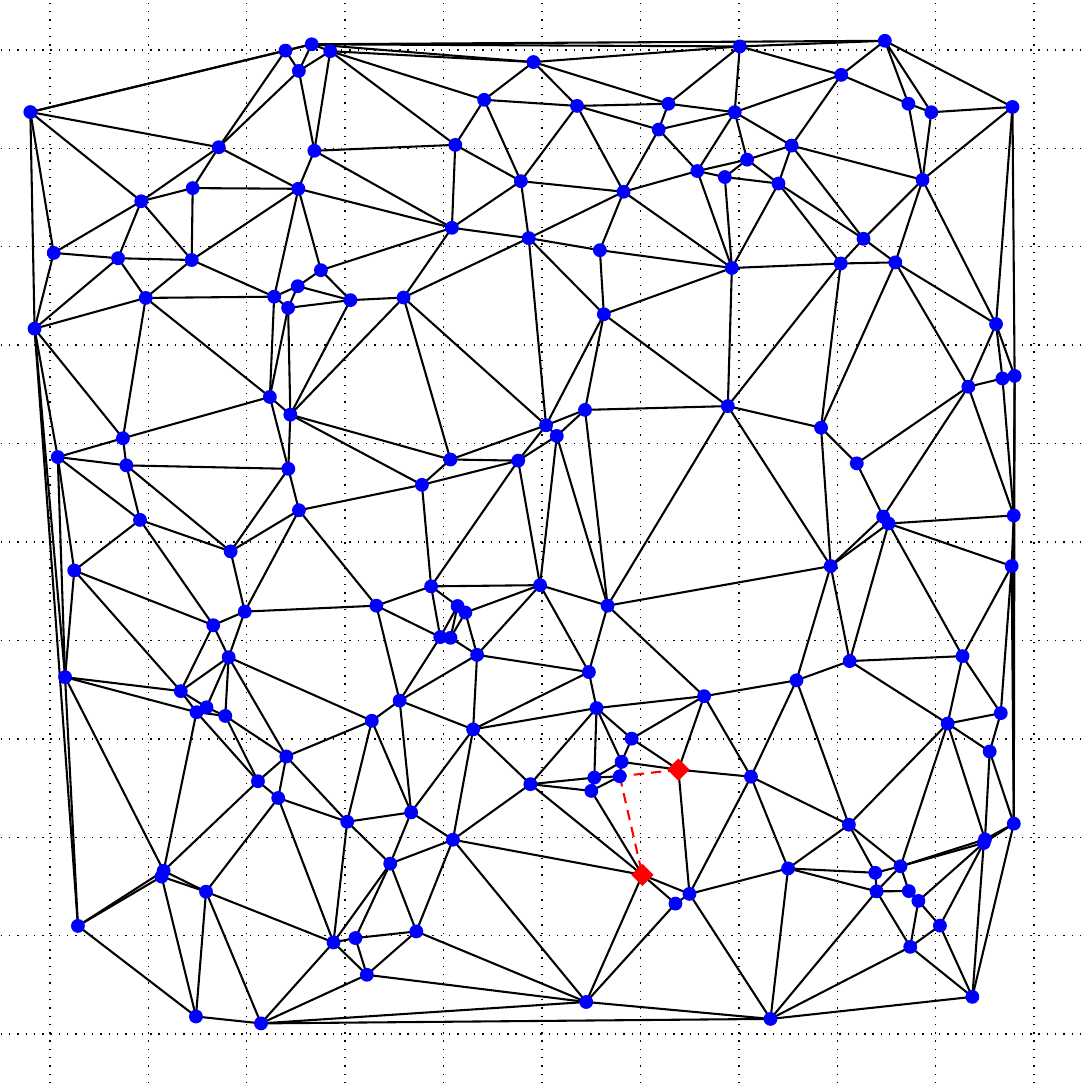}}
		\caption{The spanner generated by \texttt{KX12} on the pointset shown in Fig.~\ref{fig:n150};  degree: $8$, stretch factor: $1.440861$}
		\label{fig:demo-KX12}
	\end{minipage}

	\begin{minipage}[h]{0.46\linewidth}
		\centering
		\resizebox{\linewidth}{!}{\includegraphics{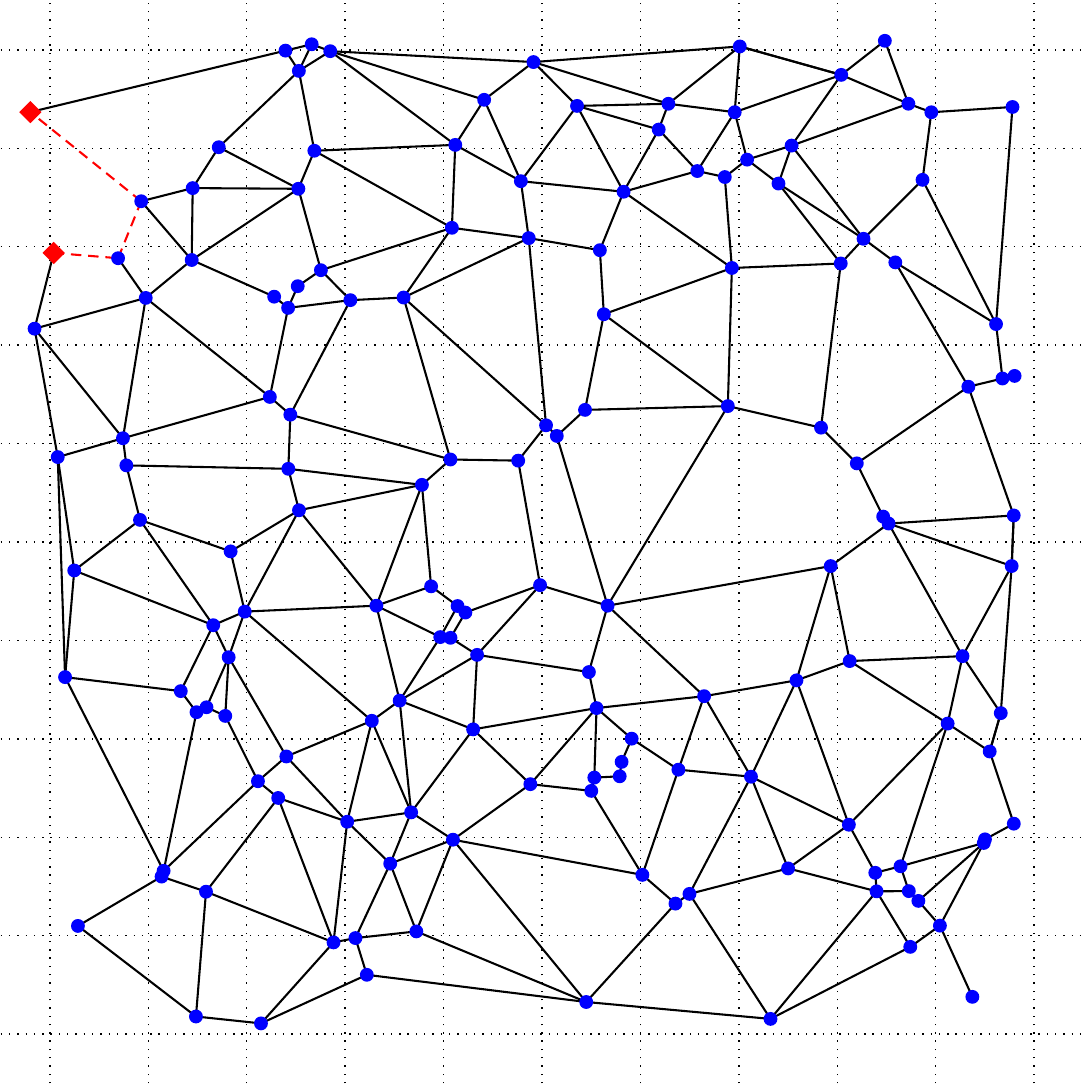}}
		\caption{The spanner generated by \texttt{BHS18} on the pointset shown in Fig.~\ref{fig:n150};  degree: $6$, stretch factor: $1.879749$}
		\label{fig:demo-BHS18}
	\end{minipage}
	\hfill
	\begin{minipage}[h]{0.46\linewidth}
		\centering
		\resizebox{\linewidth}{!}{\includegraphics{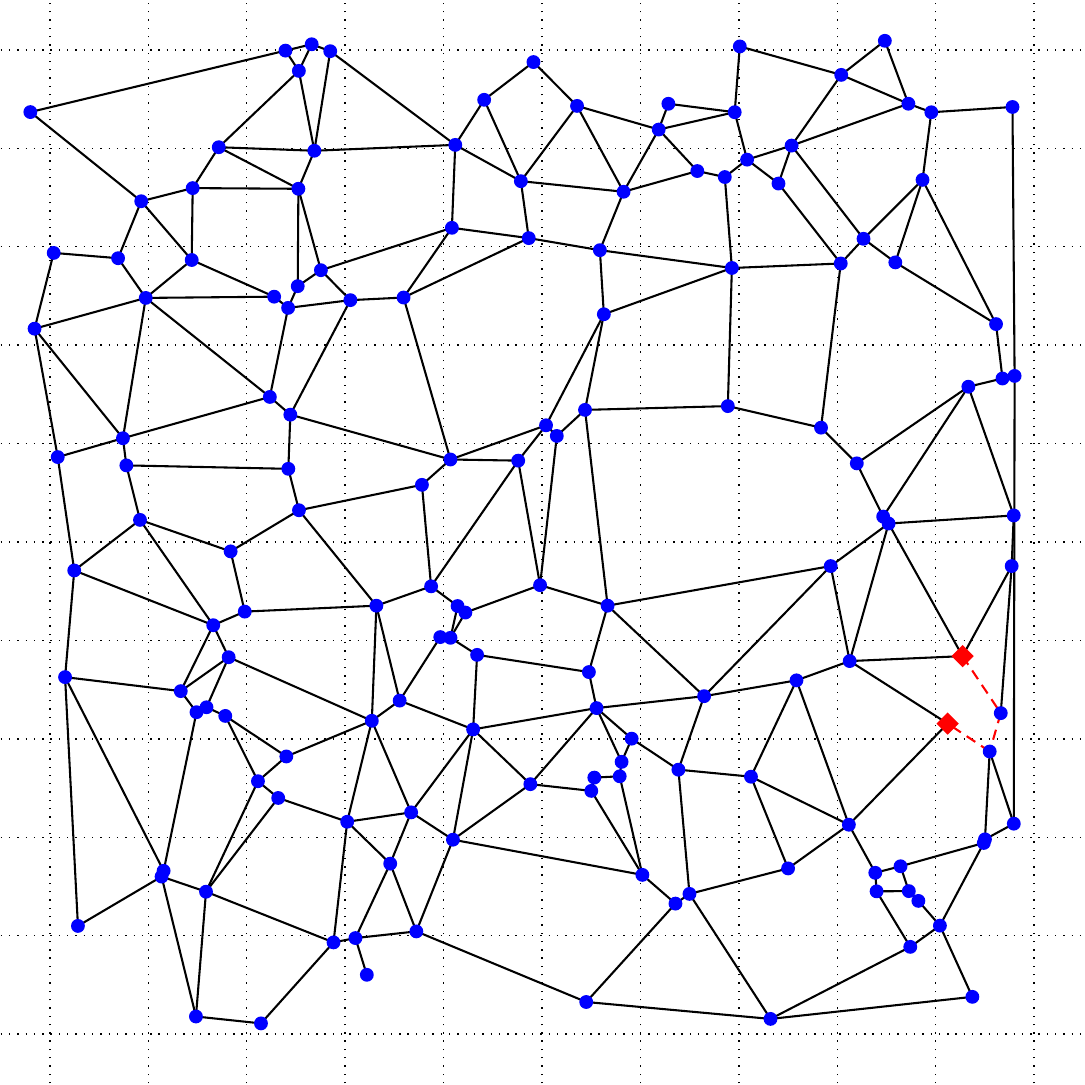}}
		\caption{The spanner generated by \texttt{BCC12-7} on the pointset shown in Fig.~\ref{fig:n150};  degree: $6$, stretch factor: $2.302473$}
		\label{fig:demo-BCC12-7}
	\end{minipage}
\end{figure}

\begin{figure}[h]
	\begin{minipage}[h]{0.46\linewidth}
		\centering
		\resizebox{\linewidth}{!}{\includegraphics{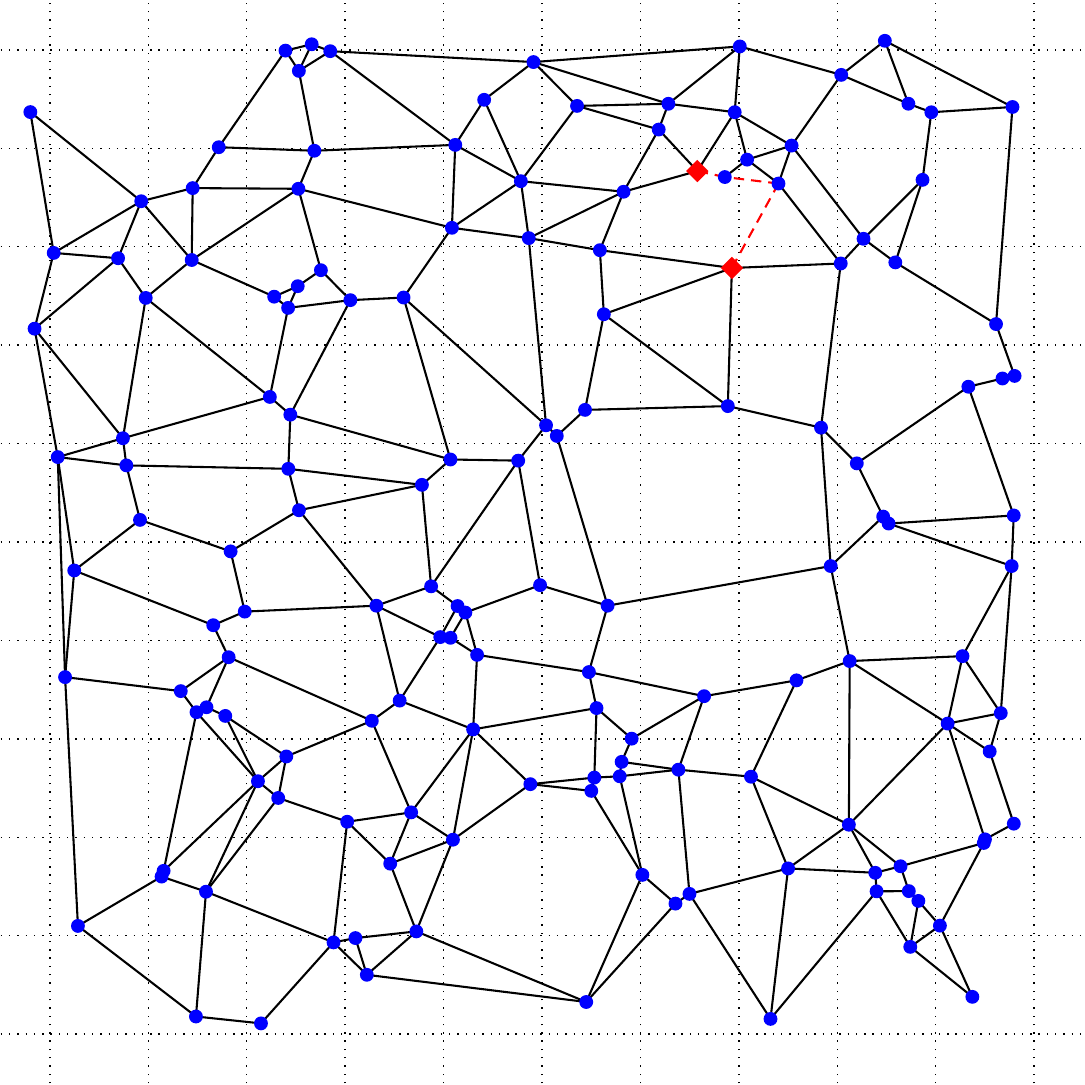}}
		\caption{The spanner generated by \texttt{BCC12-6} on the pointset shown in Fig.~\ref{fig:n150};  degree: $6$, stretch factor: $1.735716$}
		\label{fig:demo-BCC12-6}
	\end{minipage}
	\hfill
	\begin{minipage}[h]{0.46\linewidth}
		\centering
		\resizebox{\linewidth}{!}{\includegraphics{Demo-RPIS-150-BGHP2010.pdf}}
		\caption{The spanner generated by \texttt{BGHP10} on the pointset shown in Fig.~\ref{fig:n150};  degree: $6$, stretch factor: $1.817045$}
		\label{fig:demo-BGHP10}
	\end{minipage}

	\begin{minipage}[h]{0.46\linewidth}
		\centering
		\resizebox{\linewidth}{!}{\includegraphics[scale=0.75]{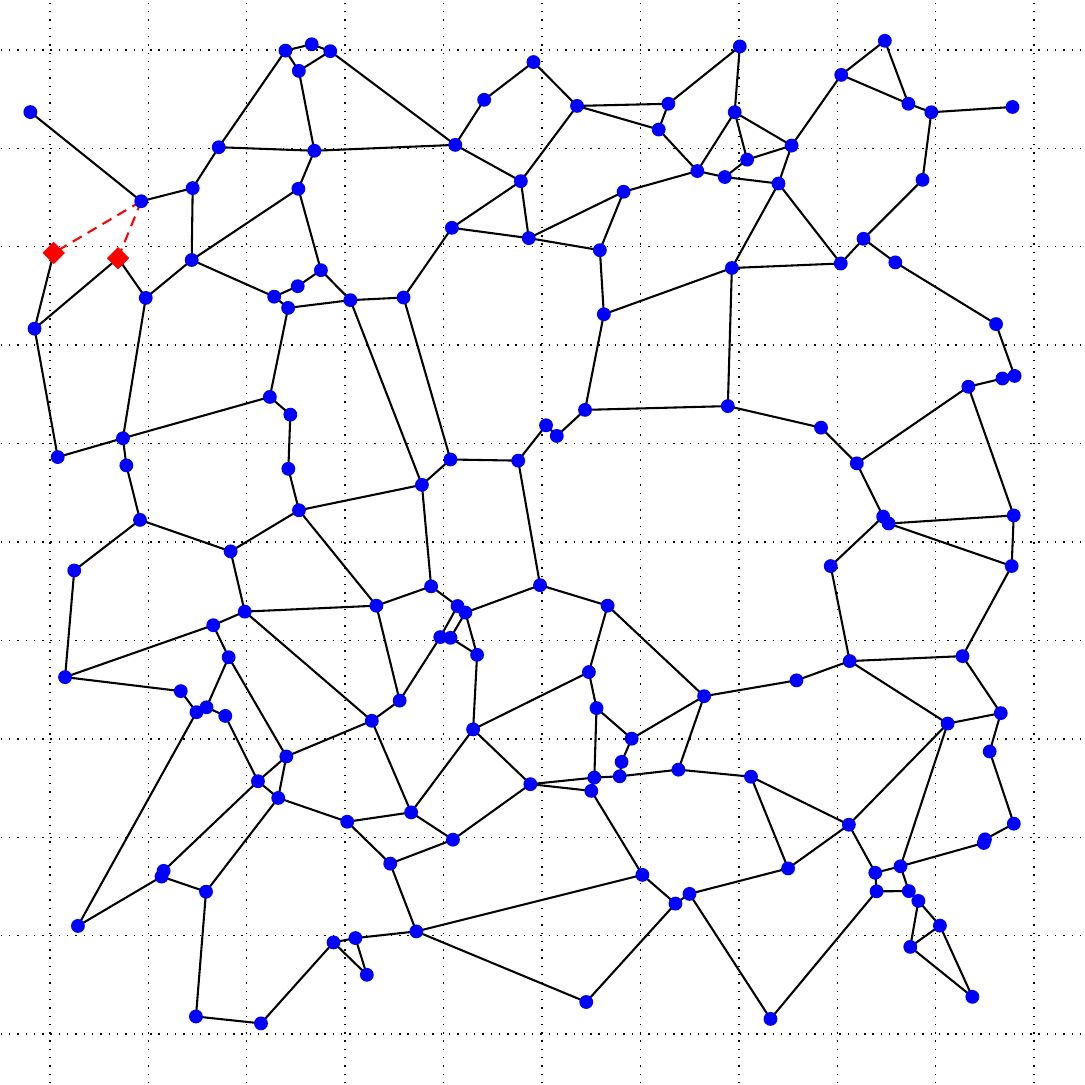}}
		\caption{The spanner generated by \texttt{BKPX15} on the pointset shown in Fig.~\ref{fig:n150};  degree: $4$, stretch factor: $2.525204$}
		\label{fig:demo-BKPX15}
	\end{minipage}
	\hfill
	\begin{minipage}[h]{0.46\linewidth}
		\centering
		\resizebox{\linewidth}{!}{\includegraphics[scale=0.75]{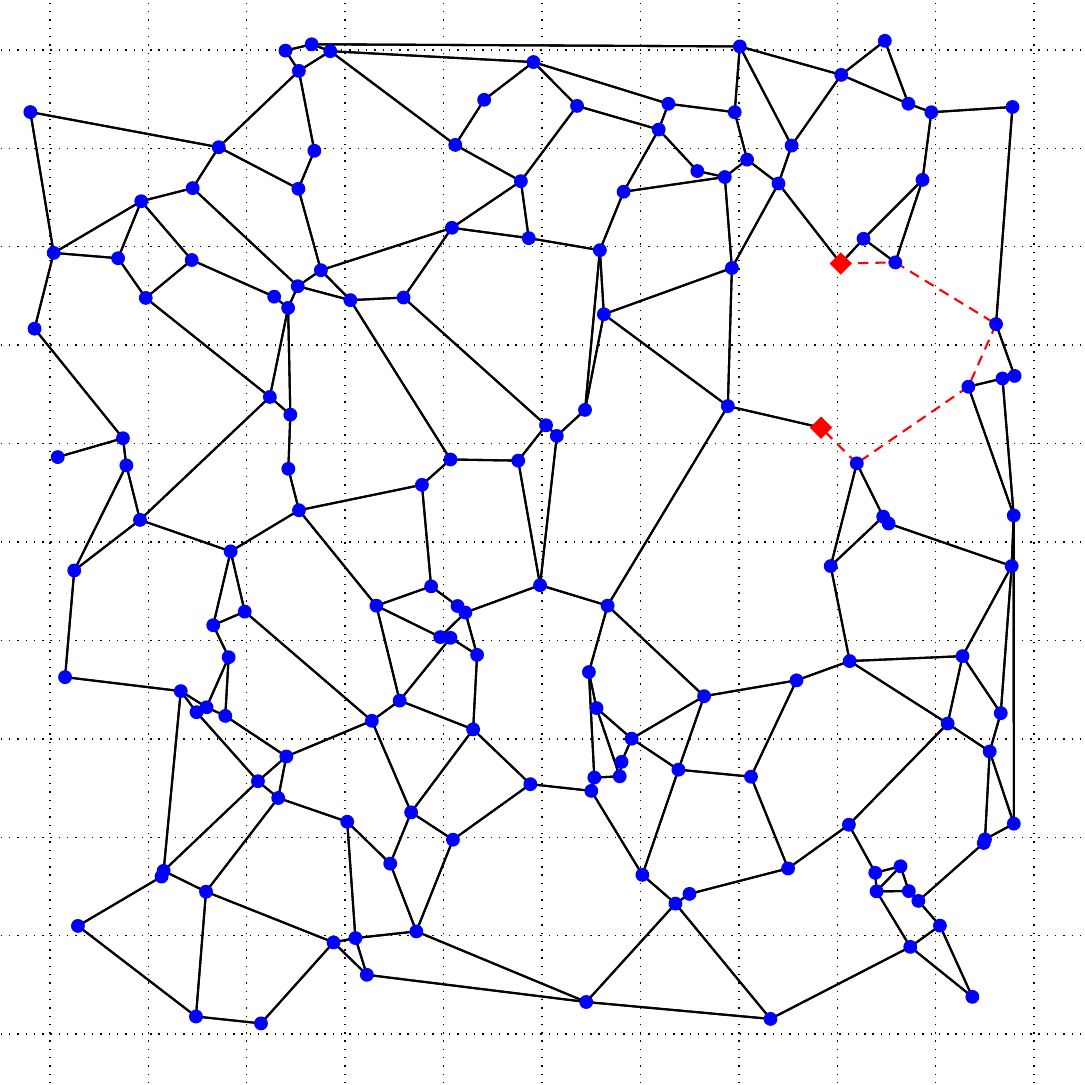}}
		\caption{The spanner generated by \texttt{KPT17} on the pointset shown in Fig.~\ref{fig:n150};  degree: $4$, stretch factor: $2.582846$}
		\label{fig:demo-KPT17}
	\end{minipage}
\end{figure}

\clearpage

\clearpage

\subsection{A counterexample for \texttt{BCC12-6}}
\label{chap:bcc_bad}

\begin{figure}[h]
	\centering
	\resizebox{\linewidth}{!}{
		\includegraphics{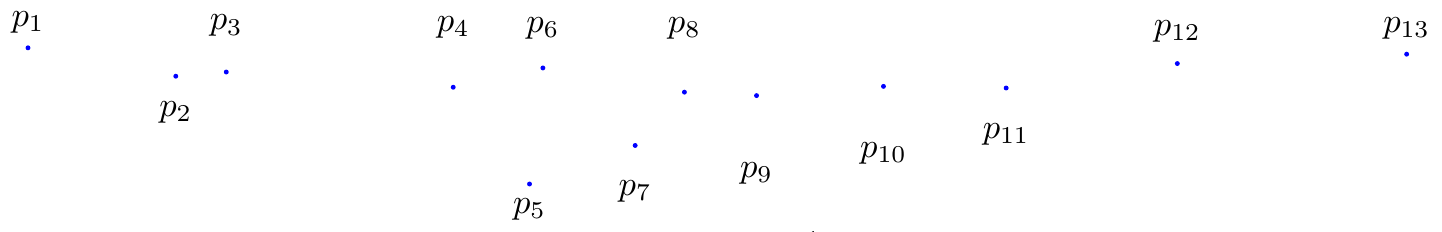}
	}
	\caption{A set $P$ of $13$ points $p_1,\ldots,p_{13}$. 	$p_1$:  $(-4.98845, 0.22414)$, 
		$p_2$: $(-4.23759, 0.08)$,
		$p_3$: $(-3.98106, 0.10125)$,
		$p_4$: $(-2.82831, 0.02396 )$, 
		$p_5$: $(-2.44066, -0.46761)$,
		$p_6$: $(-2.37275, 0.12191)$,
		$p_7$: $(-1.90395, -0.27187)$, 
		$p_8$: $(-1.65373, -0.00109)$, 
		$p_9$: $(-1.28739, -0.01854)$,
		$p_{10}$: $(-0.642516, 0.02836)$,
		$p_{11}$: $(-0.019359, 0.02)$,
		$p_{12}$: $(0.850154, 0.14431)$, 
		$p_{13}$: $(2.01517, 0.19194)$}
	\label{fig:bcc_points}
\end{figure}

In the following, we present an $13$-element pointset on which \texttt{BCC12-6} fails to construct a degree-$6$ plane spanner. Refer to Fig.~\ref{fig:bcc_points} for the pointset.

\begin{figure}[H]
	\centering
	\resizebox{\linewidth}{!}{
		\includegraphics{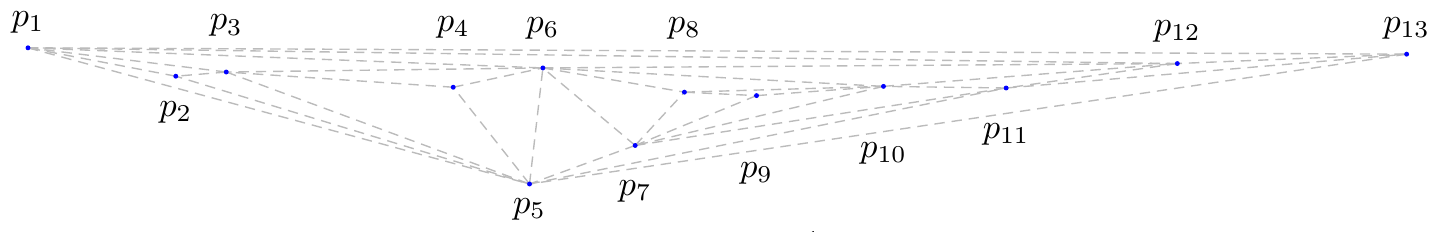}
	}
	\caption{The $L_2$-Delaunay triangulation of $P$.}
	\label{fig:bcc_delaunay}
\end{figure}

\begin{figure}[H]
	\centering
	\resizebox{\linewidth}{!}{
		\includegraphics{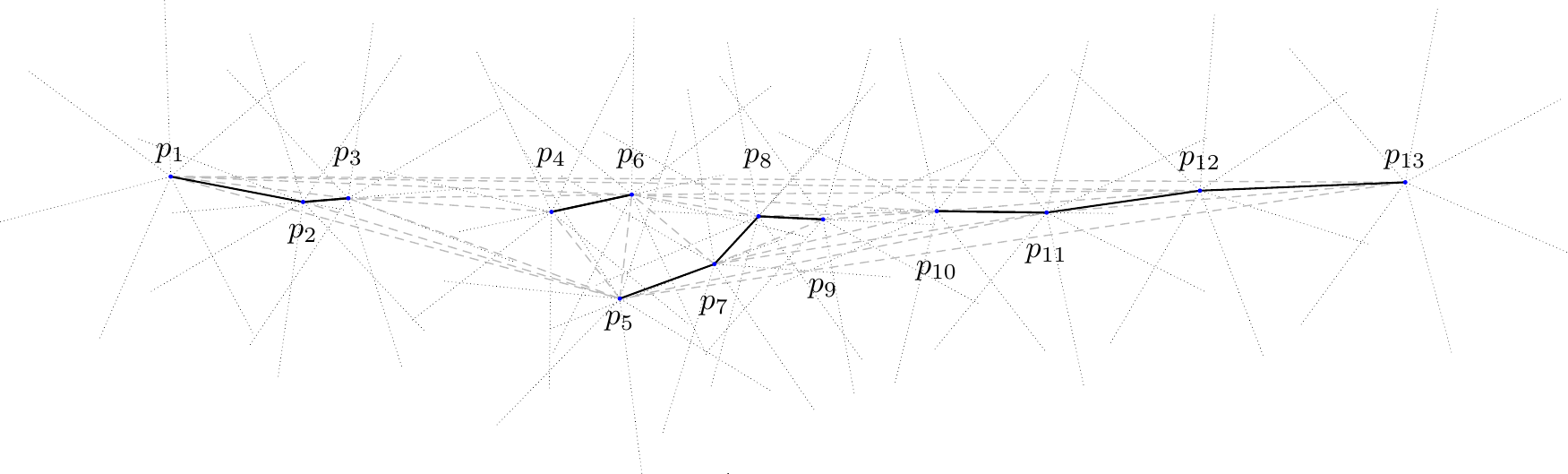}
	}
	\caption{The cones (dotted) of each point in $P$ with $\alpha = 2\pi / 7$, oriented by the shortest edge incident on that point (bold).}
	\label{fig:bcc_cones}
\end{figure}

First, \texttt{BCC12-6} creates the $L_2$-Delaunay triangulation of $P$ and initializes $7$ cones around every $p_i$, oriented such that the shortest edge incident on $p_i$ falls on a boundary. See Figs.~\ref{fig:bcc_delaunay} and \ref{fig:bcc_cones}.


\begin{figure}[p]
	\centering
	\resizebox{\linewidth}{!}{
		\includegraphics{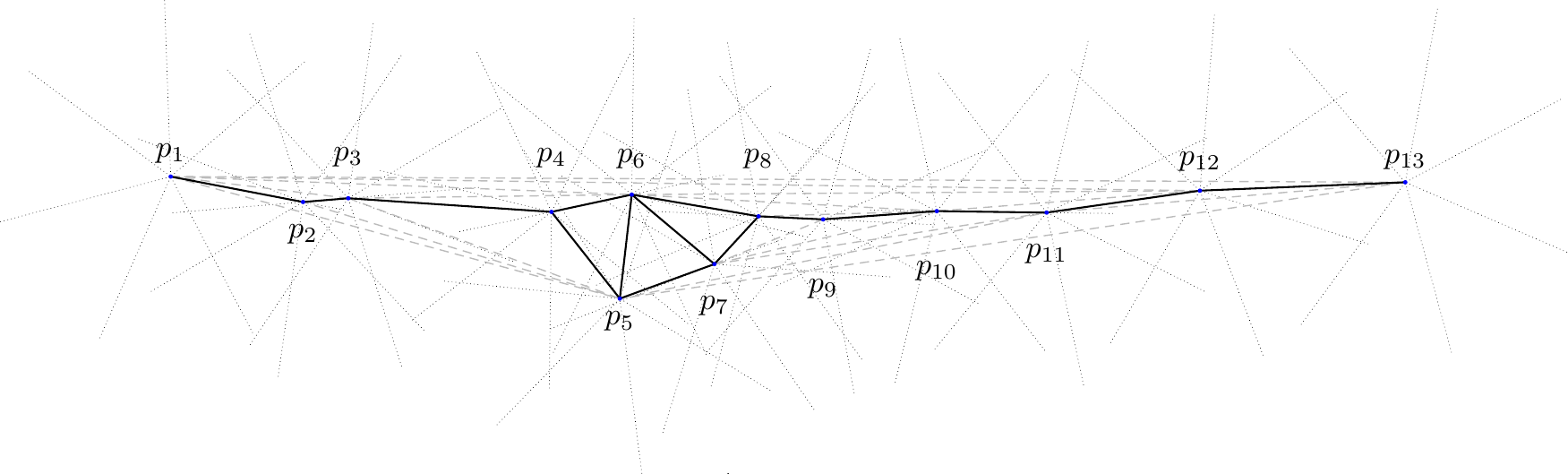}
	}
	\caption{Edges added by the main portion of \texttt{BCC12} (excluding calls to subroutine \texttt{Wedge}$_6$).}
	\label{fig:bcc_shortedges}
\end{figure}

\begin{figure}[p]
	\centering
	\resizebox{\linewidth}{!}{
		\includegraphics{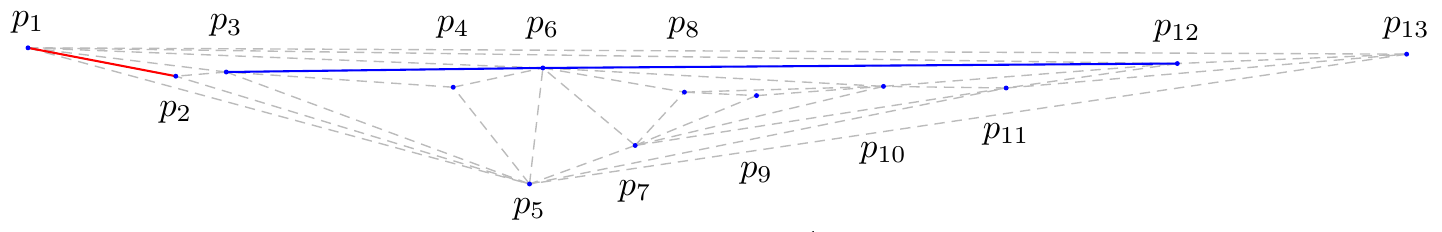}
	}
	\caption{ The edge $p_1p_2$ (shown in red) is added during the main portion of the algorithm and the call to \texttt{Wedge}$_6$($p_1,p_2$) adds the two blue edges $p_3p_6$ and $p_6p_{12}$.}
	\label{fig:bcc_wedge_p1}
\end{figure}

\begin{figure}[p]
	\centering
	\resizebox{\linewidth}{!}{\includegraphics{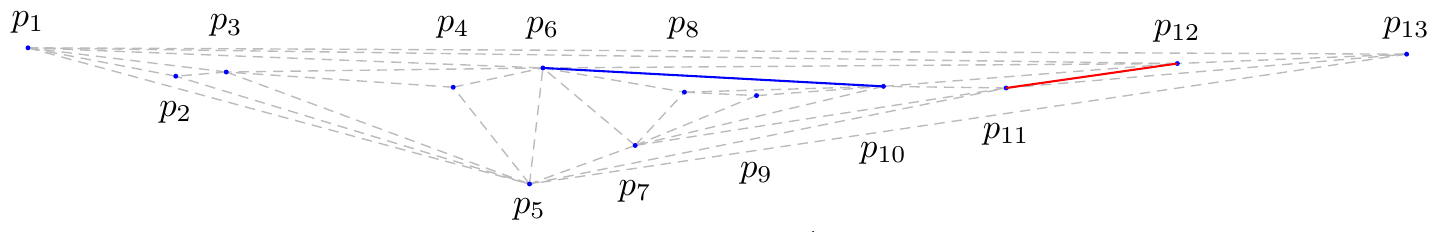}}
	\caption{The edge $p_{12}p_{11}$ (shown in red) is added during the main portion of the algorithm and the call to \texttt{Wedge}$_6$($p_{12},p_{11}$) adds the blue edge $p_6p_{10}$.}
	\label{fig:bcc_wedge_p12}
\end{figure}

\begin{figure}[p]
	\centering
	\resizebox{\linewidth}{!}{\includegraphics{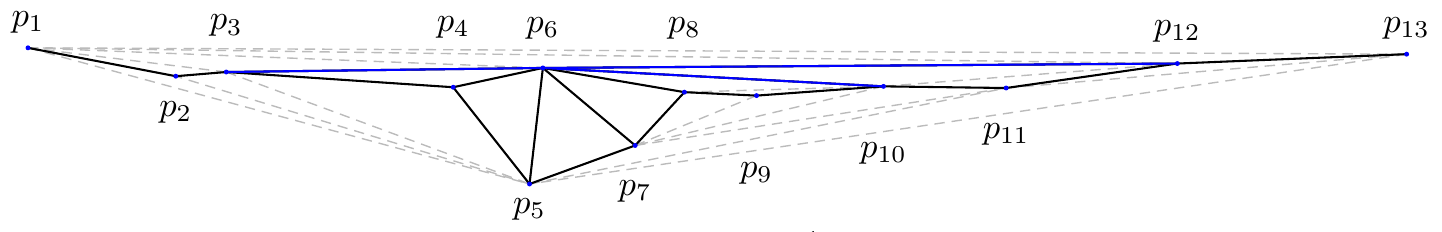}}
	\caption{The resulting graph on $P$ is a degree-$7$ plane spanner due to $p_6$ whose degree is exactly $7$. Note that this graph contains the edges shown in Fig.~\ref{fig:bcc_shortedges} alongwith the blue edges shown in Fig.~\ref{fig:bcc_wedge_p1} and \ref{fig:bcc_wedge_p12}. }
	\label{fig:bcc_output}
\end{figure}

\noindent
Next, in Fig~\ref{fig:bcc_shortedges}, we show the edges added by the main portion of the algorithm (excluding the edges added by $\texttt{Wedge}_6$ calls).  Only $\texttt{Wedge}_6(p_1,p_2)$ and $\texttt{Wedge}_6(p_{12},p_{11})$ calls add new edges to $E^*$ and thus to the final spanner as well. The former call adds the two edges $p_3p_6, p_6p_{12}$ (see Fig.~\ref{fig:bcc_wedge_p1}) and the later call adds the edge $p_6p_{10}$ (see Fig.~\ref{fig:bcc_wedge_p12}). The final spanner is shown in Fig.~\ref{fig:bcc_output}. Note that $p_6$ has degree $7$ in the spanner which violates the degree requirement of the spanners produced by \texttt{BCC12-6}.

\end{document}